\documentclass{book}
\usepackage{hyperref}
\usepackage{subfig}
\usepackage[rightcaption]{sidecap}					
\usepackage{rotating}
\usepackage{color}
\usepackage[usenames,dvipsnames]{xcolor}
\usepackage[margin=1.in]{geometry}

\usepackage{booktabs,amsmath,siunitx}
\usepackage{chngcntr}
\usepackage[justification=justified, singlelinecheck=false]{caption}

\def\d{\mathrm{d}}
\def\e{\mathrm{e}}
\def\nm{\mathrm{nm}}
\def\um{\mu\mathrm{m}}
\def\pN{\mathrm{pN}}
\def\nN{\mathrm{nN}}
\def\mN{\mathrm{mN}}
\def\uM{\mu{\rm M}}
\def\s{{\rm s}}
\def\Hz{{\rm s}^{-1}}

\newcommand{\blue}[1] {\textcolor{blue}{#1}}
\newcommand{\green}[1] {\textcolor{PineGreen}{#1}}

\providecommand\newthoughat[1]{%
   \addvspace{1.0\baselineskip plus 0.5ex minus 0.2ex}%
   \noindent\textsc{#1} 
}

\begin{document}


\title{Effect of curvature and normal forces on motor regulation of cilia}
\author{Pablo Sartori}
\maketitle



{\bf Disclaimer.} The following document is a preprint of the PhD dissertation of Pablo Sartori, defended in 2015 at the Technische Universit\"at Dresden. There are only minor differences between this preprint and the final version (formatting changes, addition of acknowledgements, resizing of figures, and an additional appendix regarding the fitting procedure). 
\clearpage
\begin{center}
{\large\it Abstract}
\end{center}

\textbf{Cilia are ubiquitous organelles involves in eukaryotic motility.} They are long, slender, and motile protrusions from the cell body. They undergo active regular oscillatory beating patterns that can propel cells, such as the algae {\it Chlamydomonas}, through fluids. When many cilia beat in synchrony they can also propel fluid along the surfaces of cells, as is the case of nodal cilia.

The main structural elements inside the cilium are microtubules. There are also molecular motors of the dynein family that actively power the motion of the cilium. These motors  transform  chemical energy in the form of  ATP into mechanical forces that produce sliding displacement between the microtubules. This sliding is converted to bending by constraints at the base and/or along the length of the cilium. Forces and displacements within the cilium can regulate dyneins and  provide  a feedback mechanism: the dyneins generate forces, deforming the cilium; the deformations, in turn, regulate the dyneins. This feedback is believed to be the origin of the coordination of dyneins in space and time which underlies the regularity of the beat pattern.
\newline

\textbf{Goals and approach.} While the mechanism by which dyneins bend the cilium is  understood, the feedback mechanism is much less clear. The two key questions are: which forces and displacements are the most relevant in regulating the beat? and how exactly does this regulation occur?

In this thesis we develop a framework to describe the spatio-temporal patterns of a cilium with different mechanisms of motor regulation. Characterizing and comparing the predicted shapes and beat patterns of these different mechanisms to those observed in experiments provides us with further understanding on how dyneins are regulated. 
\newline

\textbf{Results in this thesis.}  Chapters 1 and 2 of this thesis are dedicated to introduce cilia, chapters 3-6 contain the results, and chapter 7 the conclusions.

In chapter 1 we introduce the structure of the cilium, and discuss different possible regulatory mechanisms which we will develop along the thesis. Chapter 2 contains a quantitative description of the ciliary beat observed in experiments involving {\it Chlamydomonas}.

In chapter 3 we develop a mechanical theory for planar ciliary beat in which the cilium  can bend, slide and compress normal to the sliding direction. As a first application of this theory we analyze the role of sliding cross-linkers in static bending.

In chapter 4 we introduce a mesoscopic description of molecular motors, and show that regulation by sliding, curvature or normal forces can produce oscillatory behavior. We also show that motor regulation by normal forces and curvature bends cilia into circular arcs, which is in agreement with experimental data.

In chapter 5 we use analytical and numerical techniques to study linear and non-linear symmetric beats. We show that there are fundamental differences between patterns regulated by sliding and by curvature: the first only allows wave propagation for long cilia with a basal compliance, while the second lacks these requirements. Normal forces can only regulate dynamic patterns in the presence of an asymmetry, and the resulting asymmetric patterns are studied in chapter 6.

In chapter 6 we study asymmetric beats, which allow for regulation by normal deformations of the cilium. We compare the asymmetric beat from {\it Chlamydomonas} wild type cilia and the beat of a symmetric mutant  to the theoretically predicted ones. This comparison suggests that sliding forces cannot regulate the beat of these short cilia, normal forces can regulate them for the wild type cilia, and curvature can regulate them for wild type as well as for the symmetric mutant. This makes curvature control the most likely regulatory mechanism for the {\it Chlamydomonas} ciliary beat.

\tableofcontents





	
\chapter{Introduction}

\newthoughat{Cilia are ubiquitous} eukariotic organelles. Diverse in scale and function, they are involved in a number of different motile tasks, yet at the core of all cilia lies a common structure: the axoneme. Imaging of bending cilia, of their internal structure, and biochemical experiments, have given rise to models of the ciliary structure, as well as suggestions for their functioning. In this section we give an overview of the internal components of the cilium, its structure, and some of the suggested models for their dynamic regulation.

\section{What cilia are and where to find them}
\label{sec:table}
Cilia are long thin organelles which protrude from eukaryotic cells. They are motile structures fundamental for the functioning of cells. Indeed, immotile cilia are related with  problems in embryonal development, and lead to human diseases such as primary ciliary diskinisia \cite{afzelius_human_1976}. Although with differences among species, the core structure of cilia is highly conserved, yet they are involved in a number of diverse functions.

\begin{figure}[!ht]
\includegraphics[width=6in]{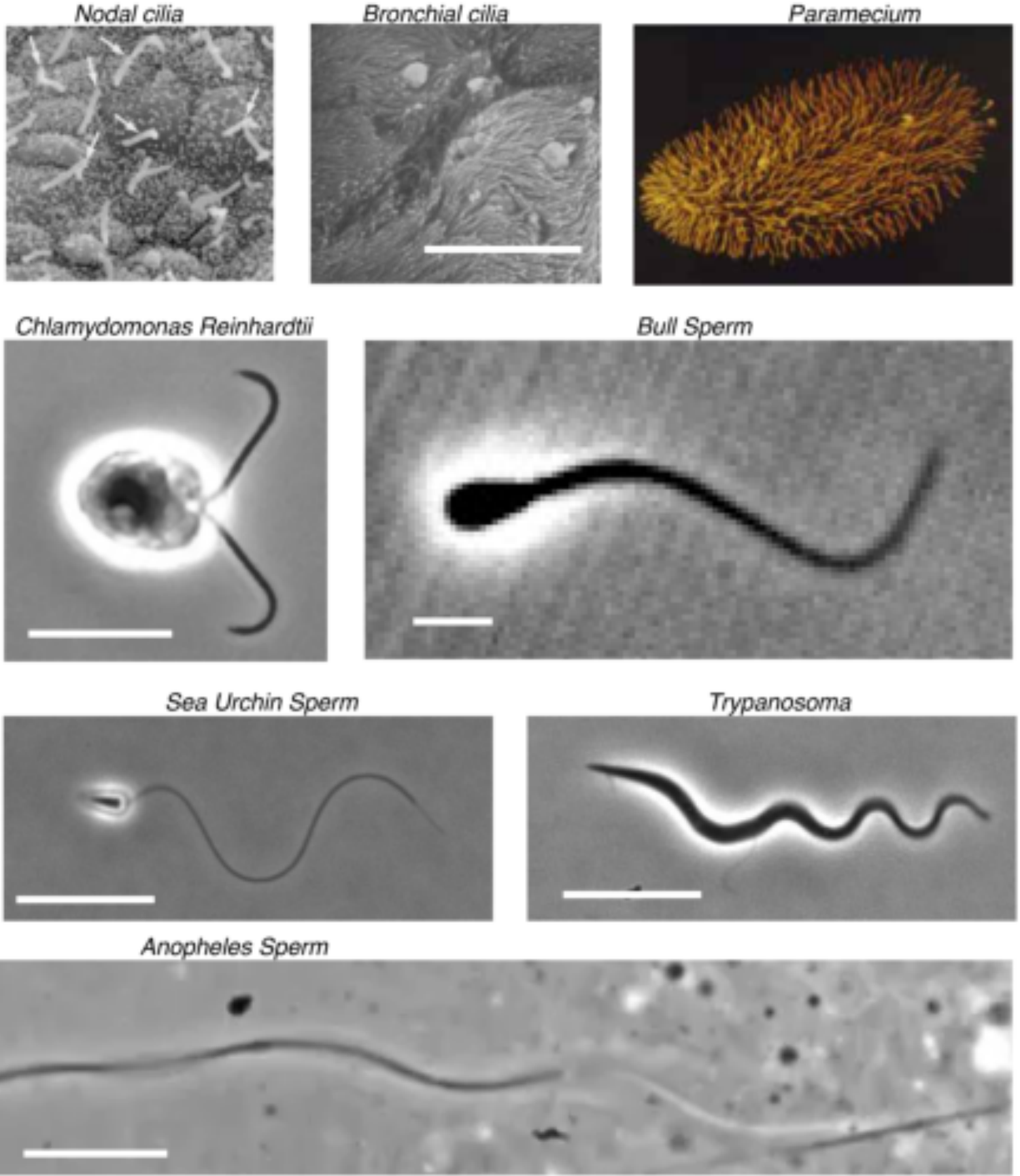}	
 	\caption{\textbf{Examples of cilia creating flows and self-propelling micro-swimmers.}  Nodal cilia \cite{nonaka_randomization_1998} and cilia in bronchi \cite{tsuda_optimum_1977} produce fluid flows. In the other examples cilia propel microorganisms. These micro-swimmers can have one cilium (as in the sperm examples shown), two (as in the case of {\it Chlamydomonas Reinhardtii}), or many, as in the case of  {\it Paramecium} \cite{CNRS}. Bars correspond to $10\,\um$. The figure is adapted from \cite{geyer_characterization_2013}.}
 	\label{fig:intro_flagellates}
\end{figure}

Cilia can exhibit a variety of periodic ondulatory motions in the presence of chemical energy. These {\it beat patters} are involved in fluid flow and propulsion of micro-swimmers. Examples of cilia involved in fluid flow are the nodal cilia, responsible for the breaking of left-right symmetry in the development of vertebrates \cite{nonaka_randomization_1998}. Another example are the cilia in the mammalian lungs, responsible for the flow of mucus \cite{sleigh_propulsion_1988} (see Fig.~\ref{fig:intro_flagellates}).

Many eukaryotic micro-swimmers use cilia to propel themselves.  Such microorganisms can have as many as hundreds of cilia (as in the case of Paramecium, see Fig.~\ref{fig:intro_flagellates}), but in this thesis we will focus on microorganisms with one or two cilia.  Examples of cells with one motile cilia are the sperm of sea urchin, bull or anopheles (see Fig.~\ref{fig:intro_flagellates}). An example of an organism with two cilia is the unicellular algae {\it Chlamydomonas}, the main model system studied in this thesis (see Fig.~\ref{fig:intro_flagellates}). 

\begin{table}
\begin{center}
  \begin{tabular}{ l  c  c  c  c }
  \toprule
  Organism & Length ($\mu{\rm m}$) & Frequency (${\rm s}^{-1}$)& Wave number & Amplitude ($\mu{\rm m}$)   \\
     \toprule 
     {\it Chlamydomonas}& 10 & 50 & 1 &1\\
     \midrule
     {\it Sea Urchin Sperm}& 60 & 30 & 2 & 5 \\
     \midrule
     {\it Trypanosome}& 30 & 20 & 3 &0,6\\
     \midrule
     {\it Bull Sperm} & 50 & 20 & 1  & 0,5\\
     \midrule
     {\it Anopheles Sperm} &$>$80 &  ? & $>$3 & $>$1,5  \\
    \bottomrule
\end{tabular}
\label{tab:flagellates}
\caption{\textbf{Characteristics of the beat patterns of some flagellates.} There is great variability among the beat of flagellates. While {\it Chlamydomonas} has a short and fast  cilium, {\it Bull Sperm} has a long slow  one.\label{tab:flagellates}}
\end{center}
\end{table}

There is great variability among the ciliary properties of the micro-swimmers in Fig.~\ref{fig:intro_flagellates}. Among them these cilia differ in size, beat frequency, wave number, and amplitudes. Estimates of these properties for the examples in Fig.~\ref{fig:intro_flagellates} are in Table~\ref{tab:flagellates}. But however different their beating properties are, underlaying all of them reside the same structural elements, which we review in the next section.


\section{Inside a cilium: the 9+2 axoneme}
\label{sec:inside}
\begin{figure}[!h]
\includegraphics[width=6in]{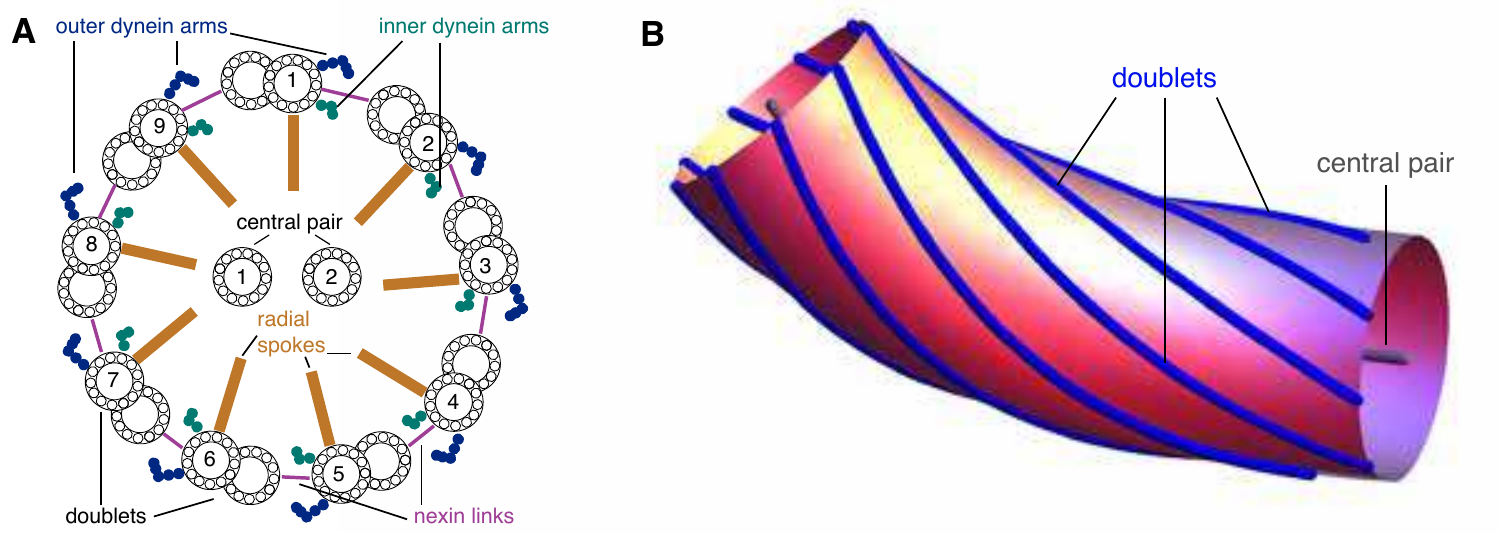}	
 	\caption{\textbf{The 9+2 axoneme.} \textbf{A}. Schematic cross-section  of the 9+2 structure showing the doublet microtubules, the  inner and outer dynein arms, the nexin links, the radial spokes and the central pair. \textbf{B}. Schematic of a bent and twisted axoneme showing the doublets (blue) and  central pair (gray). Note the relative sliding between the doublets.}
 	\label{fig:axoneme}
\end{figure}

In the core of cilia lies a cytoskeletal cylindrical structure called the axoneme. The axoneme is a cylindrical bundle of 9 parallel microtubules doublets. At its center there are 2 additional  microtubules, the central pair (see Fig.~\ref{fig:axoneme} B). They extend from the basal end, attached to the cell body, to the distal end. While there are other possible structures~\cite{feistel_three_2006, phillips_exceptions_1969}, the 9+2 axoneme is a highly  evolutionary conserved structure \cite{carvalho-santos_tracing_2011}. Here we will focus on the {\it Chlamydomonas} axoneme which is an example of the 9+2 arrangement.

The doublets and the central pair are connected by the radial spokes (see Fig.~\ref{fig:axoneme} A), responsible to keep the radius of the diameter at $\approx200{\rm nm}$ \cite{nicastro_molecular_2006}. The doublets are parallel to each other and from the radius we estimate a spacing of   $\approx30\,\nm$. Each doublet is connected to its near-neighbors by cross-linkers such as nexin, see section \ref{sec:nexin}. These   provide a resistance to the relative sliding of doublets, and to changes in the spacing between them. Doublets are also connected by dynein molecular motors \ref{sec:dyneins}, which create active sliding forces when ATP is present. All these elements are present in a highly structured manner in the axoneme, repeating along the long axis of the cylinder with a period of $\approx96\,\nm$. In the following we describe in more detail the main characteristics of the axonemal elements.

\subsection{Scaffolding elements: doublet microtubules and the central pair}
\label{sec:doublets}
\label{sec:centralpair}

\begin{figure}[!ht]
\includegraphics{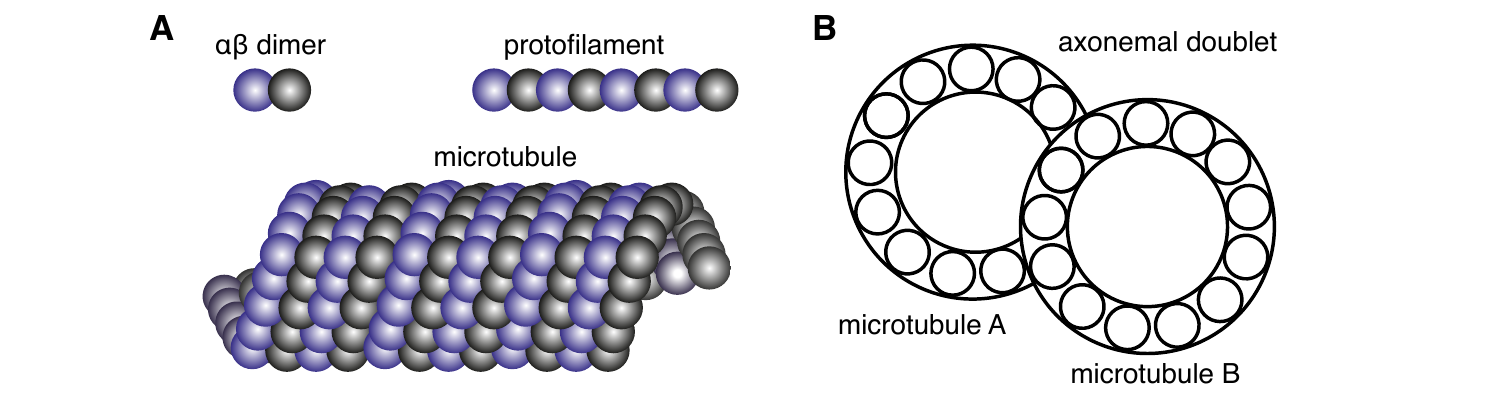}	
 	\caption{\textbf{Microtubules and doublets.} \textbf{A.} Each microtubule is composed of thirteen protofilaments, shifted with respect to each other. A protofilament is a linear assembly of $\alpha\beta$ dimers. \textbf{B.} Cross-section of a doublet. Each doublet consists of two microtubules, one complete (microtubule B, with thirteen protofilaments) and one incomplete (microtubule A, with ten protofilaments).}
 	\label{fig:microtubule}
\end{figure}

The main structural elements of the axoneme are microtubules, present in the doublets as well as in the central pair. Microtubules are filamentous protein complexes which have scaffolding functions in the eukaryotic cytoskeleton. Structurally, they are hollow cylinders with a diameter of $24\,\nm$ composed of thirteen protofilaments (see Fig.~\ref{fig:microtubule}). The basic structural element of a protofilament is the $\alpha\beta$ tubulin dimer, with a length of roughly $8\nm$. Because this dimer is polar the mircotubules are polar as well, which grow much faster from the $+$ end than from their $-$ end \cite{walker_dynamic_1988}.

The elastic properties of microtubules can be accurately accounted for using a model of incompressible semi-flexible polymers with a bending stiffness of $\sim 24\,\pN\cdotp\um^2$ \cite{howard_mechanics_2001}. While recent experiments suggest that compressibility and shearing may play a role in determining this stiffness \cite{taute_microtubule_2008, pelle_mechanical_2009}, in this thesis we will consider microtubules as inextensible and non shearing.

In the axoneme microtubules appear in the central pair and in the doublets.  The central pair consists of two singlet microtubules (i.e., with all 13 protofilaments, see Fig.~\ref{fig:microtubule}) longitudinally connected between them. But while its basic component, namely the microtubules, have well known properties, those of  the central pair are complex and elusive. For example, there is evidence suggesting that the central pair shows a large twist when extruded from the axoneme  \cite{mitchell_bend_2004}. The orientation of the central pair is used as a reference to number the doublets, as in Fig. \ref{fig:axoneme}.

The doublets are composed of one A-microtubule and one B-microtubule. The A-microtubule has all 13 protofilaments, while the B-microtubule has only 10 and a half protofilaments (see Fig.~\ref{fig:microtubule}). The stiffness of microtubule doublets has not been directly measured, but simple mechanical considerations suggest that they are roughly three times that of a single microtubule, and so we estimate their bending rigidity as $\sim 70\,\pN\,\um^2$ \cite{howard_mechanics_2001}.

The doublets and the central pair are the main structural elements involved in determining the bending stiffness of the axoneme. Using mechanical considerations one can estimate the stiffness of the axonemal bundle as being 30 times higher as that of a single microtubule, which yields  {$\approx600\,\pN\cdotp\um^2$} for the stiffness of the axoneme \cite{howard_mechanics_2001}. This number can be compared with several direct measurements of axonemal stiffness. While these vary with the experimental conditions (such as the presence of vanadate or ATP), for rat sperm \cite{schmitz-lesich_direct_2004} as well as sea urchin sperm \cite{s_flexural_1994} the value of {$\approx 4\cdotp10^3\,\pN\cdotp\um^2$}  has been measured, in good agreement to the previous estimate.

\subsection{Active force generating elements: dynein motors}
\label{sec:dyneins}
\begin{SCfigure}
\includegraphics{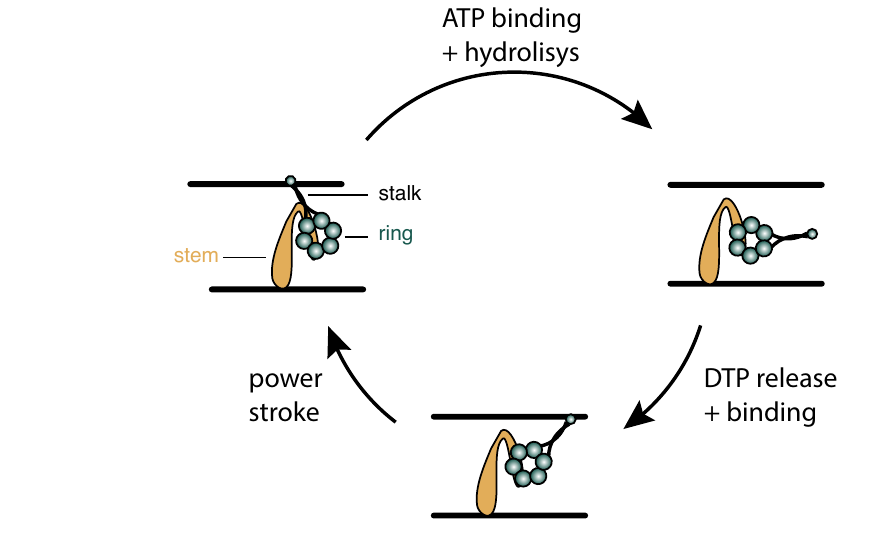}	
 	\caption{\textbf{Mechanochemical cycle of dynein.} A dynein motor is composed of a ring with 6AAA domains, a stem (in yellow) and a stalk with a MT binding domain at its end. ATP binding enhances the  detachment of dynein's stalk from the microtubule, and its hydrolysis locks it in a strained state in which the ring is rotated. DTP release aids the dynein to bind to the microtubule. Finally, once bound, the power stroke occurs, leaving the bound motor in a state with high ATP affinity.}
 	\label{fig:dynein}
\end{SCfigure}

The ciliary beat is powered by axonemal dyneins, which are a family of molecular motors (see Fig.~\ref{fig:dynein}). They convert chemical energy into mechanical work, with one full mechano-chemical cycle corresponding to the hydrolysis of one ATP molecule.  The {\it Chlamydomonas} axoneme in particular contains 14 different types of dyneins and has a total of $\sim10^4$ of these motors \cite{witman_chlamydomonas_2009} over its length of roughly $10\,\um$.  And while these dyneins can have diverse properties (such as different ATP affinities, gliding speeds, or capability of generating torques \cite{kagami_translocation_1992,smith_microtubule_1991}), we focus here on their generic properties.

Dyneins are periodically distributed along the nine microtubule doublets, with their stem (yellow in Fig.~\ref{fig:dynein}) rigidly attached to the A-microtubule, and their stalks being briefly in contact with the adjacent B-microtubule during the power stroke process \cite{howard_mechanics_2001}. Stem and Stalk are connected by a 6AAA ring, and all together the weight of a motor domain is roughly $\sim300\,{\rm kDa}$ \cite{schmidt_insights_2012,johnson_structure_1983}. Their size is roughly $30\,\nm$, and the their structure is sketched in Fig.~\ref{fig:dynein}  \cite{burgess_dynein_2003, sale_substructure_1985}.

It is known that during their power stroke, dyneins generate forces that tend to slide the axonemal doublets with respect to each other. For instance, after a protease treatment of the axoneme the presence of ATP makes the doublets slide apart \cite{summers_adenosine_1971}. Finally it is  worth noting that it has been shown  {\it in vitro} that motors such as dynein and kinesin are capable of generating oscillatory behaviors reminiscent to beat patterns in the presence of ATP~\cite{shingyoji_dynein_1998, sanchez_cilia-like_2011}. All this evidence supports the idea that dyneins generate sliding forces which regulate the beat pattern of the axoneme.


Electron micrographs have shown that a motion of the tail of up to $15\,\nm$ may occur \cite{burgess_dynein_2003}. As an upper limit  to force generation we can consider that all  the  energy contained in one ATP molecule (which is $ \sim25\,k_{\rm B}T\approx100\, \pN\cdotp \nm$, \cite{howard_mechanics_2001}) is invested in this motion, and thus we estimate a force of $\frac{100 \,\pN \cdotp\nm}{15\,\nm}=6.7\,\pN$. On the other hand, indirect estimates have resulted in motor forces of up to $\sim5\,\pN$\cite{lindemann_structural-functional_2003}, suggesting that axonemal dynein can generate as much force as cytoplasmic dynein \cite{gennerich_force-induced_2007} or kinesin \cite{carter_mechanics_2005,svoboda_force_1994}.

While the mechano-chemical dynein cycle has not been fully deciphered, much is known about its main steps \cite{johnson_pathway_1983}. ATP binding is known to occur with an affinity of roughly $4.7\,\uM^{-1}{\rm s}^{-1}$ (for ATP concentrations below $60\,\uM$) \cite{porter_transient_1983}, which dramatically accelerates detachment of the dynein tail from the microtubule. After this, ATP is hydrolyzed and a conformational change occurs which renders dynein to its pre-powerstroke configuration. This slow step occurs at roughly $\sim 0.2\,{\rm s}^{-1}$, and is thus the rate-limiting step \cite{king_dyneins:_2011}. ADP increases the affinity of dynein to microtubules (which may also be influenced by additional conformational changes \cite{lin_structural_2014}), leading to a binding state in which the powerstroke occurs, which completes the dynein cycle (see Fig.~\ref{fig:dynein}). Overall the duty ratio of dynein (fraction of cycle time which it spends bound the B-microtubule) has been estimated to be $\sim10\%$ \cite{howard_mechanics_2001}, and it is thus considered a non-processive motor.

\subsection{Passive cross-linkers: nexin and the radial spokes}
\label{sec:nexin}
\label{sec:radialspokes}

Radial spokes and nexin cross-linkers are passive structural elements involved in maintaining the structural integrity of the axoneme. Coarsely, the radial spokes (see Fig.~\ref{fig:axoneme}) help to sustain the radius of the axoneme, while the nexin cross-linkers prevent the doublets from sliding apart.  We now review some of the more intricate details of nexin and radial spokes.

The nature of nexin linkers has long been debated. Indirect evidence suggests that a protein complex must be involved in constraining the sliding of doublets, since protease treatment of axonemes results in telescoping of the doublets when ATP is present \cite{summers_adenosine_1971}. Furthermore, it is believed that such a constraint is fundamental in transforming the sliding dynein forces into axonemal bending, see section \ref{sec:mechanisms}. But while early electro microscopy identified nexin as a separate protein complex \cite{witman_chlamydomonas_1978,hakan_bozkurt_morphology_1993}, recent work suggests that it is part of the dynein regulatory complex \cite{nicastro_molecular_2006,heuser_dynein_2009,bower_n-drc_2013}. This would imply that the cross-link nexin can be closely involved in regulating the behavior of the dynein regulatory complex, and thus the flagellar beat. Furthermore, recently it has been suggested that dynein is linked to the radial spokes \cite{lin_building_2011}. In any case, it is clear that dynein provides passive resistance to sliding, as indeed it's stiffness has been directly measured to be $\sim2.0\,\pN\cdotp\nm^{-1}$ for $1\,\um$ of axoneme. This number increases five fold in the absence of ATP (when dyneins are attached), thus further indicating that dynein and nexin are intimately related may be a highly regulated passive structure  \cite{minoura_direct_1999}. Finally, since it stretches about ten times its equilibrium length, its force-displacement behavior   has been suggested to be non-linear
\cite{lindemann_counterbend_2005}.

In each $96\,\nm$ axonemal repeat there are two radial spokes per doublet in the {\it Chlamydomonas} axoneme \cite{pigino_axonemal_2012}. Recently, it has been shown that there is also an incomplete third radial spoke, which is believed to be an evolutionary vestige \cite{barber_three-dimensional_2012}. When the radius of the axoneme is reduced, radial spokes compress, which suggests that they are involved in sustaining the cross-section of the axoneme \cite{zanetti_effects_1979}. However, recent evidence has shown that  radial spokes  are connected to the same regulatory complex identified as nexin, which suggests that they may also act as regulators of the beat \cite{barber_three-dimensional_2012}. While axonemes can beat in the absence of radial spokes \cite{yagi_vigorous_2000}, it has been observed that the gliding speed of dynein increases substantially in the presence of radial spokes \cite{smith_regulation_1992}. 


\section{Structural asymmetries}
\label{sec:asymmetries}
In a coarse view the picture of the axoneme is highly symmetrical. It is periodical in its longitudinal direction, with a period of $96\,\nm$. It also has ninefold rotational symmetry around its longitudinal axes, with motors connecting all the doublets. However, recent studies have shown that this picture is not accurate. There are many important structural asymmetries in the axoneme, which can have an important role in regulating the beating dynamics.

\subsection{Axoneme polarity}
\label{sec:polarasym}
The axoneme is a polar structure, and is thus not symmetric along its length. There are several polar asymmetries in the axoneme, the most obvious one is that the microtubule doublets themselves are made of polar proteins such as the doublets. Furthermore, along each repeat the distribution of dyneins and radial spokes is not homogeneous. More importantly, there are two polar asymmetries which occur on a larger scale: the asymmetry between the ends, due mainly to the basal body; and the asymmetry along its arc-length, due to inhomogeneous distribution of motors.

The two ends of the axoneme are fundamentally different. First, the distal end (farthest from the cell body) is the + end of the doublet microtubules and is thus constantly being polymerized. Second, at the basal end of the axoneme there is a complex structure called the basal body. The basal body is the region where the axoneme attaches to the cell body, and it's composed of microtubule triplets which get transformed into doublets as the basal body transforms to the axoneme (see Fig. 30 in \cite{ringo_flagellar_1967}, also \cite{ringo_flagellar_1967}). Evidence suggests that the basal body is elastic, and can allow for doublets sliding at the base of the axoneme  \cite{vernon_basal_2004, woolley_compliance_2008,woolley_evidence_2006}. This has led to the proposal that a basal constraint is an important regulatory mechanism for regulating the beat  \cite{riedelkruse_how_2007,sanchez_cilia-like_2011}, with some evidence that it may be necessary for  beating  \cite{fujimura_requirement_2006,goldstein_motility_1981}.

But besides of differences at the ends of the axoneme,  along its length there are also asymmetries in the distribution of dyneins \cite{yagi_identification_2009,bui_polarity_2012} as well as cross-linkers \cite{pigino_comparative_2012}. The $96\,\nm$ repeats are homogoneous only in the central part of the axoneme, with repeats changing as they approach the ends. For instance, certain types of dyneins (like inner dynein arms) are missing towards the base of the axoneme \cite{bui_polarity_2012}. Yet some other types of dynein are present only near the base, the so-called minor-dyneins \cite{yagi_identification_2009}.

\subsection{Chirality}
\begin{SCfigure}
\includegraphics{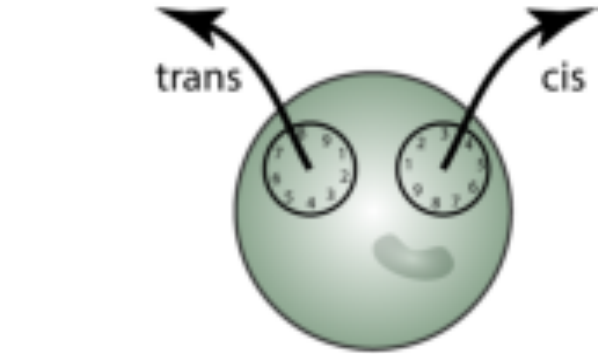}	
 	\caption[Swimming path of the axoneme.]{\textbf{Chirality of  {\it Chlamydomonas}.} Representation of {\it Chlamydomonas} with two axonemes  (polar, indicated by arrowheads)  of the same chirality (note the doublets ordering) as they are believed to be located in the cell \cite{bui_polarity_2012}. }
 	\label{fig:chirality}
\end{SCfigure}

Since the axoneme is polar, and the dyneins are bound to the A tubule and exert their power stroke on the B tubule, the axoneme is a chiral structure (compare in Fig.~\ref{fig:chirality} the two axonemes, which have the same chirality). Furthermore, to date only axonemes with one handedness have been found. One immediate consequence for {\it Chlamydomonas} is that {\it it swims with two left arms}. That is, unlike with human arms, its two flagella are not mirror-symmetric with respect to each other (see Fig.~\ref{fig:chirality}). But beyond this intrinsic chirality, detailed studies on the axonemal  structure have revealed further chiral asymmetries which we now discuss.

Certain adjacent doublets are connected by various structures: the doublets~1 and~2 are connected by the so-called~1-2 bridge, believed to be important in setting the beating plane \cite{bui_polarity_2012,hoops_outer_1983}. Additional cross-linkers have recently been identified the doublets~9 and~1, 5~and~6; as well as 1~and~2 \cite{pigino_comparative_2012}. All these structures have been suggested to constrain the sliding of the doublets they link, thus defining the beating plane as that in which the non-sliding doublets lay (which is roughly that of doublets 1 and 5, see Fig.~\ref{fig:axoneme}). Another asymmetry in the beating plane are the presence of beaks inside the B tubules, filamentous structures are all along  doublets 1, 5 and 6 \cite{hoops_outer_1983,pigino_comparative_2012}.

Several species of inner dyneins are also missing in certain doublets, for example in the doublet 1 \cite{bui_polarity_2012}. Interestingly, the same dynein arms that are missing in the basal region are also missing all along the axoneme in the doublet 9, thus coupling the polar asymmetry with the chiral asymmetry.

Further contribution to this chiral asymmetry can come from the inherent twist observed in the central pair, whose interaction with the doublets remains  unclear \cite{kamiya_extrusion_1982}. Furthermore, evidence in {\it Sea Urchin sperm} has shown that the axoneme itself can take spiral-like shapes, which are intrinsically chiral \cite{woolley_study_2001}.

\section{The mechanism of ciliary beat}
\label{sec:mechanisms}
How the beating patterns of cilia occur is a subject of intense research, and the main topic we address in this thesis. It is generally believed that  the beat is generated by alternating episodes of activation of opposing sets of dynein  \cite{spungin_dynein_1987}. Which dyneins are activated and which are deactivated is believed to be regulated by a mechano-chemical feedback. The beat is thus believed to be a self-organized process, with no {\it a priori} prescription of dynein activity: dyneins regulate the beat, and the beat regulates the dyneins. We now go over the key points of this process, reviewed in \cite{woolley_flagellar_2010,lindemann_flagellar_2010,brokaw_thinking_2009}.

\begin{itemize}
\item {\bf Dyneins produce active sliding forces:} Since dyneins can slide doublets \cite{smith_microtubule_1991} and slide axonemes apart \cite{summers_adenosine_1971}, it is believed that they produce sliding forces between the doublets, see section \ref{sec:dyneins}. In a rotationally symmetric axoneme all these forces balance exactly and there is no net sliding force, a situation termed tug-of-war \cite{howard_mechanical_2009}. In this case fluctuations can create a small imbalance in force, which is then amplified by the motors and finally produces a significant net sliding force. Alternatively, chiral asymmetries (see \ref{sec:asymmetries}) in the axoneme can also result in a net sliding force.

\item {\bf Passive cross-linkers constrain sliding and convert it into bending:} The net sliding force produced when the dyneins on one side of the axoneme win the tug-of-war leads to relative motion between the doublets \cite{brokaw_direct_1989}. Thanks to cross-linkers along the axoneme \cite{lindemann_counterbend_2005,bower_n-drc_2013}, and possibly also at the base \cite{aoyama_cyclical_2005,riedelkruse_how_2007,brokaw_flagellar_1972}, this sliding is constrained and converted into bending.

\item {\bf A mechano-chemical feedback can amplify the bending and produce switching}. In the tug-of-war scenario, when one set of dyneins wins and the axoneme bends, the other set of dyneins becomes gradually less active while the winning side becomes more active, thus the bending is increased \cite{howard_mechanical_2009}. However, possibly due to a delay in the mechano-chemical cycle by the dyneins \cite{machin_wave_1958,riedelkruse_how_2007,howard_mechanical_2009}, eventually the switch occurs and the opposing dyneins become active. The bend is thus reversed, giving rise to an oscillation.  
\end{itemize}

Of the three processes described above, the one which is the least understood is that of the mechano-chemical feedback. It is not clear {\it what} mechanical cue is sensed by the dyneins, and {\it how} it is sensed. We now review some of the proposals.

\subsection{Sliding control}
In sliding control, the mechano-chemical feedback responsible for  the switch occurs by the dynein motors sensing the sliding displacement between doublets. Motor models where  response to sliding can induce oscillations of a group of motors have a long history \cite{brokaw_molecular_1975,grill_theory_2005,julicher_cooperative_1995,julicher_spontaneous_1997}. The way the feedback can appear varies from one model to another, typically for processive motors it is believed that sliding forces induce detachment of motors \cite{grill_theory_2005}, while for non-processive motors like dynein ratchet mechanisms have been suggested \cite{julicher_cooperative_1995,julicher_spontaneous_1997}

An axoneme bends because of the sliding forces generated by the opposing dyneins in the plane perpendicular to the bending plane. For example, if the beating plane is formed by the doublets~1 and~5, the opposing dyneins are those in doublets~3 and~8 (see Fig. \ref{fig:raxoneme} and \cite{hilfinger_chirality_2008}). In this case, the sliding experienced by one set of motors is positive, in the sense that it favors their natural displacement to the + end of the axoneme. The displacement of the opposing set of motors is negative. It is believed that this acts as a regulatory mechanism of the motors force, with the dyneins that slide to the positive direction creating an active force and the others a resisting force.

\begin{figure}[!h]
\includegraphics{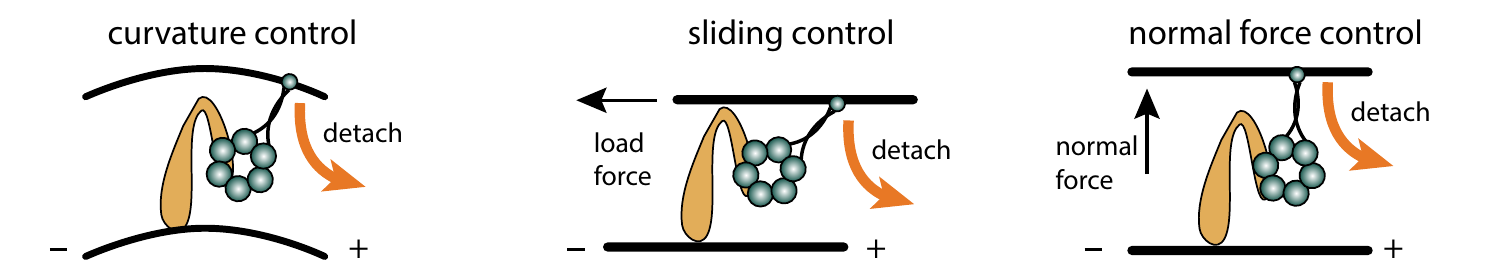}	
 	\caption{\textbf{Three mechanism of motor regulation.} In curvature control bending of the doublets enhances their detachment (left panel). In sliding control sliding of the doublets to the minus end produces a load force with triggers detachment (center panel). In normal force control the increase in spacing of the doublets produces a normal force that tends to detach dyneins.}
 	\label{fig:regular}
\end{figure}

This mechanism has been shown to produce beating patterns \cite{brokaw_molecular_1975, camalet_self-organized_1999}, that under certain special conditions are very similar to the beat of  {\it Bull Sperm} \cite{riedelkruse_how_2007}. Furthermore, there is also direct observation of sliding oscillations in straight axonemes \cite{kamimura_high-frequency_1992,yagi_novel_1995}.

\subsection{Curvature control}
In curvature control, the switching of molecular motors is regulated by the curvature of the flagellum. This mechanism was initially proposed in \cite{machin_wave_1958}, where the moment generated by the motors was suggested to be controlled by a delayed reaction to curvature. Later works have used the motor moment density as the quantity to be affected by curvature \cite{brokaw_bend_1971,brokaw_computer_2002,riedelkruse_how_2007}.

Notice that there is a crucial difference between curvature and sliding control. In sliding control oscillations can occur for straight filaments (as, for example, occurs in actin fibers in the presence of myosin \cite{placais_spontaneous_2009}). This is impossible in curvature control. Models of the axoneme where curvature regulates the beat operate by opposing motors being activated or inactivated when a critical value of the curvature is reached, together with a delay \cite{brokaw_bend_1971,brokaw_flagellar_1972}. One reason why opposing motors may react differently to curvature is that active and inactive motors, perpendicular to the beating plane, experience different concavity of the adjacent doublet (see Fig. \ref{fig:regular}).

This model is supported by the fact that some flagellar waves seem to show a traveling wave of constant curvature \cite{brokaw_bending_1983}. Furthermore, while the study in  \cite{riedelkruse_how_2007} favored a sliding control model, curvature control  also showed good agreement with the {\it Bull Sperm} beat. However, there is no accepted mechanism of curvature sensing by dynein motors. Direct geometrical sensing is unlikely, given the small size of dynein compared to curvatures in the axoneme  \cite{riedelkruse_how_2007}. Indeed, a radius of curvature of $2\,\um$ corresponds to the bending of an $8\,\nm$ tubulin dimer through an angle of only 0.025 degrees which is two orders of magnitude smaller than the curved-to-straight conformation associated with the straightening of a free GDP-bound tubulin subunit needed for its incorporation into the microtubule wall \cite{ravelli_insight_2004,mukundan_motor_2014}. An alternative explanation is provided by the geometric clutch \cite{lindemann_geometric_2007}, reviewed in the next section.

\subsection{Geometric clutch}
In the geometric clutch the key factor regulating the switch of dynein motors is the spacing between the doublets, or equivalently the corresponding transverse force ($t-$force) between them \cite{lindemann_geometric_1994, lindemann_geometric_2007}. The hypothesis is that there are two contributions to the $t-$force coming from curvature and sliding. The contribution of the curvature is termed global $t-$force, and tends to detach motors. This global $t-$force will be termed normal force in our description. The sliding contributes to the  local $t-$force, which tends to attach motors and is much smaller than the global $t-$force \cite{lindemann_structural-functional_2003}. Thus the geometric clutch combines effects of sliding and curvature control. There is an additional element required in the geometric clutch, which is a distributor of the $t-$force between the opposing sides of the axoneme. Radial spokes have been suggested to fulfill this role
  \cite{lindemann_structural-functional_2003}.

Computer models using the geometric clutch have successfully replicated the beat of {\it Chlamydomonas} \cite{lindemann_geometric_2002}. Furthermore, there is direct experimental evidence suggesting that bent axonemes show higher  spacing in bent regions, where the global $t-$force increases \cite{lindemann_evidence_2007}. However, it is so far unclear how the $t-$force is distributed in a cross-section of the axoneme \cite{ lindemann_geometric_2007}. This implies that it is not known how it gets distributed between the opposing motors. The same problem is present in curvature control, but not in sliding control: there opposing motors experience opposite sliding displacement.

\section{Physical description of axonemal beat}
\label{sec:basictheo}

To gain insight into ciliary beat, in this thesis we combine tools of non-linear dynamics, elasticity, and fluid mechanics. This will allow us to provide a full description of the self-organized dynamics of a cilia propelled by molecular motors \cite{machin_wave_1958, camalet_generic_2000}. We now give some brushstrokes on the main elements of the theory.

First, to derive the mechanical forces we will use a variational approach. That is, we will construct an energy functional $G[{\bf r}]$, where ${\bf r}(s)$ is the position of the arc-length point $s$ of the cilium, which collects all the elastic properties of the cilium. By performing variations, we will obtain the mechanical forces.  Second, to describe the effect of molecular motors, we will write down a dynamic equation for the motor force which tends to slide the doublets apart. This dynamic equation will depend on the internal strains and stresses of the cilium, thus coupling the system. Third, to model the fluid we will use resistive force theory \cite{gray_propulsion_1955}, in which the fluid force is characterized by two  friction coefficients $\xi_{\rm n}$ and $\xi_{\rm t}$ (corresponding to normal and tangential motion). This simplification is possible because the Reynolds number of a cilium is small, which allows us to neglect non-linear effects \cite{purcell_life_1976}; and because the cilium is a slender body, with a diameter $a$ much smaller than its length $L$.

Putting all these elements together, we will obtain a dynamic equation for the shape of the cilium. If we parametrize the cilium by its local tangent angle $\psi(s,t)$ at arc-length $s$ and time $t$, we will have  to linear order the following force balance equation 
\begin{align}
\label{eq:spermold}
\xi_{\rm n}\partial_t\psi = -\kappa\partial_s^4\psi-a\partial_s^2f\quad.
\end{align}
This equation includes the effects of fluid friction, elasticity (with $\kappa=EI$,  $E$ the young modulus and $I$ the second moment of inertia), and the motor force $f$. The full description of the system still requires an equation that couples the motor force dynamics to an internal strain in the cilium. For example, if the motor force responds to sliding velocity (as has been suggested \cite{brokaw_molecular_1975,camalet_self-organized_1999}), we have to linear order
\begin{align}
f(s,t)=\int_0^t\chi(t-t')\Delta(s,t')\d t'
\end{align}
where $\chi(t-t')$ is the linear response function, and $\Delta$ is the local sliding between doublets. The two equations above define a linearized dynamical system which can undergo oscillatory instabilities and go to a limit cycle (which, to be described, requires to take into account non-linearities).  Alternatively, the motor force could respond to changes in the local curvature $\partial_s\psi$ \cite{machin_wave_1958,brokaw_bend_1971} or the doublets spacing $a$ \cite{lindemann_geometric_1994}. In this thesis we characterize the beat patterns corresponding to these three different different regulatory mechanisms, and compare the results to the experimentally observed ciliary patterns described in  the next chapter.

\section{Conclusions}
\begin{itemize}
\item The core element of the cilium is the axoneme: a cylindrical bundle of microtubules connected by dynein molecular motors.
\item The beat of cilia is a self-organized process powered through  microtubule sliding forces that are produced by dynein motors.
\item Self-organization arises through motor regulation via one (or several) of the following mechanisms: regulation by microtubules sliding, by their curvature, or by the normal force arising between them.
\end{itemize}

\chapter{Characterization of the {\it Chlamydomonas} beat}

{\textit{ Chlamydomonas} cilia can be isolated from the cell} and, in the presence of ATP, beat periodically. In this section we mathematically describe the beat pattern of intact as well as disintegrated cilia. We show that the observed beat pattern of an intact cilium is well characterized by its static and fundamental harmonics. While for wild-type the zeroth harmonic is very important as the beat is asymmetric, this is not the case for a symmetric mutant which we analyze. We also show that in disintegrated cilia pairs of doublets can interact with each other, reaching a static equilibrium in which the shape is a circular arc. All of the data appearing in this section was taken in the laboratory of Jonathon Howard: the data from the first section by Veikko Geyer, and that of the second by Vikram Mukundan.

\section{The beat of isolated {\it Chlamydomonas} cilia}
\label{sec:experchlam}
The cilia of {\it Chlamydomonas} can be isolated from the cell body \cite{sanders_centrin-mediated_1989}. When their membrane is removed and ATP is added to the solution, they exhibit periodic beat patters \cite{bessen_calcium_1980}. This is  evidence that the beat of the axoneme is a self-organized property, as it occurs independent of the cell body and with ATP homogeneously present around the axoneme.

\begin{figure}[!ht]
\centerline{\includegraphics[width=\textwidth]{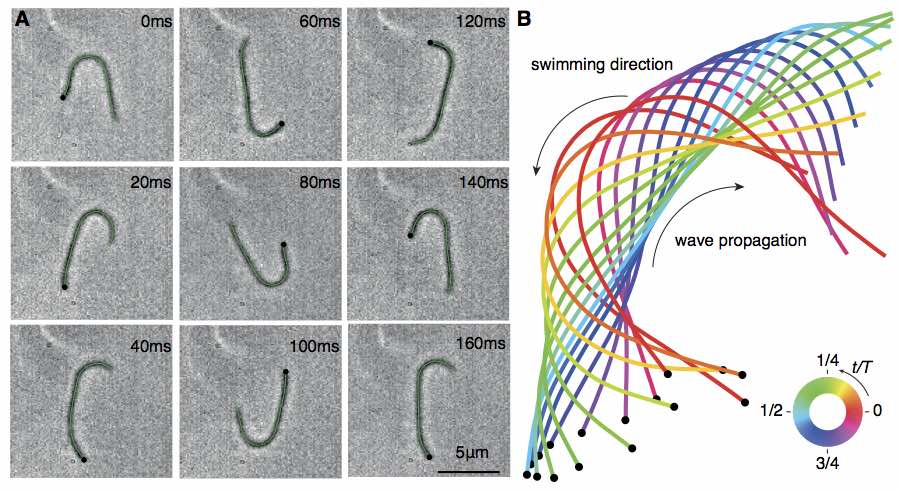}}
\caption{\textbf{Time series of a beating cilium.} \textbf{A.} An isolated {\it Chlamydomonas} cilium has  an asymmetric beat pattern which makes it swim counter-clockwise. The tracking is noted with a green line, and the base with a black circle (adapted from \cite{geyer_characterization_2013}). \textbf{B.} Beat pattern over one period of duration $T$ obtained after tracking, showing forward wave propagation (from base, indicated with a black circle, to tip) and thus counter-clockwise swimming.  As time progresses over one period the cilium changes color following the circular legend counter-clockwise }
\label{fig:tracktrace}
\end{figure}

Our collaborators in the group of Jonathon Howard, in particular Veikko Geyer, have imaged the beat of isolated {\it Chlamydomonas} axonemes using high speed phase contrast microscopy with high spatial and temporal resolution. The pixel-size was of $139 \times 139\,\nm ^2/{\rm pixel}$, and the frame-rate $10^3\,\Hz$. In comparison, the typical size of the axonemes was $
L\sim 10\,\um$, and the characteristic beat frequency $f\sim 50\,\Hz$. Nine frames from a sample beat pattern appear in Fig.~\ref{fig:tracktrace}, the tracking of the axoneme appears as a green line, the basal end of the axoneme is marked by a black circle. The tracking was performed using the {\it fiesta} code published in \cite{ruhnow_tracking_2011}. As one can see, the beat of the axoneme is asymmetric, and because of this the axoneme swims in circles. While the typical beat frequency of the axoneme is $f\sim 50\,\Hz$, the rotational frequency of the axoneme is ten times slower, on the order of $f_{\rm rot}\sim 5\,\Hz$.

\begin{SCfigure}
\includegraphics{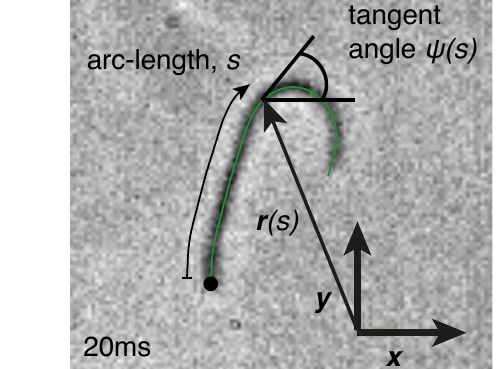}	
 	\caption{\textbf{ Tracking the tangent angle.}  The tangent angle $\psi(s)$ at arc-length $s$ measured from the basal end, is defined with respect to the lab frame ${\bf xy}$ using the pointing vector ${\bf r}(s)$.}
 	\label{fig:tangent}
\end{SCfigure}

The tracked cilium is characterized at each time $t$ by a set of two dimensional pointing vectors ${\bf r}(s,t)$ from the fixed laboratory reference frame to each point along the axonemal arc-length $s\in[0,L]$, with $L$ the length of the cilium (see Fig.~\ref{fig:tangent}). To describe its shape  decoupled from the swimming  we use an angular representation. That is, for each point along the cilium we calculate the tangent angle $\Psi(s,t)$ with respect to the horizontal axis ${\bf x}$, see Fig.~\ref{fig:tangent}. Note that this angular representation contains less information than the vectorial representation, in particular the swimming trajectory is lost. However  $\Psi(s,t)$ still retains information about the rotation velocity of the axoneme $f_{\rm rot}$. To obtain a pure shape description we also subtract this rotation, that is
\begin{align}
\psi(s,t) = \Psi(s,t)-f_{\rm rot} t\quad,
\end{align} 
where the tangent angle $\psi(s,t)$ now only describes the shape of the axoneme.

\begin{figure}[ht]
\centerline{\includegraphics[width=\textwidth]{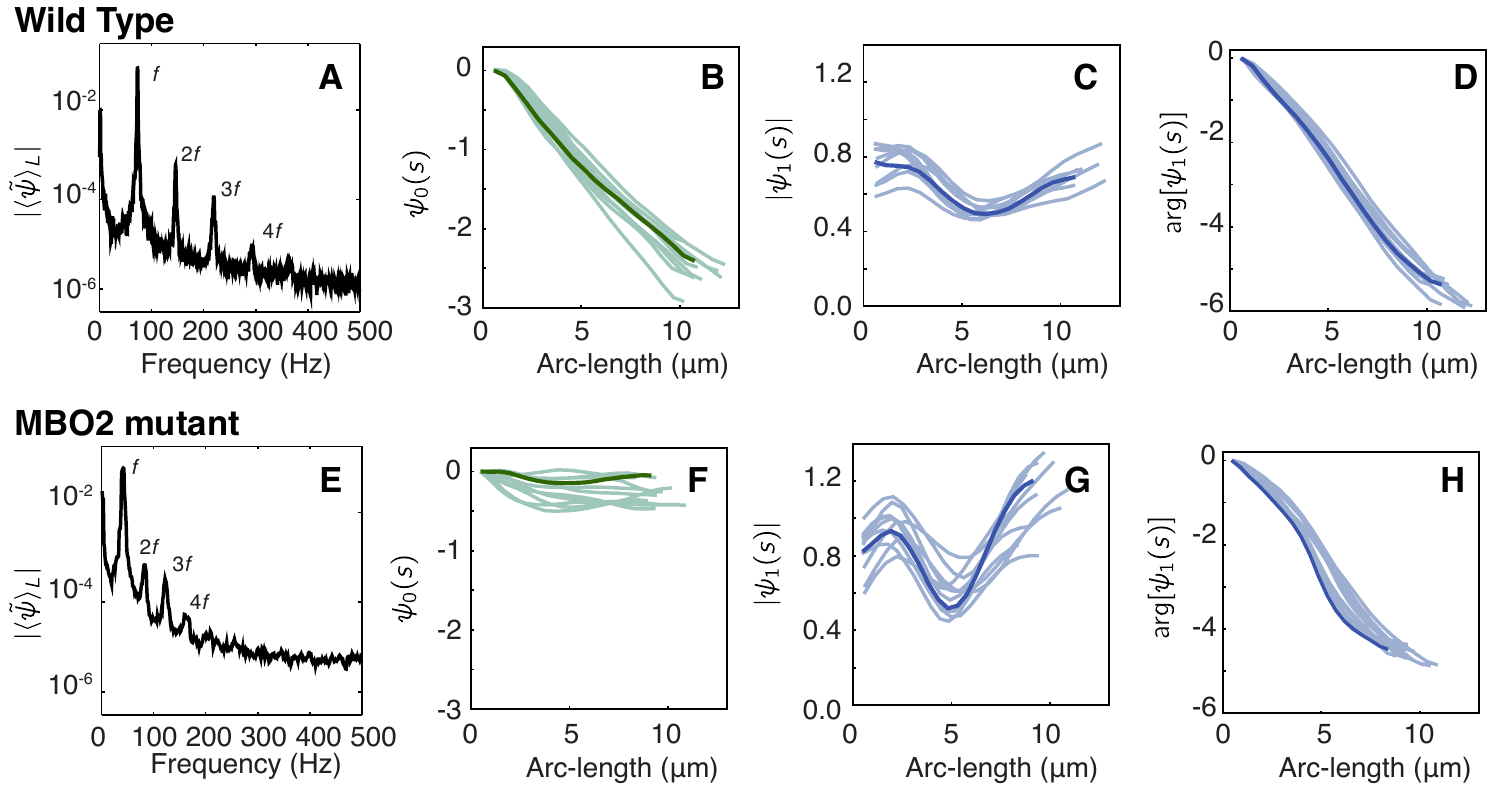}}
\caption{\textbf{Characterization of the {\it Chlamydomonas} beat.} Top, characterization of  Wild Type beat. \textbf{A.} The power spectrum of the angle at mid-length shows up to seven harmonics or modes. \textbf{B.} The zeroth mode of the angle grows linearly over arc-length, resulting in a circular arc of constant curvature. \textbf{C.}  The amplitude of the first mode is roughly constant over arc-length. \textbf{D.}  The phase of the first mode drops $\sim2\pi$ over the length of the cilium. Bottom, characterization of mbo2 beat. The main difference resides in F, mbo2 shows a small zeroth mode since its beat is symmetric. Each line corresponds to an individual cilium, and the bolder one corresponds to the same cilium in panels B, C and D for WT, and in panels F, G and H for mbo2.}
\label{fig:wtchar}
\end{figure}

Since the beat of {\it Chlamydomonas} is periodic in time, we analyze the shape information $\psi(s,t)$ by transforming its temporal coordinate to frequency space. This can be done using a fast Fourier transform, and the result for the mid-point of the axoneme is  shown in Fig.~\ref{fig:wtchar} A. There are peaks around frequencies multiple of the fundamental frequency, and a background of noise. Neglecting the noise, the beat can thus be  approximated by the following Fourier decomposition
\begin{align}
\psi(s,t)\approx\sum_{n=-\infty}^{n=+\infty}\psi_{n}(s)\exp(i2\pi n f t)
\end{align}
Where $f$ is the fundamental frequency of the beat, which in the example considered in Fig.~\ref{fig:wtchar} A is $f=55.4\,\Hz$. The modes $\psi_{n}(s)$ are complex functions of $s$, and their amplitude decreases as the mode increases, see Fig.~\ref{fig:wtchar} A. Because the angle is a real quantity, its modes satisfy $\psi_{-n}=\psi^{\rm *}_n$.

Importantly, the above description includes the zeroth mode $\psi_{0}(s)$, which corresponds to the average shape of the cilium and defines its asymmetry. In Fig.~\ref{fig:wtchar} B this mode is shown for an example. As one can see, besides a flattening at the ends, the tangent angle decreases monotonically along the axoneme. The amplitude and phase profiles of the first, second and third dynamic modes are shown in Fig~\ref{fig:wtchar} B and C. As one can see, the amplitude of the first mode is much larger than that of all higher modes, and roughly constant along the arc-length with a small dip in the middle. The phase profile of the first mode is shown in Fig.~\ref{fig:wtchar} D, and is monotonically decreasing. Over the full length of the axoneme the phase decreases about $2\pi$, which corresponds to a wave-length equal to the length of the axoneme.

\begin{figure}[h]
\centerline{\includegraphics[width=\textwidth]{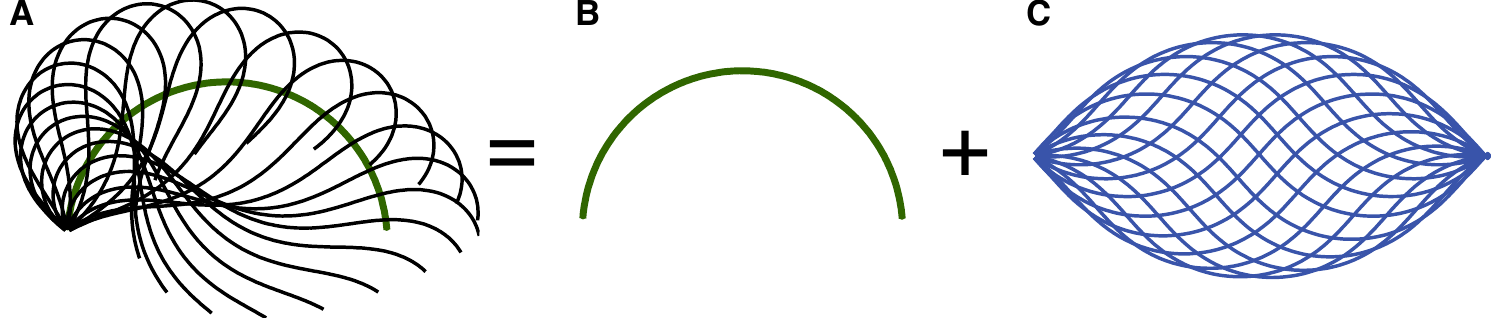}}
\caption{\textbf{Schematic of asymmetric beat.} An asymmetric beat (A) can be described as a circular arc (constant mean curvature, B) plus a plane angular wave (C). Figure adapted from \cite{geyer_characterization_2013}.}
\label{fig:takehome}
\end{figure}

The curvature is the derivative of the tangent angle with respect to the arc-length. This means that the constant slope of the angle in the zeroth mode corresponds to a constant mean curvature of roughly $\sim 0.25\,{\rm rad}\cdotp\um^{-1}$. Furthermore, since the amplitude of the first mode is approximately constant and its phase decreases with a constant slope, we conclude that a good approximation of the {\it Chlamydomonas} ciliary beat, shown in Fig.~\ref{fig:takehome} A, is the superposition of an average constant curvature, Fig.~\ref{fig:takehome} B, and plane wave of its angle, Fig.~\ref{fig:takehome} C. Thus, the beat of {\it Chlamydomonas} is fundamentally asymmetric, and its asymmetry is well characterized by  a constant mean curvature.

It is important  to note that this  asymmetry is not present during the phototropic response of {\it Chlamydomonas} and in certain mutants that Move Backwards Only (MBO). The MBO 1-3 mutants beat symmetrically \cite{segal_mutant_1984, geyer_characterization_2013}. In particular the mutant MBO 2 has a very small static component in the beat,  see Fig.~\ref{fig:wtchar} A. Interestingly, the amplitude profile of the first mode is very similar to that of the wild type cilium, as one can see by comparing Fig.~\ref{fig:wtchar} B and Fig.~\ref{fig:wtchar} C. The same is true of the phase profile, which appears in Fig.~\ref{fig:wtchar} C.  Thus, these mutants have a beat that, while lacking the asymmetry, are in the rest very similar to that of the wild-type cilium. Finally, it is worth noting that also wild-type {\it Chlamydomonas} can exhibit symmetric beat patterns in the presence of Calcium, although these are three-dimensional beats fundamentally different from those of mbo2 \cite{bessen_calcium_1980,geyer_characterization_2013}.

\section{Bending of disintegrated axonemes into circular arcs}
\label{sec:circles}

\begin{figure}[h]
\centerline{\includegraphics[width=6in]{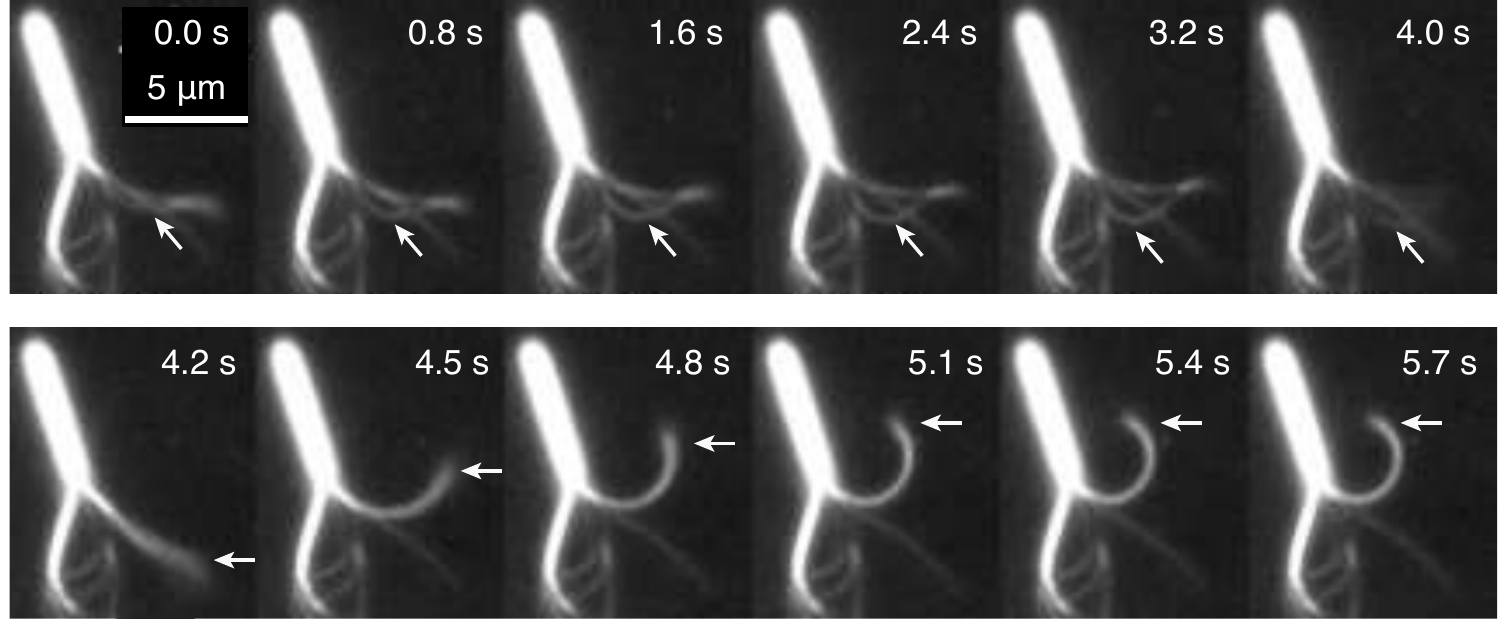}}
\caption{\textbf{Sliding of adjacent doublets in a split axoneme} One doublet slides along another and then dissociates (0.0 to 4.0 s). After reassociating (4.2 s), the two doublets remain in close apposition and bend into a circular arc (4.5 s to 5.7 s) before dissociating again (not shown). Figure adapted from \cite{mukundan_motor_2014}.}
\label{fig:bendseries}
\end{figure}

In the presence of a protease treatment, axonemes  partially loose their cross-linkers. When this occurs, two doublets can interact via the dyneins of one of them. This provides a minimal system, which has been reported to produce sliding and bending waves \cite{aoyama_cyclical_2005,mukundan_motor_2014}. In particular, at low concentrations of ATP, pairs of filaments associate, and propagate small bending waves towards the basal end as one filament slides along the other (Fig.~\ref{fig:bendseries}, first row, arrows). Furthermore, in some occasions the two filaments re-associate along their entire length and bend into a circular arc (Fig.~\ref{fig:bendseries}, second row). The system then becomes unstable and the filaments separate again.

\begin{SCfigure}[\sidecaptionrelwidth][h!]
\includegraphics[width=3in]{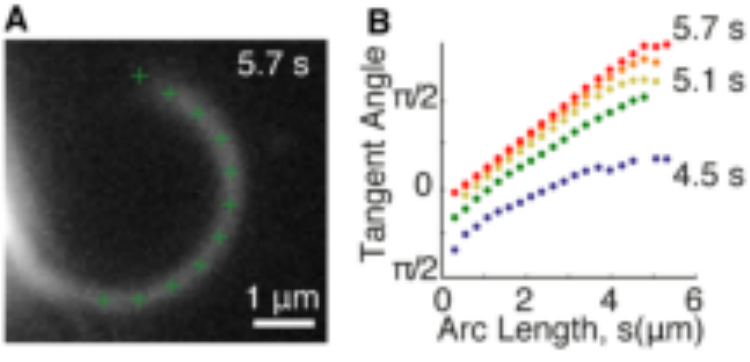}	
 	\caption{\textbf{Tracking the doublets.} \textbf{A.}  The shape was characterized by tracking the centerline of intensity along the filament contour marked by green crosses. \textbf{B.}  The tangent angles are plotted as a function of arc-length. The tangent angle increased linearly with arc-length (times $5.1-5.7\,\s$), indicating that the shape of the doublets is nearly a circular arc. The curvature of the arc is increasing over time and approaches a final, quasi-static shape at $5.7\,\s$. Figure adapted from \cite{mukundan_motor_2014}.}
 	\label{fig:bendtrack}
\end{SCfigure}

To analyze the bending process in detail, the shapes of the filaments pairs are digitized, as shown in Fig.~\ref{fig:bendtrack} A. From this one can calculate the tangent angle as a function of arc-length in successive frames as the filaments become more and more bent, see Fig.~\ref{fig:bendtrack} B. Importantly, the filament pair approaches a steady-state shape in which the tangent angle increases linearly with arc length, except at the very distal end where it flattens (Fig.~\ref{fig:bendtrack}, $5.1\,\s$ to $5.7\,\s$). Such a linearly increasing tangent angle implies that the steady-state shape is approximately a circular arc. This static constant curvature is analogous to that observed in the wild-type {\it Chlamydomonas} cilia as was described in section \ref{sec:experchlam}.

\begin{SCfigure}[\sidecaptionrelwidth][h!]
\includegraphics[width=3in]{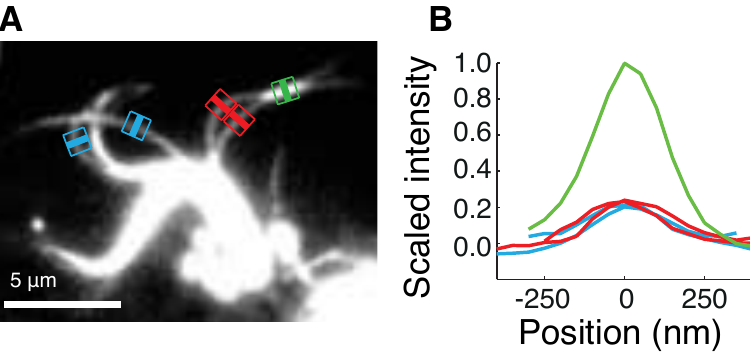}	
 	\caption{\textbf{Intensity analysis of filaments.} \textbf{A.}  Frame of a bending event with line-scans of filaments interacting  (red), non-interacting  (blue) and overlapping (green). \textbf{B.}  Corresponding intensity profiles. The interacting and non-interacting filaments have similar intensity, while the intensity in the overlap is 4 times higher than the non-overlap regions, as corresponds to dark-field microscopy. Figure adapted from \cite{mukundan_motor_2014}.}
 	\label{fig:intensities}
\end{SCfigure}

There is evidence that the two interacting filaments are two doublet microtubules and not two singlet microtubules or one doublet microtubule interacting with the central pair. The two individual singlet microtubules that comprise the central pair will each have a lower intensity than a doublet. However, the interacting filaments (Fig.~\ref{fig:intensities} A and B, red) have the same intensities as the non-interacting filaments (Fig.~\ref{fig:intensities} A and B, blue). This implies that the interacting filaments are not two singlet microtubules (which would both be much dimmer than the non-interacting filaments) or a singlet and a doublet (one of the two interacting filaments would be much dimmer than the other).

One of the advantages of this experiment is that the shapes reach a quasi-static limit, which can be analyzed assuming mechanical equilibrium. The assumption of mechanical equilibrium means that the frictional forces due to motion through the fluid can be ignored, and applies if the relaxation time of the bent beam in the fluid is much smaller than the typical time of the bending observed. The relaxation time of a beam in a fluid is $\sim\xi_{\rm n} L^4/\kappa$. Using $\kappa=120\,\pN\cdotp\um^{2}$ for the doublets stiffness, $\xi_{\rm n}=0.0034\,\pN\cdotp\s\cdotp\um^{-2}$ for the fluid friction, and $L=5\,\um$  for the length, we obtain a relaxation time of $\sim0.01\,\s$, which is much smaller than the typical duration of the process $\sim1\,\s$.

\section{Conclusions}
\begin{itemize}
\item The beat of the {\it Chlamydomonas} cilium is well described by its static and fundamental mode.
\item In a wild-type cilium the static mode is large and has the shape of a circular arc, producing an asymmetry in the beat. In an mbo2 cilium the static mode is small, and the resulting beat symmetric.
\item Pairs of doublets are statically bent into circular arcs due to motor regulation.
\end{itemize}

%

	
\chapter{Force balance of planar cilia}

{The planar beat of a cilium can be described as a pair of opposing filaments.} In this section we introduce a two-dimensional representation of the axoneme as a pair of opposing inextensible filaments. We consider the sliding as well as variable spacing between these filaments. By balancing mechanical and fluid forces, we derive the general non-linear dynamic equations of a cilium beating in a plane. As a first application of this theory, we study the role of sliding cross-linkers in the static bending of a cilium.

\section{Planar geometry of the axoneme}
\label{sec:geo}
\begin{figure}
\includegraphics{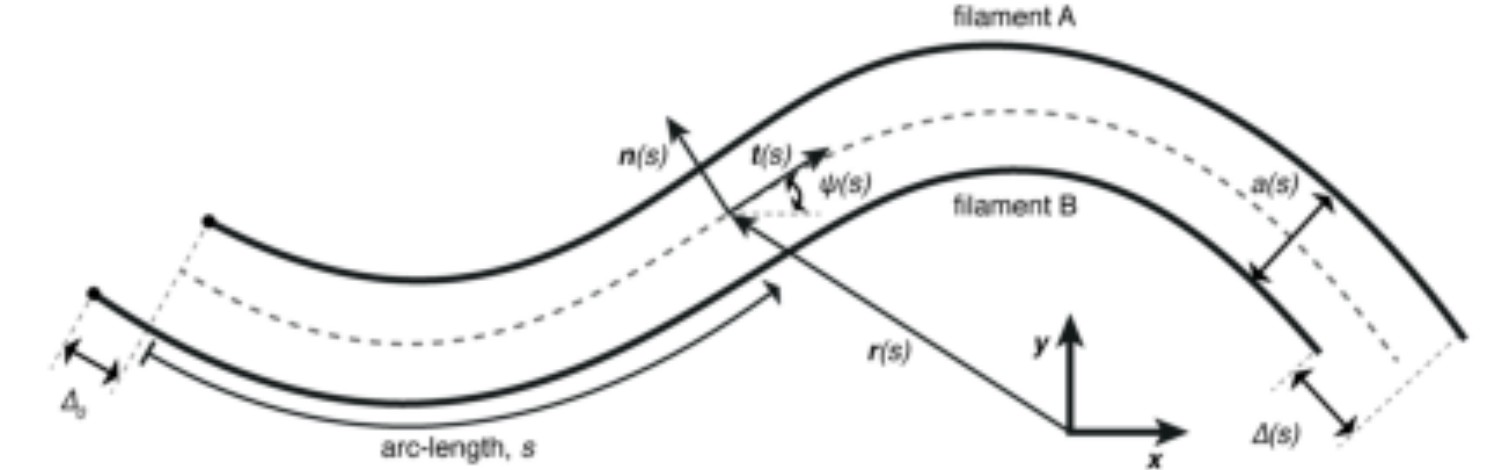}	
 	\caption{\textbf{Planar geometry of a cilium.} Scheme of a cilium as a pair of opposing filaments (labeled A and B) separated by a spacing $a(s)$ with depends on the arc-length $s$ (measured from the base, black circles). The relative sliding of the filaments $\Delta(s)$ is shown, as well as the sliding at the base $\Delta_0$. Each position along the centerline (dashed line) is characterized by a position vector ${\bf r}(s)$ with origin at the laboratory frame ${\bf xy}$. From the position vector we can obtain the tangent vector field ${\bf t}(s)$, as well as the normal ${\bf n}(s)$ field through Eq.~\ref{eq:tannor}. The tangent angle $\psi(s)$ can be calculated at each point (see Eq.~\ref{eq:posang}). }
 	\label{fig:2dgeometry}
\end{figure}

We describe the axoneme by a pair of opposing filaments, which we label A and B (see Fig.~\ref{fig:2dgeometry}). Each filament is parametrized by the arc-length $s$ of the centerline, which ranges from 0 at the base to $L$ at the tip. The filaments are separated by a distance $a(s)$ which can depend on the arc-length, and they can slide with respect to each other at every point.

The geometry of the centerline is characterized by ${\bf r}(s)$, a two-dimensional pointing vector from the laboratory frame. At any given point of the centerline we can define the local tangent vector ${\bf t}(s)$ and the local normal vector ${\bf n}(s)$, as shown in Fig.~\ref{fig:2dgeometry}. The tangent vector is given by
\begin{align}
{\bf t}= \frac{\partial {\bf r}}{\partial s} =\dot {\bf {r}}\quad,
\end{align}
where in the last expression we have introduced a notation in which upper dots denote arc-length derivatives. This notation will be kept throughout the rest of this thesis, with the number of dots denoting the order of the arc-length derivative. The normal vector is defined simply as normal to ${\bf t}(s)$, with orientation such that ${\bf t}\times{\bf n}$ points out of the ${\bf x\,y}$ plane. Using the tangent vector we can also define the local tangent angle $\psi(s)$ between the tangent vector ${\bf t}(s)$ and the ${\bf x}$-axis. The relationship between the tangent angle and the pointing vector of the centerline is
\begin{align}
\label{eq:posang}
{\bf r}(s) = {\bf r}_0 + \int_0^s\begin{pmatrix}\cos(\psi(s'))\\\sin(\psi(s'))\end{pmatrix}\d s'\quad ,
\end{align}
where ${\bf r}_0={\bf r}(s=0)$ is the position of the base of the centerline. The local curvature of the centerline $C(s)$ is given by the arc-length derivative of the tangent angle, $C(s)=\dot{\psi}(s)$. The geometry of the centerline is thus given by the set of equations
\begin{align}
\label{eq:tannor}
{\bf \dot{r}}={\bf t}\quad\quad ;\quad\quad {\bf \dot{t}}=\dot{\psi}{\bf {n}}\quad\quad ;\quad\quad{\bf \dot{n}}=-\dot{\psi}{\bf {t}}
\end{align}
which are  the Frenet-Serret formulas for the special case of a planar geometry in the absence of torsion.

Having introduced the geometry of the centerline we relate  it to that of the pair of opposing filaments. Since each of the filaments is at a distance $a(s)/2$ from the centerline, we can write
\begin{align}
{\bf r}_{\rm A}(s)= {\bf r}(s) + \frac{a(s)}{2}{\bf n}(s)\quad \quad ,\quad\quad{\bf r}_{\rm B}(s)= {\bf r}(s) - \frac{a(s)}{2}{\bf n}(s)\quad .
\end{align}
Note that $s$ does not parametrize the arc-length of either of the filaments, but that of the centerline. The arc-length along each of the filaments is given by
\begin{align}
s_{\rm A}(s) = \int_0^s|{\bf \dot{r}}_{\rm A}(s')|\d s' \quad \text{ and }\quad s_{\rm B}(s) = \int_0^s|{\bf \dot{r}}_{\rm B}(s')|\d s'\quad.
\label{eq:rabdot}
\end{align}
Using the corresponding Frenet-Serret frame for each of the filaments, we  obtain the curvature of the filaments (see Appendix A). To lowest order these curvatures are given by
\begin{align}
C_{\rm A}(s)\approx\dot{\psi}(s)+\frac{\ddot{a}(s)}{2}\quad\quad,\quad\quad {\rm and}\quad\quad C_{\rm B}(s)\approx\dot{\psi}(s)-\frac{\ddot{a}(s)}{2}
\label{eq:curvAB}
\end{align}
where geometric non-linearities have been neglected.

The sliding of one filament  with respect to the other at centerline arc-length position $s$ is given by the mismatch in arc-length along one filament with respect to the other plus the  reference sliding at the base. We thus have
\begin{align}
\Delta_{\rm A}(s)=\Delta_{0}+s_{\rm B}(s)-s_{\rm A}(s)\quad \quad {\rm and\; correspondingly}\quad\quad \Delta_{\rm B}(s)=-\Delta_{\rm A}(s)\quad ;
\end{align}
where $\Delta_0$ is the basal sliding of filament A with respect to B. From now on we take as reference filament B, and define the local sliding as that of A, we thus have $\Delta(s) =\Delta_{\rm A}(s)$. The explicit expression of the local sliding can be calculated using Eqs.~\ref{eq:rabdot} and  is given to cubic order by
\begin{align}
\Delta(s)&\approx\Delta_0+\int_0^s a(s')\dot{\psi}(s')\d s'\quad.
\label{eq:generalslide}
\end{align}
Note that Eqs.~\ref{eq:curvAB} and \ref{eq:generalslide} take a particularly simple form in the limit of homogeneous spacing $a\to a_0$, in which $\Delta(s)=\Delta_0+a_0(\psi(s)-\psi(0))$, and $C_{\rm A}=C_{\rm B}=\dot{\psi}$.

\section{Static balance of forces}
\label{sec:stat}
As discussed in section 2, the axoneme is composed of passive and active mechanical elements. The passive elastic elements such as the doublets and nexin cross-linkers provide structural integrity to the axoneme, and tend to restore it to a straight configuration without sliding. On the other hand the active elements create sliding forces between the doublets which ultimately bend it.

\begin{SCfigure}[\sidecaptionrelwidth][h!]
\includegraphics{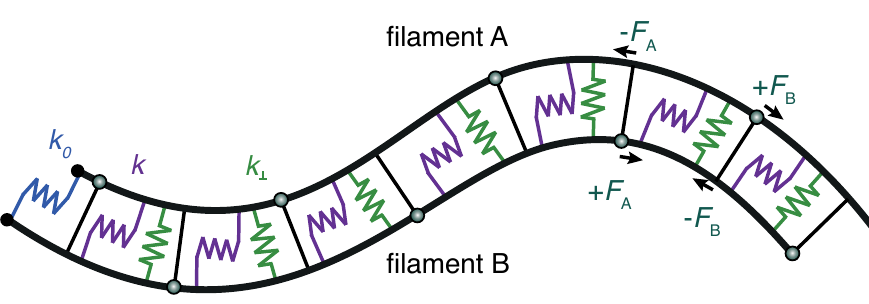}	
 	\caption{\textbf{Mechanics inside a cilium:} Opposing motors in filaments A/B (green circles) produce forces $F_{\rm A/B}$, and reactions $-F_{\rm A/B}$. Normal compression is constrained by springs (green) of stiffness $k_{\perp}$. Sliding is limited by bulk springs (purple, stiffness density $k$) and a basal spring (blue, stiffness $k_0$).\label{fig:mech}}
\end{SCfigure}

All the key elements of our ciliary description are provided in Fig.~\ref{fig:mech}. To characterize their mechanical properties, we introduce the following work functional:
\begin{align}
G[\Delta_0,a,{\bf r}] &= \int_0^L\left[\frac{\kappa_0}{2}(C_{\rm A}^2 +C_{\rm B}^2)+ \frac{k}{2}\Delta^2 -f_{\rm m}(s)\Delta+ \frac{k_\perp}{2}(a-a_0)^2 +\frac\Lambda2({\bf \dot{r}}^2-1)\right]\d s+\frac{k_0}{2}\Delta_0^2
\label{eq:workfunc}
\end{align}
where the explicit expressions of curvatures and sliding are given in Eqs.~\ref{eq:curvAB} and \ref{eq:generalslide}. The integral contains the energy density of the bulk of the axoneme, and the  last term is the energetic contribution of the base.  The first term in the integral is the bending energy characterized by the bending rigidity $\kappa_0$ (in ${\rm pN}\cdotp\mu{\rm m}^2$) of each filament, which favors straight shapes. The second term is the energy of elastic linkers of stiffness density $k$  (in ${\rm pN}\cdotp\mu{\rm m}^{-2}$)  which are stretched by sliding (purple springs in Fig.~\ref{fig:mech}). The stiffness (in ${\rm pN}\cdotp\mu{\rm m}^{-1}$) of basal linkers is denoted $k_0$ (blue spring in Fig. \ref{fig:mech} ). We have denoted by $k_\perp$ (in ${\rm pN}\cdotp\mu{\rm m}^{-2}$) the stiffness to normal deformations relative to the reference spacing $a_0$ (green springs in Fig.~\ref{fig:mech}). The work performed by motors which generate relative force $f_{\rm m}$ (in ${\rm pN}\cdotp\mu{\rm m}^{-1}$) between the two filaments is given by the contribution $-f_{\rm m}\Delta$. We have also introduced a Lagrange multiplier $\Lambda$  (in ${\rm pN}$) to ensure the inextensibility of the centerline.

To obtain the equations of static equilibrium of the axoneme under external forces we use the  virtual work principle. We consider as reference a straight configuration ({\bf r}(s)=s{\bf x}), with no basal sliding ($\Delta_0=0$) and homogeneous spacing ($a=a_0$). The  virtual work principle establishes that the internal virtual work $\delta W_{\rm i} = \delta G$ performed by a variation $\{\delta{\bf r},\delta a,\delta\Delta_0\}$ , is equal to the external work $\delta W_{\rm e}$ performed by external forces against this variation. For the forces applied on the bulk of the axoneme we have
\begin{align}
 \delta W_{\rm i}= \delta G &=\delta W_{\rm e}\quad,\nonumber\\
  \int_0^L\d s\left( \frac{\delta G}{\delta {\bf r}}\cdotp\delta{\bf r}+ \frac{\delta G}{\delta a}\delta a\right)+ \frac{\delta G}{\delta \Delta_0}\delta\Delta_0&=  \int\d s \,{\bf f}^{\rm ext} \cdotp\delta{\bf r}\quad.
\end{align}
Where  ${\bf f}^{\rm ext}$  are  the bending force density  applied externally in each point with a specified direction in the plane, and we are assuming that there are no external basal forces (conjugate of $\delta\Delta_0$) or compressive forces (conjugate of $\delta a$).  At the boundaries, the same principle applies, and we have
\begin{align}
{\bf F}_{0}^{\rm ext}\cdotp\delta {\bf r}(0)=\frac{\delta G}{\delta {\bf r}(0)}\cdotp\delta{\bf r}(0)\quad &,\quad {\bf F}_{L}^{\rm ext}\cdotp\delta {\bf r}(L)=\frac{\delta G}{\delta {\bf r}(L)}\cdotp\delta{\bf r}(L)\quad,\nonumber\\
T_{0}^{\rm ext}\delta\psi(L)=\frac{\delta G}{\delta {\bf \dot{r}}(L)}\cdotp\delta{\bf \dot{r}}(0)\quad &,\quad T_{L}^{\rm ext}\delta\psi(L)=\frac{\delta G}{\delta {\bf \dot{r}}(L)}\cdotp\delta{\bf \dot{r}}(L)\quad,\nonumber\\
0=\frac{\delta G}{\delta a(0)}\delta a(0)\quad &,\quad 0=\frac{\delta G}{\delta a(L)}\delta a(L)\quad,\nonumber\\
0=\frac{\delta G}{\delta \dot{a}(0)}\delta \dot{a}(0)\quad &,\quad 0=\frac{\delta G}{\delta \dot{a}(L)}\delta \dot{a}(L)\quad.
\label{eq:mechbound}
\end{align}
Where we have now introduced the external forces ${\bf F}_{0}^{\rm ext}$ and ${\bf F}_{L}^{\rm ext}$ at the boundaries. We have also allowed the presence of external torques $T_{0}^{\rm ext}$  and $T_{L}^{\rm ext}$  at the boundaries.

To obtain the balance of forces in the axoneme the first step is to calculate the corresponding functional derivatives. Doing standard variation calculus as detailed in Appendix A we obtain 
\begin{align}
\frac{\delta G}{\delta {\bf r}}&=\partial_s\left[ (\kappa\ddot{\psi} -\dot{a} F + a f){\bf n}-\tau{\bf t} \right]\nonumber\quad,\\
\frac{\delta G}{\delta a}&=k_\perp(a-a_0)+\kappa\ddddot{a}/2-F\dot{\psi}\nonumber\quad,\\
\frac{\delta G}{\delta \Delta_0}&=-F(0)+F_0\quad .
\label{eq:mechvar}
\end{align}
Where $\kappa=2\kappa_0$ is the stiffness of both filaments, and  we have introduced the static sliding force density $f$  and the basal force $F_0$ as
\begin{align}
f(s)&=f_{\rm m}(s)-k\Delta(s)\quad,\nonumber\\
F_0&= k_0\Delta_0\quad;
\label{eq:statfor}
\end{align}
and the integrated force $F(s)=\int^L_sf(s')\d s'$. In the first equation we have introduced the  Lagrange multiplier $\tau=\Lambda+\kappa\dot{\psi}^2-F\dot{\psi} a$ to replace $\Lambda$. We prefer $\tau$  as it can be interpreted as the tension of the centerline. This can be seen by the following relation 
\begin{align}
\tau(s) = \tau(0)-{\bf t}(s)\cdotp\int_0^s\frac{\delta G}{\delta {\bf r}}(s')\d s'\quad,
\label{eq:tensiondef}
\end{align}
which can be derived directly  from Eq.~\ref{eq:mechvar}. 

The static equilibrium balance equations are thus given by
\begin{align}
{\bf f}^{\rm ext}&=\partial_s\left[ (\kappa\ddot{\psi} -\dot{a} F + a f){\bf n}-\tau{\bf t} \right]\nonumber\quad,\\
k_\perp(a-a_0)&=F\dot{\psi} -\kappa \ddddot{a} /2\nonumber\quad,\\
 k_0\Delta_0&=F(0)\quad .
 \label{eq:allstat}
\end{align}
 The forces and torques balances at the ends are obtained by using the boundary terms of the variations. This yields
\begin{align}
{\bf F}_{0}^{\rm ext}&=(\kappa\ddot{\psi} -\dot{a} F + a f){\bf n}-\tau{\bf t}\quad,&{\bf F}_{L}^{\rm ext}&=(\kappa\ddot{\psi}-\dot{a} F + a f){\bf n}-\tau{\bf t}\quad,\nonumber\\
T_{0}^{\rm ext}&=-\kappa\dot{\psi}+ak_0\Delta_0\quad,&T_{L}^{\rm ext}&=\kappa\dot{\psi}\nonumber\quad,\\
0&=\dot{a}=\ddot{a}\quad,&0&=\dot{a}=\ddot{a}\quad;
\label{eq:fullbcend}
\end{align}
where the equations to the left correspond to those at the basal end $s=0$ and those to the right at the distal end $s=L$. Note from Eq.~\ref{eq:allstat} that the coupling between the spacing $a$ and the angle $\psi$ is non-linear, since $F$ is of order $\psi$. This implies that for small bending the  change in spacing is negligible.

\section{Dynamics of  a cilium in fluid}
\label{sec:fluiddyn}
So far we have considered the static equilibrium of a cilium, where the mechanical passive and active elements of the axoneme are balanced by time-independent external forces. In the case in which the cilium is immersed in a fluid, the external forces to which it is subject are fluid forces. This is the basis of the description introduced in \ref{sec:basictheo}, and can be summarized by
\begin{align}
{\bf f}^{\rm ext}(s) ={\bf f}^{\rm fl}(s) \quad ,
\end{align}
where ${\bf f}^{\rm fl}$ are the forces that the fluid exerts on the axoneme along its length.

\begin{SCfigure}[\sidecaptionrelwidth][h!]
\includegraphics[]{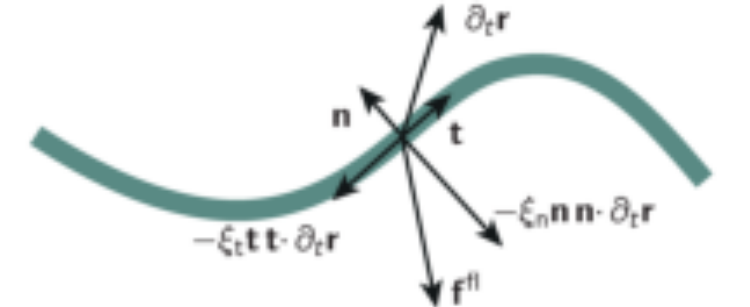}	
 	\caption{\textbf{Sketch of Resistive Force Theory:} The motion of a fragment of axoneme is characterized by the local velocity $\partial_t{\bf r}$, which can be projected  in the tangential ${\bf t}$ and  normal ${\bf n}$ directions with respect to the arc-length. The local fluid drag force opposes these components with corresponding drag coefficients $\xi_{\rm t}$ and $\xi_{\rm n}$. Since $\xi_{\rm n}>\xi_{\rm t}$ for slender rods,  ${\bf f}^{\rm fl}$ and $\partial_t{\bf r}$ are not parallel, making net propulsion possible.\label{fig:rft}}
\end{SCfigure}

In general, obtaining the fluid forces acting on a moving object is not an easy task, as it involves solving the non-linear Navier-Stokes equation. However at low Reynolds number and in the limit of slender filaments (which applies to freely swimming cilia, see discussion in section \ref{sec:basictheo}), the fluid forces take the simple form
\begin{align}
{\bf f}^{\rm fl} = -( {\bf nn}\xi_{\rm n} + {\bf tt}\xi_{\rm t})\cdotp\partial_t{\bf r}\quad .
\end{align}
This corresponds to forces opposing the moving filament with a drag coefficient $\xi_{\rm n}$  in the normal direction, and a coefficient $\xi_{\rm t}$ in the tangential direction (see Fig.~\ref{fig:rft}).

This description of the fluid-cilium interaction is known as Resistive Force Theory (RFT), and was introduced in a seminar paper by Hancock \cite{hancock_self-propulsion_1953}.  The effective friction coefficients can be related  \cite{hancock_self-propulsion_1953,gray_propulsion_1955} to the fluid viscosity $\mu$ as
\begin{align}
\xi_{\rm n}=\frac{4\pi\mu}{\log(2L/a)+1/2}\quad {\rm and}\quad\xi_{\rm t}=\frac{2\pi\mu}{\log(2L/a)-1/2}\quad.
\end{align}
Crucially, the friction coefficients are different and satisfy $\xi_{\rm n}>\xi_{\rm t}$. This asymmetry allows that, without exerting any net force on the fluid, the cilium can propel itself. We note here that the validity of RFT for ciliary beat has  been verified in the literature  for several swimming micro-organisms \cite{johnson_flagellar_1979,friedrich_high-precision_2010}, and is also verified  in Appendix A for some examples of freely swimming {\it Chlamydomonas} cilia.

\label{sec:gendym}

To collect the effect of the fluid viscosity and other viscous components inside the axoneme, we  introduce the following Rayleigh dissipation functional
\begin{align}
\label{eq:rayleigh}
R[\partial_t\Delta_0,\partial_ta,\partial_t{\bf r}] &= \int_0^L \left[\frac{\xi_{\rm t}}{2}({\bf t}\cdotp\partial_t{\bf r})^2+ \frac{\xi_{\rm n}}{2}({\bf n}\cdotp\partial_t{\bf r})^2 + \frac{\xi_{\rm i}}{2}(\partial_t\Delta)^2+\frac{\xi_\perp}{2}(\partial_t a)^2\right]\d s+\frac{\xi_0}{2}(\partial_t\Delta_0)^2
\end{align}
The first two terms inside the integral refer to fluid friction. We have also included internal sliding friction $\xi_{\rm i}$ and compressive friction $\xi_{\perp}$, both measured in  $\pN\cdotp \s\cdotp \um^{-2}$. Finally, the base also contributes to the internal sliding friction with  a coefficient $\xi_0$ (in  $\pN\cdotp \s\cdotp \um^{-1}$).

Just as the mechanical forces are obtained through variations of the mechanical work functional $G$ with respect to the corresponding fields $\{\Delta_0,a,{\bf r}\}$, the dissipative  viscous forces are obtained through variations of the Rayleigh functional with respect to their time derivatives. Thus, ignoring inertia, the force balance is now established as
\begin{align}
\frac{\delta G}{\delta {\bf r}} + \frac{\delta R}{\delta \partial_t{\bf r}} =0\quad,\quad\frac{\delta G}{\delta { a}} + \frac{\delta R}{\delta {\partial_t a}} =0\quad,\quad\frac{\delta G}{\delta {\Delta_0}} + \frac{\delta R}{\delta {\partial_t\Delta_0 }} =0\quad.
\end{align}
From this we  obtain the dynamics of the centerline ${\bf r}(s,t)$, the spacing $a(s,t)$ and the basal sliding $\Delta_0(t)$. Using the functional derivatives in Eq.~\ref{eq:mechvar}, and computing those of the Rayleigh functional we obtain the following set of dynamic equations
\begin{align}
(\xi_{\rm n}{\bf n}{\bf n}+\xi_{\rm t}{\bf t}{\bf t})\cdotp\partial_t{\bf r}&=-\partial_s\left[ (\kappa\ddot{\psi} -\dot{a} F + a f){\bf n}-\tau{\bf t} \right]\quad,\nonumber\\
\xi_\perp\partial_t a&=F\dot{\psi}-k_\perp(a-a_0)-\kappa\ddddot{a}/2\quad,\nonumber\\
\xi_0\partial_t\Delta_0&=F(0)-k_0\Delta_0\quad;
\label{eq:dynfull}
\end{align}
where we have omitted the dependences on time and arc-length,  $f$ denotes the sliding force density, and $F=\int_s^Lf(s')\d s'$ its integral. Analogously to $f$, we can define the normal force density $f_\perp$ and the basal force $F_0$. We thus have
\begin{align}
f&=f_{\rm m}-k\Delta -\xi_i\partial_t\Delta\quad,\nonumber\\
f_\perp&=k_\perp(a-a_0)+\xi_\perp\partial_ta\quad,\nonumber\\
F_0&=k_0\Delta_0+\xi_0\partial_t\Delta_0\quad .
\label{eq:dynfor}
\end{align}
Note that while all three forces contain viscous and elastic components, only the sliding force $f$ has a contribution of active motor forces. 

The equations above depend on the tension $\tau$, which is a Lagrange multiplier.  The standard procedure to obtain its value is to use the corresponding constraint equation, which in this case is $\dot{\bf r}(s,t)^2=1$. Since the dynamic equations do not directly involve ${\bf r}(s,t)$, but instead involve  its time derivative $\partial_t{\bf r}(s,t)$, we calculate the time derivative of the constraint and obtain  $\partial_t\dot{\bf {r}}^2=0$. This equation can alternatively be written as $\partial_t\dot{\bf {r}}^2=2{\bf t}\cdotp\partial_t\dot{\bf r}=0$, from which we  obtain the tension equation
\begin{align}
\ddot{\tau}-\frac{\xi_{\rm t}}{\xi_{\rm n}}\dot{\psi}^2\tau = -\frac{\xi_{\rm t}}{\xi_{\rm n}}\dot{\psi}\partial_s(\kappa\ddot{\psi} -\dot{a} F + a f)-\partial_s\left[\dot{\psi}(\kappa\ddot{\psi} -\dot{a} F + a f)\right]\quad.
\label{eq:tenfull}
\end{align}
Together with the boundary conditions given in Eq.~\ref{eq:fullbcend}, the set of Eqs.~\ref{eq:dynfull}-\ref{eq:tenfull}  constitute a complete set of integro-differential equations. Provided a prescription for the motor force $f_{\rm m}$ and the external boundary forces these equations can be solved to obtain dynamic shapes of the cilium. 

\subsection{Constrain of constant filament spacing}
\label{sec:consta}
The dynamic equations of a cilium take a particularly simple form when it is imposed that the spacing $a$ between the filaments is constant along the arc-length, that is $a(s)\to a_0$. This requires that $k_\perp\to\infty$, while keeping the normal force $f_\perp=k_\perp(a -a_0)$ finite, such that it acts as a Lagrange multiplier. The resulting dynamic  equations are
\begin{align}
(\xi_{\rm n}{\bf n}{\bf n}+\xi_{\rm t}{\bf t}{\bf t})\cdotp\partial_t{\bf r}&=-\partial_s\left[ (\kappa\ddot{\psi} +a_0 f){\bf n}-\tau{\bf t} \right]\label{eq:dynpos}\quad,\\
\xi_0\partial_t\Delta_0&=F(0) - k_0\Delta_0\label{eq:basaldyn}\quad .
\end{align}
Note, that in this limit the sliding is determined by the shape through
\begin{align}
\Delta = \Delta_0+ a_0(\psi(s)-\psi(0))\quad .
\label{eq:sliddyn}
\end{align}
Finally, the equation for the Lagrange multipliers $\tau$ (tension) and $f_\perp$ (normal force) become 
\begin{align}
\frac{\xi_{\rm n}}{\xi_{\rm t}}\ddot{\tau}-\dot{\psi}^2\tau &= -\dot{\psi}(\kappa\dddot{\psi}+ a_0 \dot{f})-\frac{\xi_{\rm n}}{\xi_{\rm t}}\partial_s[\dot{\psi}(\kappa\ddot{\psi} + a_0 f)]\quad,\label{eq:teninc}\\
f_\perp&=F\dot{\psi} \label{eq:normfor}\quad;
\end{align}
which have to be solved at every time. Note that, while $\tau$  explicitly appear in the dynamic equation, $f_\perp$ does not. Another important fact is that the expression of $f_\perp$ is quadratic, which means that for small changes in the shape the normal force is negligible. This is analogous to the non-linear coupling between spacing and bending seen before.

\begin{figure}[!ht]
\centering
\includegraphics[width=\textwidth] {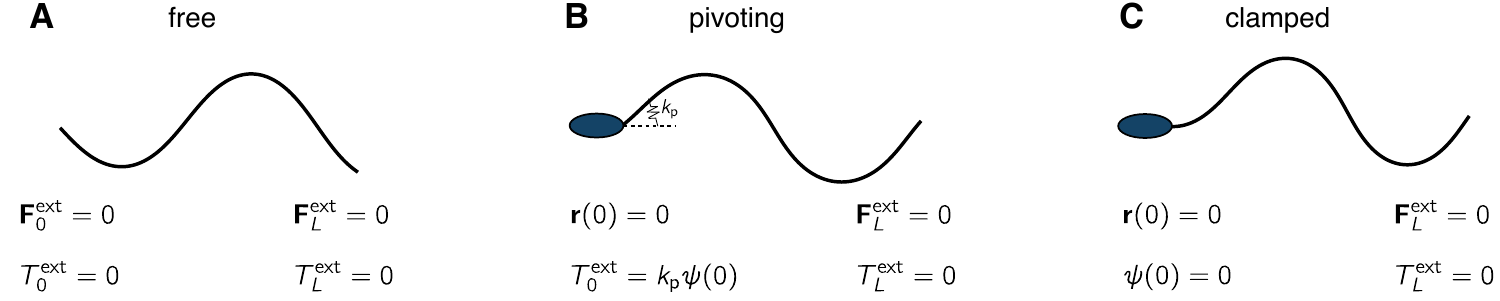}
\caption{\textbf{Schematic of boundary conditions:} A free cilium (A) is not subject to any external forces or torques. For a pivoting cilium (B) the force balance at the base is replaced by the conditions of no basal motion, and there is an additional basal torque coming from the pivoting spring. In the clamped case (C) the base does not move, and the pivot is so stiff that the tangent angle at the base remains fixed.}
\label{fig:boundaries}
\end{figure}

The boundary equations in \ref{eq:fullbcend} simplify to the following four equations
\begin{align}
\label{eq:bcdyn}
{\bf F}_{0}^{\rm ext}=(\kappa\ddot{\psi} + a_0 f){\bf n}-\tau{\bf t}\quad &,\quad{\bf F}_{L}^{\rm ext}=(\kappa\ddot{\psi} + a_0 f){\bf n}-\tau{\bf t}\nonumber\quad,\\
T_{0}^{\rm ext}=-\kappa\dot{\psi}+a_0(k_0\Delta_0+\xi_0\partial_t\Delta_0)\quad &,\quad T_{L}^{\rm ext}=\kappa\dot{\psi}\quad.
\end{align}
With adequate choices of the external  forces and torques at the basal and distal ends (${\bf F}^{\rm ext}_{0/L}$ and $T^{\rm ext}_{0/L}$), these provide the six necessary boundary conditions. In this work, we will focus on three types of boundary conditions summarized in Fig.~\ref{fig:boundaries}. These are free ends (A), in which the cilium is not subject to external forces or torques; pivoting base (B), in which the base of the cilium is held fixed, and it's constrained from rotating with a stiffness $k_{\rm p}$ while the distal end is free; and clamped base (C), in which the base is held fixed and $k_{\rm p}\to\infty$ enforces that $\psi(0)=0$ with the distal end free.

In section \ref{sec:experchlam} we described the beat of {\it Chlamydomonas} cilia using an angular representation. Motivated by this, we introduce an angular representation of the dynamic equation above by using $\partial_t\dot{{\bf r}}={\bf n}\partial_t\psi$. This directly gives
\begin{align}
\label{eq:angnonlin}
\partial_t\psi&=\xi_{\rm n}^{-1}(-\kappa\ddddot{\psi} -a_0 \ddot{f} +   \dot{\psi}\dot{\tau} + \tau\ddot{\psi}) +\xi_{\rm t}^{-1} \dot{\psi}(\kappa\dot{\psi}\ddot{\psi}+a_0f\dot{\psi}+\dot{\tau})\quad,
\end{align}
which, to linear order, is  Eq.~\ref{eq:spermold} of the introduction. Together with the tension equation and the boundary conditions, this equation provides a mechanical description of the angles of all points along the cilium.  Eq.~\ref{eq:posang} relates angle and position, and shows that the angular representation does not contain the trajectory of the basal point ${\bf r}_0(t)$. It can, however, be obtained by inserting the solution of the tangent angle in the right hand side of Eq.~\ref{eq:dynpos} and integrating ${\bf r}_0(t)$ over time.

\section{Requirements for static bending}
\label{sec:minreq}
In order to understand the mechanism regulating the shape of the cilium, we begin by analyzing the static limit. In  absence of external forces and torques,  Eqs.~\ref{eq:allstat} can be integrated. Constraining the spacing to be homogeneous (see \ref{sec:consta}), the force balances simply become
\begin{align}
\kappa\ddot{\psi}(s) &=- a_0f(s)\quad,\nonumber\\
k_0\Delta_0&= \int_0^Lf(s')\d s' \quad;
\label{eq:stat}
\end{align}
with $f=f_{\rm m}-k\Delta$ the sliding force density. The tension in this case is  null (i.e., $\tau(s)=0$), and  the normal force is given by 
\begin{align}
f_\perp= \kappa \dot{\psi}^2/a_0\quad,
\end{align}
which as indicated before is a second order term. The normal force is  always positive, indicating that filaments  tend to split apart. Finally, there are two boundary conditions
\begin{align}
\kappa\dot{\psi}(0)=a_0k_0\Delta_0\quad {\rm and}\quad \psi(0)=0\quad ,
\label{eq:bcstatuse}
\end{align}
which allow to calculate static shapes $\psi(s)$.


In the static regime  the force produced by the motors in either filament is the stall force, that is $F_{\rm A}=F_{\rm B}=F_{\rm st}$. Considering motor densities $\rho_{\rm A}$ and $\rho_{\rm B}$, we have that $f_{\rm m}=(\rho_{\rm A}-\rho_{\rm B})F_{\rm st}$. If  both filaments have the same densities of motors then $\rho_{\rm A}=\rho_{\rm B}$. In this case we have that the opposing forces balance each other and $f_{\rm m}=0$, thus the cilium doesn't bend ($\psi(s)=0$) nor does it slide ($\Delta_0=0$). When $\rho_{\rm A}\neq\rho_{\rm B}$, then  Eq.~\ref{eq:stat} gives non-trivial solutions.

\begin{figure}[!ht]
\centerline{\includegraphics{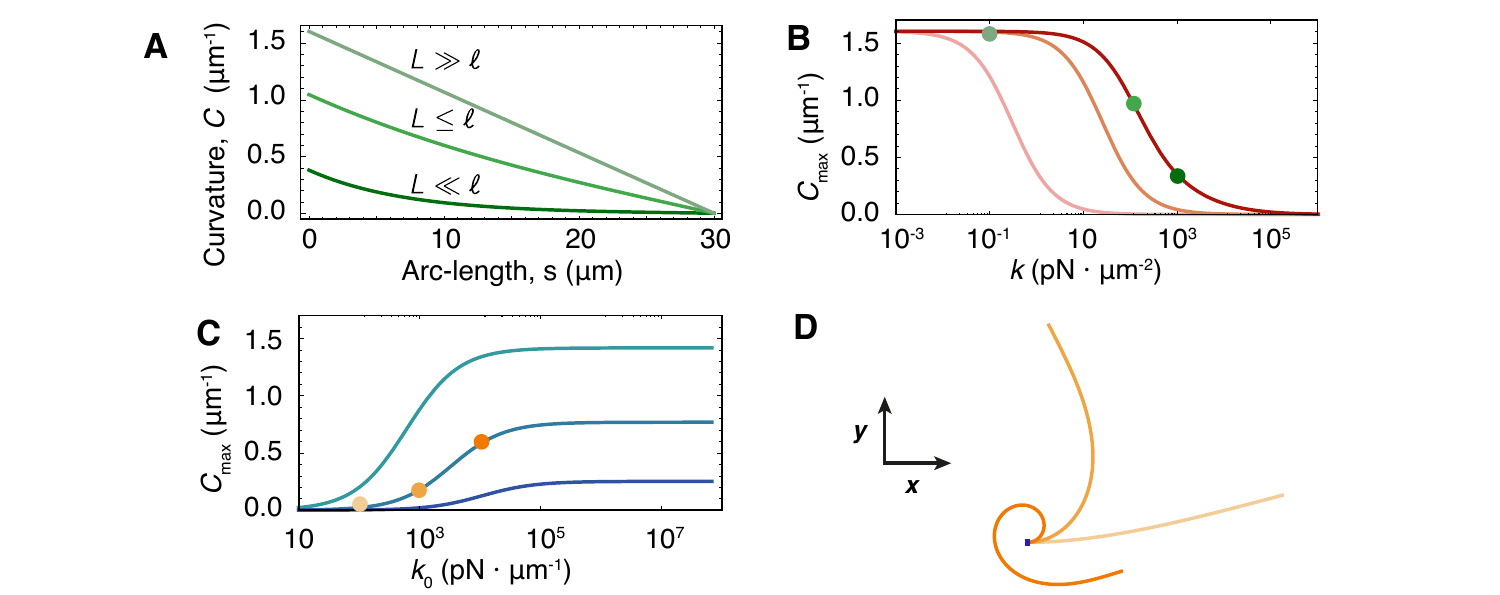}}
\caption{\textbf{Role of cross-linkers in bending.} \textbf{A.} For weak cross-linkers we have  $L\gg\ell$, and the resulting shape has a linearly decreasing curvature. As cross-linkers get stronger (values for increasing shades of green are $k=10^{-1}\,\pN/\um^2$, $k=10^2\,\pN/\um^2$ and $k=10^3\,\pN/\um^2$), $L$ becomes comparable and eventually bigger than $\ell$, which flattens the resulting shape. \textbf{B.} Dependence of maximal curvature with cross-linker stiffness for three values of basal compliance. The softer the cross-linkers, the larger the curvature. Lighter shades of red correspond to lower values of $k_0$ ($k_0=92352\,\pN/\um$, $k_0=923.5\,\pN/\um$ and $k_0=9.2\,\pN/\um$), and the dots correspond to the shapes in A. \textbf{C.} Dependence of maximal curvature with basal stiffness for three decreasing cross-linker stiffness (decreasing shades of blue correspond to $k=2280\,\pN/\um^2$, $k=228\,\pN/\um^2$ and $k=22.8\,\pN/\um^2$). Stiffer bases result in higher curvatures. \textbf{D.} Three examples of shapes corresponding to the dots in C. Other parameters are  $\kappa= 1730\,\pN\cdotp\um^2$; $L = 30\,\um$; $a = 0.185\,\um$;  $f_{\rm m}=500\,\pN/\um$ and, unless otherwise specified, $k_0 =92352\,\pN/\um $ and $k = 228\,\pN/\um^2$, which results in $\ell\approx 15$.}
\label{fig:statminreq}
\end{figure}

We can analytically solve Eq.~\ref{eq:stat} for the case of a constant motor force  along the arc-length, and obtain
\begin{align}
\psi(s) = \frac{a\ell k_0\Delta_0}{\kappa} \frac{\cosh[L/\ell]-\cosh\left[{(L-s)/\ell}\right]}{\sinh\left[L/\ell\right]}\quad ,
\label{eq:solpsi}
\end{align}
where we have used the two boundary conditions in Eq.~\ref{eq:fullbcend}, and defined the characteristic length 
\begin{align}
\ell=\sqrt{\kappa/(k a^2)}
\label{eq:charlen}
\end{align}
beyond which the sliding compliance of the cross-linkers becomes significant. The basal sliding  $\Delta_0$ is obtained  via  Eq.~\ref{eq:sliddyn}, and is
\begin{align}
\Delta_0 = \frac{ f_{\rm m}}{k + (k_0/\ell)\coth[L/\ell] }\quad .
\label{eq:solD0}
\end{align}
Note that due to the presence of static cross-linkers with stiffness $k$ the net sliding force $f=f_{\rm m}-k\Delta$ does depend on arc-length even if the motor force does not.

To characterize the static shapes given by Eq.~\ref{eq:solpsi} we first consider the limit in which the role of cross-linkers is small. According to Eq.~\ref{eq:charlen} in this case we have $L\ll \ell$. We thus expand in $L/\ell$ and to lowest order obtain
\begin{align}
\psi(s)\approx-\frac{a_0 f_{\rm st}}{\kappa}\left(\frac{s^2}{2}-sL\right)\quad .
\end{align}
This parabolic solution implies that for filaments of length $L\ll\ell$ the curvature $\dot{\psi}$ decreases linearly along the arc-length from its maximum value at the base to the minimum at the tip, as is shown in Fig.~\ref{fig:statminreq} A. As filaments get longer relative to $\ell$ the effect of cross-linkers becomes more prominent, which  makes the curvature decrease sub-linearly, and also reduces the maximum curvature at the base (see Fig.~\ref{fig:statminreq} A, darker shades of green).

A key parameter that characterizes the bent cilium is thus its maximal curvature $C_{\rm max}=\dot{\psi}(0)$, which is shown in Fig.~\ref{fig:statminreq} B as a function of $k$ for several values of  $k_0$. The maximal curvature  decreases monotonically as the cross-linkers become stiffer, and saturates for vanishing cross-linker stiffness. At the same time, the maximal curvature increases as the base of the cilium becomes stiffer ($k_0$ grows), and eventually saturates in the limit of an incompressible base for which $\Delta_0=0$ when $k_0\gg kL$, see Fig. \ref{fig:statminreq} C (three examples of shapes appear in Fig.~\ref{fig:statminreq} D). Clearly in this limit the maximal curvature still depends on the bulk cross-linker stiffness  $k$.  In particular, for $k_0=0$ there is a sliding $\Delta=f_{\rm m}/k$, but the cilium remains straight, $\psi=0$. Our analysis thus indicates that for the static bending of cilia a basal stiffness is necessary, and the stronger it is the higher the bend. Biologically, this basal stiffness can arise from the distinct properties of the basal body. Thus severing the basal body and other basal constrains is expected to straighten actively bent cilia.

\begin{SCfigure}[\sidecaptionrelwidth][h!]
\includegraphics[]{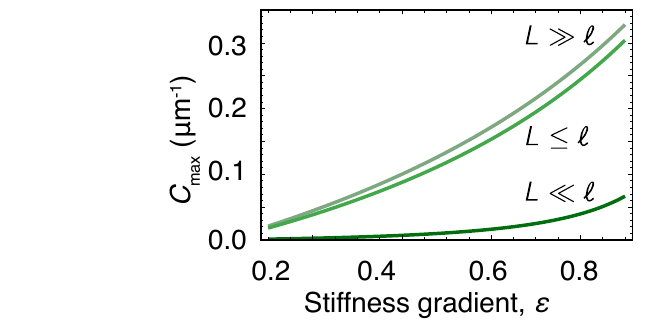}	
 	\caption{\textbf{Bending with a cross-linker gradient.} The maximal curvature grows with the  cross-linker slope $\epsilon$. Values are as in Fig.~\ref{fig:statminreq}, with increasing shades of green corresponding to $\tilde{k}= 10^{-1}\,\pN/\um^2$, $\tilde{k}=10^2\,\pN/\um^2$ and $\tilde{k}=10^4\,\pN/\um^2$.  $\ell$ is defined with $k(s=0)=\tilde{k}$. \label{fig:grad}}
\end{SCfigure}

Until now we have shown that bending requires a lateral asymmetry ($\rho_{\rm A}\neq\rho_{\rm B}$)  and a polar asymmetry which we have identified as the basal stiffness ($k_0\neq0$). An alternative way of introducing the polar asymmetry is to remove the basal constrain ($k_0=0$) but consider an inhomogeneous cross-linker stiffness. One possibility is a linear profile $k(s) = \tilde{k}(1-\epsilon s/L)$, with $\epsilon$  the slope of the linear gradient and $\tilde{k}$ the cross-linker stiffness at the base. In this case, one recovers a flat solution when $\epsilon=0$. As the gradient slope grows (see Fig.~\ref{fig:statminreq} C), a bend develops and the maximal curvature increases (see Fig.~\ref{fig:grad}). Since a linear gradient is a smaller inhomogeneity than a basal stiffness, the maximal curvature in the limit $L\gg\ell$ is smaller than in the case where a basal stiffness is present (compare Fig.~\ref{fig:grad} to Fig.~\ref{fig:statminreq} A).

\section{Conclusions}
\begin{itemize}
\item We obtained the equations of motion for a pair of inextensible filaments with a variable spacing between them that are immersed in a fluid and subject to active sliding forces. These equations show that the normal force between the filaments is a non-linear effect.

\item The static bending of a filament pair requires: {\it (i)} the presence of a lateral asymmetry, such as a motor density higher in one filament than in the other; {\it (ii)} a polar asymmetry, such as basal constraint; {\it (iii)} a filament length larger than the characteristic length $\ell$ defined by the cross-linkers.

\end{itemize}

\chapter{Dynamics of collections of motors}

{The beat of cilia is powered by the action of molecular motors.} In this chapter we introduce a mesoscopic description of the force generated by the dynein motors, and show how oscillatory instabilities  emerge as a collective property of these. We also introduce a minimal stochastic biochemical motor model in which motors are regulated through their detachment rate. We finally demonstrate that if the detachment rate is regulated by normal forces or curvature, the static shape of the cilium corresponds to a circular arc. This is in agreement with experimental data obtained for the shapes of interacting doublet pairs.

\section{Nonlinear response of the motor force}
\label{sec:motors}

Consider that a cilium is dynamically bent. Its mechanical strains and stresses, such as the sliding $\Delta$,  curvature $\dot{\psi}$, and normal force $f_\perp$, will depend on time. In such a scenario, the motor force can be generically characterized by its non-linear response to these strains and stresses. This  reflects the  idea that motors can dynamically respond to the mechanics of the cilium. For example, in the case in which motors respond to the sliding of the doublets, we will have 
\begin{align}
\label{eq:nonlinres}
f_{\rm m}(s,t)&=F^{(0)}+\int_{-\infty}^{\infty}F^{(1)}(t-t')\Delta(s,t')\d t'+\int_{-\infty}^{\infty}F^{(2)}(t-t',t-t'')\Delta(s,t')\Delta(s,t'')\d t'\d t''\nonumber\\
&+\int_{-\infty}^{\infty}F^{(3)}(t-t',t-t'',t-t''')\Delta(s,t')\Delta(s,t'')\Delta(s,t''')\d t'\d t''\d t'''+\ldots\quad ,
\end{align}
where $F^{(h)}$ is the response kernel of order $h$, and terms of order higher than cubic have not been included. Note that the $F^{(h)}$ can in principle depend on arc-length, which would account for inhomogeneities of the motors along the cilium. In this thesis however we will consider them to be arc-length independent.

For periodic dynamics with fundamental frequency $\omega$, the response equation in Fourier space is
\begin{align}
f_{{\rm m},j} = F^{(0)}+F^{(1)}_{j}(\omega)\Delta_j+ F^{(2)}_{k,j-k}(\omega)\Delta_k\Delta_{j-k}+ F^{(3)}_{k,l,j-k-l}(\omega)\Delta_k\Delta_l\Delta_{j-k-l}+\ldots\quad,
\label{eq:nonlinfou}
\end{align}
with the nonlinear response coefficients $F^{(h)}_{i,j,k,\ldots}(\omega)$ being the $h-$dimensional Fourier transform of the corresponding $h-$order kernel. The  coefficients $F^{(h)}_{i,j,k,\ldots}$ completely characterize the motor-filament interaction, and  examples of them are given in sections \ref{sec:effmot} and \ref{sec:biochem}. The linear response coefficient is of particular importance, and we use a distinct notation for it:
\begin{align}
\lambda(j\omega) = F^{(1)}_{j}(\omega)\quad,
\end{align}
where the static response is $\lambda(0)$ and the linear response to the fundamental mode is $\lambda(\omega)$. The linear dynamic response then becomes
\begin{align}
f_{{\rm m},1}=\lambda(\omega)\Delta_1\quad.
\end{align}

Analogous relations can be written for regulation via normal forces and curvature, and for each case we will use a different greek letter for the linear response coefficient. Thus for curvature response we have
\begin{align}
f_{{\rm m},1}=\beta(\omega)\dot{\psi}_1\quad,
\label{eq:curres}
\end{align}
where in this case $\beta(\omega)=F^{(1)}_{1}(\omega)$, with $F^{(1)}_j$ obtained from a relation analogous to Eq.~\ref{eq:nonlinfou} involving curvature. Finally, for normal force response we use 
\begin{align}
f_{{\rm m},1}=\gamma(\omega) f_{\perp,1}\quad,
\end{align}
with  $\gamma(\omega)=F^{(1)}_{1}(\omega)$.


\begin{SCfigure}[\sidecaptionrelwidth][ht]
\includegraphics{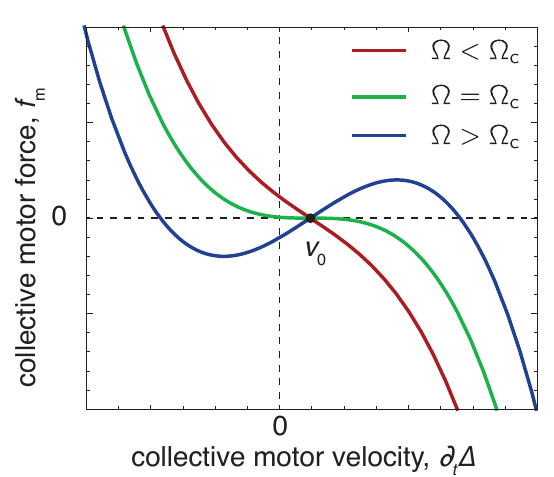}	
 	\caption{\textbf{Force-velocity curve of a collection of motors.} For values of the control parameter $\Omega$ in the subcritical region $\Omega<\Omega_{\rm c}$ the force velocity curve is monotonic. In this regime, in the absence of any force, motors move at the collective velocity $v_0>0$. At the critical value $\Omega=\Omega_{\rm c}$ an instability occurs, and  in the region where $\Omega>\Omega_{\rm c}$ the system is unstable and there are two possible collective velocities. Coupling the motors with an elastic element will then give rise to oscillations \cite{julicher_cooperative_1995}.   For  Eq.~\ref{eq:moteff} we have $\Omega=\alpha_1$, and $v_0=0$.}
 	\label{fig:forvel}
\end{SCfigure}

\section{Oscillatory instability of sliding filaments with motors}
\label{sec:effmot}
A collection of molecular motors can cooperate to produce collective behavior such as  spontaneous motion \cite{julicher_cooperative_1995} or oscillatory instabilities \cite{julicher_spontaneous_1997}. To show this, we consider a simple motor model in which the motor force responds to the local sliding velocity of filaments. In particular, we consider the following model
\begin{align}
\partial_t f_{\rm m} = -\frac{1}{\tau_0}(f_{\rm m} - \alpha_1\partial_t \Delta + \alpha_3(\partial_t\Delta)^3)\quad ,
\label{eq:moteff}
\end{align}
where the model parameters are the relaxation time $\tau_0$, the response parameter $\alpha_1$, and the saturation strength $\alpha_3>0$. These parameters do not depend on the arc-length  $s$, corresponding to a homogeneous distribution of motors along the cilium. In this description of the motor force there are no quadratic terms, which is a consequence of time-reversal symmetry. The response coefficients of this model can be easily calculated, and are
\begin{align}
\label{eq:slidresp}
F^{(1)}_{j}(\omega)=\frac{\alpha_1i\omega j}{1+i\omega j \tau_0}\quad {\rm and}\quad F^{(3)}_{j,k,l}(\omega)=\frac{\alpha_3i\omega^3jkl}{1+i\omega\tau_0(j+k+l)}\quad,
\end{align}
with $\omega=2\pi/T$ the fundamental angular frequency and $T$ the period of the oscillation.

At the steady state in which the sliding velocity  $\partial_t\Delta$ is stationary, this model is characterized by a non-linear force velocity relation schematized in Fig.~\ref{fig:forvel}, in which $\Omega=\alpha_1$ is the control parameter. While for values $\alpha_1\le0$ the force-velocity is stable, for  $\alpha_1>0$ there is a region of instability of the motor force (see Fig.~\ref{fig:forvel}). This instability may give rise to beat patterns of the cilium, which we will study in the next chapters of this thesis. To illustrate this it is helpful to  consider a very stiff cilium, with $\kappa\to\infty$. In this limit bending and fluid forces are not relevant, and setting the basal stiffness and viscosity to zero, the filament will remain straight (thus $\dot{\psi}=0$). The internal sliding forces are then balanced, and we have 
\begin{align}
f_{\rm m}-k\Delta-\xi_i\partial_t\Delta=0\quad.
\label{eq:forbal}
\end{align}
Together with the motor model  in Eq.~\ref{eq:moteff}, this equation defines a dynamical system. To see how this system can exhibit an instability, consider as an {\it ansatz} small perturbations of force and sliding which are $\propto \exp(\sigma t)$ \cite{julicher_cooperative_1995,julicher_spontaneous_1997}. In general $\sigma=\tau^{-1}+i\omega$ is complex, with  $\tau$  the characteristic relaxation time of the perturbation, and   $\omega$ the frequency of the perturbation. Using this {\it ansatz}  in the equations one obtains for $\sigma$ the following second order characteristic equation
\begin{align}
\sigma=(k+\xi_{\rm i}\sigma)(1+\tau_0\sigma)/\alpha_1\quad,
\label{eq:motchar}
\end{align}
which we now discuss.

\begin{figure}
\includegraphics[width=\textwidth]{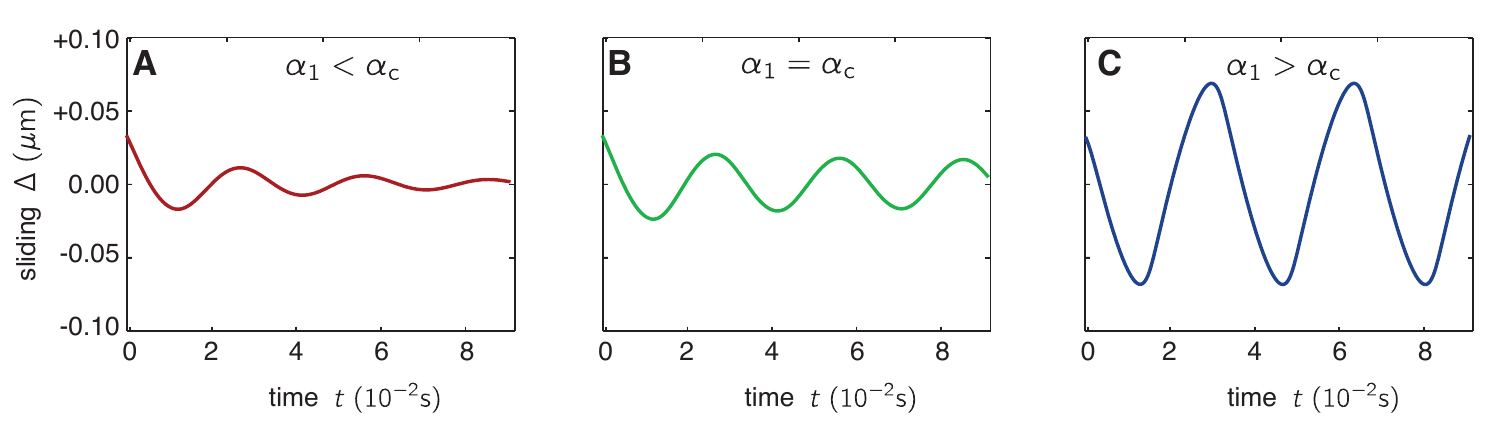}	
 	\caption{\textbf{Oscillatory instability of motor force.} \textbf{A}. For values of the control parameter $\alpha_1<\alpha_{\rm c}$ the system is stable and shows damped oscillations. \textbf{B}. At the critical point $\alpha_1=\alpha_{\rm c}$ sinusoidal oscillations occur. \textbf{C}. Far from the critical point $\alpha_1>\alpha_{\rm c}$ non-linear oscillations occur, with higher harmonics and a different frequency to that of critical sinusoidal oscillations. In this example, triangular waves. The parameters used for this numerical integration of Eqs.~\ref{eq:moteff} and \ref{eq:forbal} are $k=1812\,\pN/\um^{2}$, $\xi_{\rm i}=9.76\,\pN\s/\um^2$, $\tau=0.00378\,{\rm s}$, $\alpha_1=\{15,17.6,30\}\,\pN\,\s/\um^2$ and $\alpha_3=4\cdotp10^{-3}\,\pN\,\s^3/\um^4$. The same initial sliding was used in all cases, and the axes of all three panels are equal.}
 	\label{fig:critmot}
\end{figure}

Using the coefficient $\alpha_1$ as control parameter, we  study the stability of this dynamical system. In particular, note that for $\alpha_1=0$ the system is clearly stable and non-oscillatory, and so $\tau<0$  while $\omega=0$. As $\alpha_1$ increases the system is still  stable (i.e. $\tau<0$), but since $\omega\neq0$ it will exhibit damped oscillations as shown in Fig~\ref{fig:critmot} A. Eventually, the control parameter may reach a critical value $\alpha_{\rm c}$ for which one of the two solutions to Eq.~\ref{eq:motchar} becomes critical, that is $\tau\to\infty$. In this case small amplitude sinusoidal oscillations appear with frequency $\omega$, as can be seen in Fig.~\ref{fig:critmot} B. Finally, for values larger than the critical one (the region $\alpha_1>\alpha_{\rm c}$) non-linear oscillations occur. Their  amplitude is set by the nonlinear saturation term $\alpha_3$, see Fig.~\ref{fig:critmot} C.

Similar models can be worked out for curvature and normal force control. In such cases, however, the dynamics of the motor model cannot be studied independently from those of the filaments. It is indeed this coupling between the motor force and filaments, together with the dynamic instability of the motor force, which underlies the ciliary beat. An example of a mechanism in which the motor force responds to changes in curvature is 
\begin{align}
\partial_t f_{\rm m} = -\frac{1}{\tau_0}(f_{\rm m} - \alpha_1\dot{\psi} + \alpha_3\dot{\psi}^3)\quad .
\label{eq:moteffcc}
\end{align}
Here  $\tau_0$ corresponds to a delay, $\alpha_1$ to the linear response, and $\alpha_3$ to the saturation which controls the amplitude. The response coefficients in this case are given by \begin{align}
\label{eq:slidresp}
F^{(1)}_{j}(\omega)=\frac{\alpha_1 j}{1+i\omega j \tau_0}\quad {\rm and}\quad F^{(3)}_{j,k,l}(\omega)=\frac{\alpha_3\omega^2jkl}{1+i\omega\tau_0(j+k+l)}\quad.
\end{align}

\section{A stochastic biochemical model of ciliary motors}
\label{sec:biochem}
So far we have considered an effective description of the motor force, we now introduce a minimal stochastic biochemical model capable of giving active response to sliding, curvature, and normal force. To do so, we first note that in the description of the axoneme as a pair of opposing filaments, motors are present on filament A and filament B (see Fig.~\ref{fig:mech}). The motors rigidly attached through their stem to filament B exert with their stalks a force per motor $F_{\rm B}$ on filament A during their power stroke, and a reaction force $−F_{\rm B}$ in filament B. The analogous holds for motors with their stem attached to A. Since motors can stochastically attach and detach from the filaments, there are probabilities $p_{\rm A}$ and $p_{\rm B}$ of a motor being attached to either filament. Thus, given line densities of  motors  $\rho_{\rm A}$ and $\rho_{\rm B}$ along the two filaments, the total motor force density exerted on filament A is
\begin{align}
f_{\rm m} = \rho_{\rm B}p_{\rm B}F_{\rm B}-\rho_{\rm A}p_{\rm A}F_{\rm A}\quad ,
\end{align}
and analogously filament B is subject to a force $-f_{\rm m}$. 

In steady state where motors are stalled the motor force is $F_{\rm st}=F_{\rm A}=F_{\rm B} $, where $F_{\rm st}$ is the stall force. When the motors are moving at a certain velocity, they are characterized by a force-velocity relationship, which we consider for simplicity to be linear with slope $\alpha$. Thus for the motors in filament A we have 
\begin{align}
F_{\rm A} = F_{\rm st} -\alpha\partial_t\Delta_{\rm A}\quad,
\label{eq:singmotofr}
\end{align}
where we note that $\alpha\ge0$ acts as an active friction. Analogously, the force generated by the antagonistic motors is  $F_{\rm B} =F_{\rm st} + \alpha\partial_t\Delta_{\rm B}$. Since $\Delta_{\rm A}=\Delta$ and $\Delta_{\rm B}=-\Delta$ (see \ref{sec:geo}), we can write for the net motor density the following relation
\begin{align}
f_{\rm m} = (\rho_{\rm A}p_{\rm A}-\rho_{\rm B}p_{\rm B})F_{\rm st}+ (\rho_{\rm A}p_{\rm A}+\rho_{\rm B}p_{\rm B})\alpha\partial_t\Delta \quad .
\label{eq:biomot}
\end{align}
Considering the densities of motors to be homogeneous along the filaments, the description of the motor force is completed with the binding kinetics of the motors.

Since motors can either be bound and exert force or unbound, we have for the motors in filament A the following equation
\begin{align}
\frac{\d p_{\rm A}}{\d t} = -k_{\rm off,A}p_{\rm A} + k_{\rm on}(1-p_{\rm A})\quad,
\end{align}
and an analogous equation for the binding probabilities of motors in filament B. It is known that the detachment rate of a molecular motor can be enhanced by the forces which they sustain, and the dependence is exponential according to Bell's law \cite{bell_models_1978}. If we generalize this to also allow curvature dependent detachment, we have
\begin{align}
k_{\rm off,A} = k_0 \exp\left[\frac{F_{\rm A}}{F^{\rm c}} + \frac{f_\perp}{f_{\perp}^{\rm c}}+ \frac{C_{\rm A}}{C^{\rm c}}\right]\quad ,
\label{eq:bell}
\end{align}
where $F^{\rm c}$ is the characteristic sliding detachment force, $f_\perp$ is the normal force as defined  in Eq.~\ref{eq:dynfor}, $f_{\perp}^{\rm c}$  the characteristic normal spacing detachment force, and $C^{\rm c}$ the characteristic detachment curvature. Together Eqs.~\ref{eq:biomot} to \ref{eq:bell} form a minimal biochemical description of the motor force. This description, alternative to the generic approach used  in the previous section, also gives rise to oscillatory instabilities as can be seen in Fig.~\ref{fig:critmot}. The nonlinear response coefficients of the generic approach corresponding to this biochemical model can be obtained by expanding these equations in powers of the strains and stresses. 

\begin{figure}
\includegraphics[width=\textwidth]{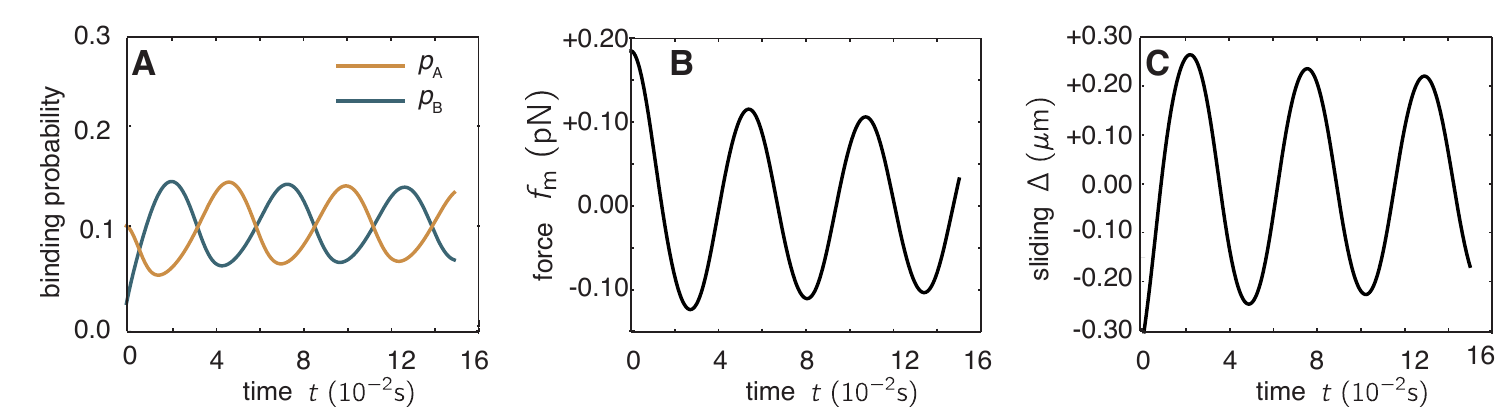}	
 	\caption{\textbf{Sliding oscillations of antagonistic biochemical motors.} \textbf{A}. The  probabilities of the motors in the rigid filaments A and B to be bound oscillate with a phase shift. \textbf{B}. The mismatch in binding probability generates a net active motor force $f_{\rm m}$, which is also oscillatory. \textbf{C}. This finally results in sliding oscillations. The parameters used for this numerical integration of Eqs.~\ref{eq:biomot} and \ref{eq:bell} are $k=1812\,\pN/\um^{2}$, $\xi_{\rm i}=9.76\,\pN\s/\um^2$, $k_{\rm on}=14\,\s^{-1}$, $k_0=2.3\,\s^{-1}$, $\alpha=0.04$, $\rho_{\rm A}=\rho_{\rm B}=10^3\,{\rm motors}/\um$, $F^{\rm c}=1\,\pN$, and $F_{\rm st}=4\,\pN$. }
 	\label{fig:critmot}
\end{figure}

In the static regime, the motors are stalled, and the motor force density takes the value
\begin{align}
f_{\rm m}=(\rho_{\rm A}-\rho_{\rm B})p_{\rm 0}F_{\rm st}\quad,
\label{eq:motstat}
\end{align}
where $p_0$ is the stall attachment probability which in the limit $C_{\rm A}=C_{\rm B}=\dot{\psi}$ of constant spacing (see Eq.~\ref{eq:curvAB}) is the same for motors on both filaments. This static attachment probability is given by 
\begin{align}
p_0 = \left(1+\frac{k_0}{k_{\rm on}}\exp\left[\frac{F_{\rm st}}{F^{\rm c}} + \frac{f_{\perp,0}}{f_{\perp}^{\rm c}}+ \frac{\dot{\psi}_0}{C^{\rm c}}\right]\right)^{-1}\quad ,
\label{eq:motstat}
\end{align}
and is independent of the static sliding, but depends on the static curvature and normal force making static regulation of bent shapes possible. Importantly, the static motor force vanishes for the case $\rho_{\rm A}=\rho_{\rm B}$. Thus, in this model static active force can only be generated when lateral symmetry  is broken. 

Following the path outlined in  the previous section, the response coefficients of the motor force to sliding, curvature, and normal forces can be calculated. In particular, the linear coefficients are
\begin{align}
\label{eq:motres}
\lambda(\omega)&= -(\rho_{\rm A}+\rho_{\rm B})p_0(1-p_0)\frac{i\omega +\omega^2p_0/k_{\rm on}}{1+(\omega  p_0/k_{\rm on})^2}\frac{F_{\rm st}}{F^{\rm c}}\alpha+(\rho_{\rm A}+\rho_{\rm B})p_0\alpha i\omega\quad, \nonumber\\
\beta(\omega)&=(\rho_{\rm A}-\rho_{\rm B})p_0(1-p_0)\frac{1-i\omega  p_0/k_{\rm on}}{1+(\omega p_0/k_{\rm on})^2}\frac{F_{\rm st}}{C^{\rm c}}\quad,\nonumber\\
\gamma(\omega)&=(\rho_{\rm A}-\rho_{\rm B})p_0(1-p_0)\frac{1-i\omega  p_0/k_{\rm on}}{1+(\omega  p_0/k_{\rm on})^2}\frac{F_{\rm st}}{f_\perp^{\rm c}}\quad.
\end{align}
Importantly, for the case $\rho_{\rm A}=\rho_{\rm B}$ in which lateral symmetry is preserved, the response coefficients to curvature and normal force vanish ($\beta(\omega)=0$ and $\gamma(\omega)=0$), but the one to sliding does not (this is only true in the limit $a\to a_0$, otherwise curvature control would produce a response to spacing). On the other hand, in the case that $\rho_{\rm A}\neq\rho_{\rm B}$ this model exhibits static regulation if it is sensitive to curvature and normal forces, since according to Eq.~\ref{eq:motstat} $p_0$ will be  a function of arc-length. In this case the linear response coefficients above will not be homogeneous along the arc-length. Finally, we note that in the sliding response coefficient only the first term is active, with the second one corresponding to protein friction.

\section{Motor regulation produces circular bends}
\label{sec:circlesmot}
In section \ref{sec:minreq} we saw that the shapes resulting from a constant motor force in the cilium are spirals, which correspond to a linear decrease in curvature. The presence of cross-linkers makes the decay of curvature even stronger (see Fig.~\ref{fig:statminreq}). However, we described in section \ref{sec:circles} experiments where the shapes of statically bent doublets were circular arcs. That is, they were characterized by a linearly growing tangent angle and thus a constant curvature (see Fig.~\ref{fig:bendtrack}). This suggests  that motor regulation as introduced in this chapter may be responsible for producing a non-homogeneous distribution of bound motors leading to the observed shapes of bent cilia.

\begin{figure}[h]
\centerline{\includegraphics[width=\textwidth]{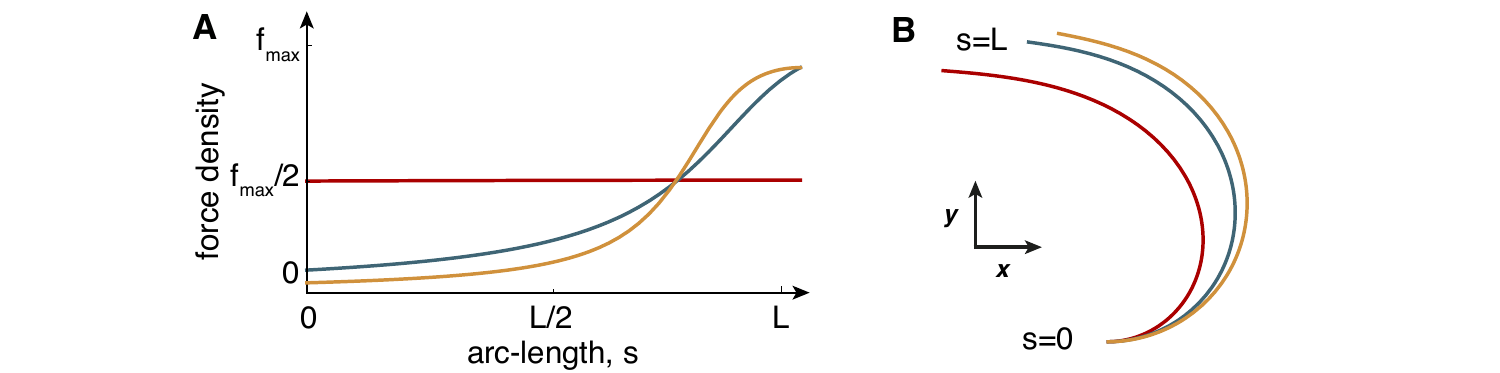}}
\caption{\textbf{Formation of arcs under force regulation.}  A uniform force distribution (panel A, red curve) produces a spiral shape (panel B, red curve). Tip-concentrated force distributions (panel A, yellow and blue curves) produce nearly circular arcs (panel B, yellow and blue curves).}
\label{fig:circsketch}
\end{figure}

We can infer the distribution of  active motors from the shape using the moment balance equation (Eq.~\ref{eq:stat}), which after one integration is $F(s)=\kappa C(s)/a_0$. This balance equation implies that the motor activity is concentrated at the distal end of the filaments pair, $s=L$. This follows because a circular arc (i.e., a shape with constant curvature) requires a constant total force, which in turn requires the sliding force density $f(s)$ to be zero except at the distal end. Indeed, in Fig.~\ref{fig:circsketch} A we show three examples of motor force distributions, with the corresponding shapes in Fig.~\ref{fig:circsketch} B obtained from numerically solving Eq.~\ref{eq:stat}. As the motor force accumulates more at the tip of the cilium, the shape becomes closer to a circular arc.

To obtain such circular shapes we consider the biochemical model from the previous section in the static limit. In this case there is no sensitivity to sliding, as $\partial_t\Delta=0$, but motors can sense curvature and normal forces. The moment balance Eq.~\ref{eq:stat} together with Eq.~\ref{eq:motstat} for the motor binding probability results in
\begin{align}
\kappa\ddot{\psi}(s) = -\frac{ \rho_{\rm A} A_0F_{\rm st} }{1+u\exp\left[\left(\frac{\dot{\psi}(s)}{\sqrt{a_0F^{\rm c}_\perp/\kappa}}\right)^2+\frac{\dot{\psi}(s)}{C^{\rm c}}\right]}\quad .
\label{eq:vikram}
\end{align}
where we have used that the normal force is given by $f_\perp=\dot{\psi}^2/a_0$ and that $\rho_{\rm B}=0$, since in the doublet experiment only the motors of one filament are exerting active forces. We have also defined the parameter $u = k_0\exp[F_{\rm st}/F^{\rm c}]/k_{\rm on}$, which is the fraction of time that the motor spends unbound. This equation leads to force concentration at the distal end, as can be seen by the following argument.

Motor sliding forces cause bending, which results in a normal force that tends to separate the filament pair. When the normal force $f_\perp(s)$ exceeds the characteristic normal force density $F_{\rm c}$, the motors detach, resulting in a decrease in sliding force. Only near the distal end, at which the curvature decreases to zero (according to the boundary condition in Eq.~\ref{eq:bcstatuse}), will the normal force fall below its critical value and the motors will remain attached. Thus, this motor regulation mechanism results in feedback: as the doublet starts to bend, the higher curvature at the base (Fig.~\ref{fig:circsketch}) causes basal motors to detach, and as the bend develops there will be a wave of detachment that only stops at the distal end. Note that this argument is equally valid for normal force and curvature regulation.

\subsection{Comparison to disintegrated doublet experiments}
\begin{figure}[h]
\centerline{\includegraphics[width=\textwidth]{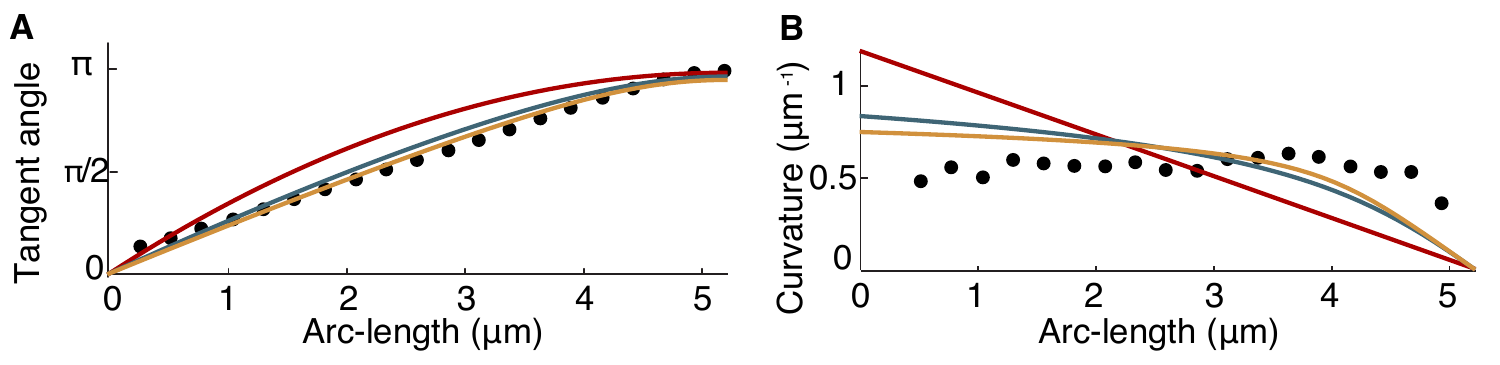}}
\caption{\textbf{ Comparison of model and tracked data.} \textbf{A.} Tangent angle vs arc-length for several control mechanisms. Red line corresponds to no control (or sliding control), yellow to normal force control,  and blue to curvature control. Black circles correspond to tracked data, note that no control departs significantly from the data. \textbf{B.} Curvature vs arc-length for three control mechanisms. The data, which is noisier than in A, shows a roughly constant curvature that decreases towards the end. Normal force control (yellow) and curvature control (blue)  show the same behavior, but no control (in red) does not.}
\label{fig:bendreg}
\end{figure}

To compare quantitatively the predictions of this model with the experimental data, we numerically integrated Eq.~\ref{eq:vikram} using the boundary conditions in Eq.~\ref{eq:bcstatuse}. For regulation via normal force, which is shown in yellow in Fig.~\ref{fig:bendreg}, we use as  parameters a stall motor force $F_{\rm st}=5\,\pN$, a motor density $\rho_{\rm A}=200\,\um^{-1}$, a bending stiffness $\kappa=120\,\pN\,\um^{2}$, a critical-motor force $f^{\rm c}_\perp=200\,\pN/\um$ and no curvature control $C^{\rm c}\to\infty$.  For the solution under curvature control which is shown in blue, we use the same parameters but a critical curvature of $C^{\rm c}=0.18\um^{-1}$ and  no normal force regulation by setting $f^{\rm c}_\perp\to\infty$. In both cases we considered the fraction of unbound time to be $u=0.1$.

As we can see in Fig.~\ref{fig:bendreg}, these two regulatory mechanisms (in blue and yellow) give rise to shapes in which the tangent angle increases roughly linearly, as corresponds to a constant curvature. They both show good agreement with the experimental data (black circles), and are clearly different from the case of no regulation (red lines). We mention now that the forces and shapes in Fig.~\ref{fig:circsketch} correspond to these solutions. That is, normal-force and curvature regulation concentrates the forces at the distal tip. In this region where motors are attached, there is a sharp decrease of the curvature, which is zero at the end of the filament (Fig.~\ref{fig:bendreg} B).

If there is no regulation by normal force or curvature, then all motors will have the same probability of being attached, resulting in a constant force per unit length. As already discussed, this results in a linearly decreasing curvature (see Figs.~\ref{fig:circsketch} and \ref{fig:bendreg}, red lines), which is not consistent with the experimental data. The sliding control model, where detachment is proportional to the sliding force experienced by the motors (given by Eq.~\ref{eq:singmotofr}), also leads to a constant force density because, at steady state, the shear force experienced by all the motors is the same, and corresponds to the stall force. Thus, the sliding control model also has a constant shear force per unit length, and, like the unregulated case, leads to a non-circular shape, inconsistent with the observations.

\begin{SCfigure}[\sidecaptionrelwidth][h!]
\includegraphics[width=.5\textwidth]{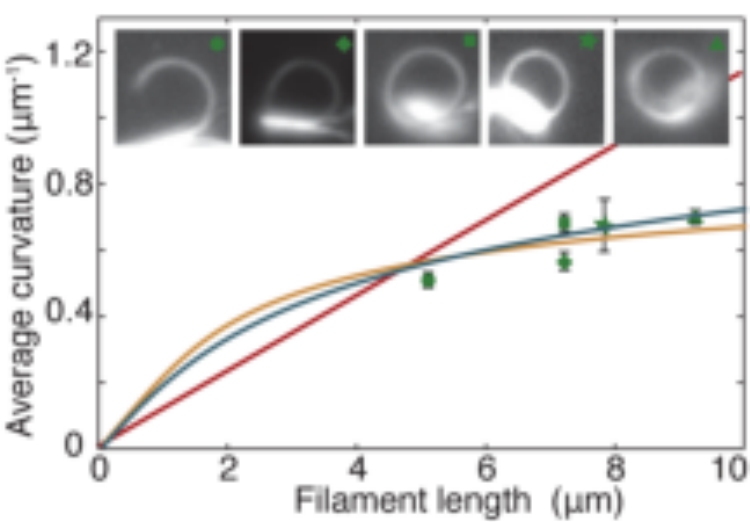}	
 	\caption[Swimming path of the axoneme.]{\textbf{Dependence of average curvature on filament length.} The curvatures of the five tracked split axonemes (insets) show a weak dependence on their length (green symbols). Curvature control (blue line) and normal-force control (yellow line) both predict the observed weak length dependence, without introducing additional parameters. In the absence of regulation, the predicted linear scaling vastly deviates from the data (red line).}
 	\label{fig:curlen}
\end{SCfigure}

To determine how curvature depends on filament length, we analyzed five pairs of microtubule doublets that showed the arcing behavior whose lengths ranged from 5 to 9 $\um$. A total of 24 arcing events were observed, with up to 8 events for a single pair of filaments. All bent into nearly circular arcs (Fig.~\ref{fig:curlen}, upper images). The average curvature (excluding the last 1 $\um$ from the distal tip) increased only weakly with the length of the doublets (Fig.~\ref{fig:curlen}, filled points). The same parameters used in Fig.~\ref{fig:bendreg} were then used to fit the curvature vs. length data for all five doublet pairs, without any additional parameters. Both the curvature control and the normal-force control models were in very good agreement with the data (Fig.~\ref{fig:curlen}, blue and yellow curves respectively). The models predict a weak dependence of the curvature on filament length because it is only the most-distal motors that generate the bending forces. By contrast, if the density of active motors were constant along the doublets, as in the sliding control model, then the average curvature would be proportional to length (integrating Eq. 3), which is inconsistent with the data (Fig.~\ref{fig:curlen}, red line).

\section{Conclusions}
\begin{itemize}

\item We developed an effective description for the non-linear dynamic response of a collection of motors to changes in curvature, sliding, and normal forces; and showed that it can give rise to oscillatory instabilities.

\item Regulation of motor detachment by normal forces or curvature gives rise to tip accumulated forces in the static limit, thus bending pairs of filaments into circular arcs. A sliding control mechanism produces homogeneous force distribution and thus spiral shapes.

\item Circular arcs similar to those obtained by the theory are observed for pairs of doublets, which supports regulation by normal forces or curvature instead of sliding.

\end{itemize}
%
%

	
\chapter{Non-linear dynamics of the cilia beat}

{Cilia exhibit a great variety of beating patterns.} In this chapter we analyze how different mechanisms of motor regulation give rise to non-linear beat patterns which allow ciliary propulsion. The critical beats are analyzed analytically to linear order. Finite amplitude beats, which are non-linear, are supercritical. We investigate them by numerically integrating the time-domain equations of motion.

We first consider sliding regulation, and show that wave propagation is only possible for long cilia and in the presence of a boundary asymmetry. This is the case of single-flagellates such as {\it Bull Sperm}, where the cilium is long and the head provides the required asymmetry. We further show that under sliding control the boundaries play a fundamental role in determining the direction of wave propagation.

We then move on to curvature control, where we show that the direction of wave propagation is determined by the motor parameters instead of the boundaries. Furthermore, under curvature control wave propagation is possible for short cilia such as those of {\it Chlamydomonas}.

\section{Bending waves under sliding control}

It was shown in section \ref{sec:effmot} that motors regulated by sliding of the filament pair can become dynamically unstable. Consider a motor model characterized by its response coefficients $\lambda_{i,j,\ldots}^{(h)}$ to sliding. Further, we assume that the motor does not  couple to curvature or normal forces. In such a case, we say that motors are {\it controlled} by sliding.

\begin{figure}[!ht]
\centering
\includegraphics[width=\textwidth] {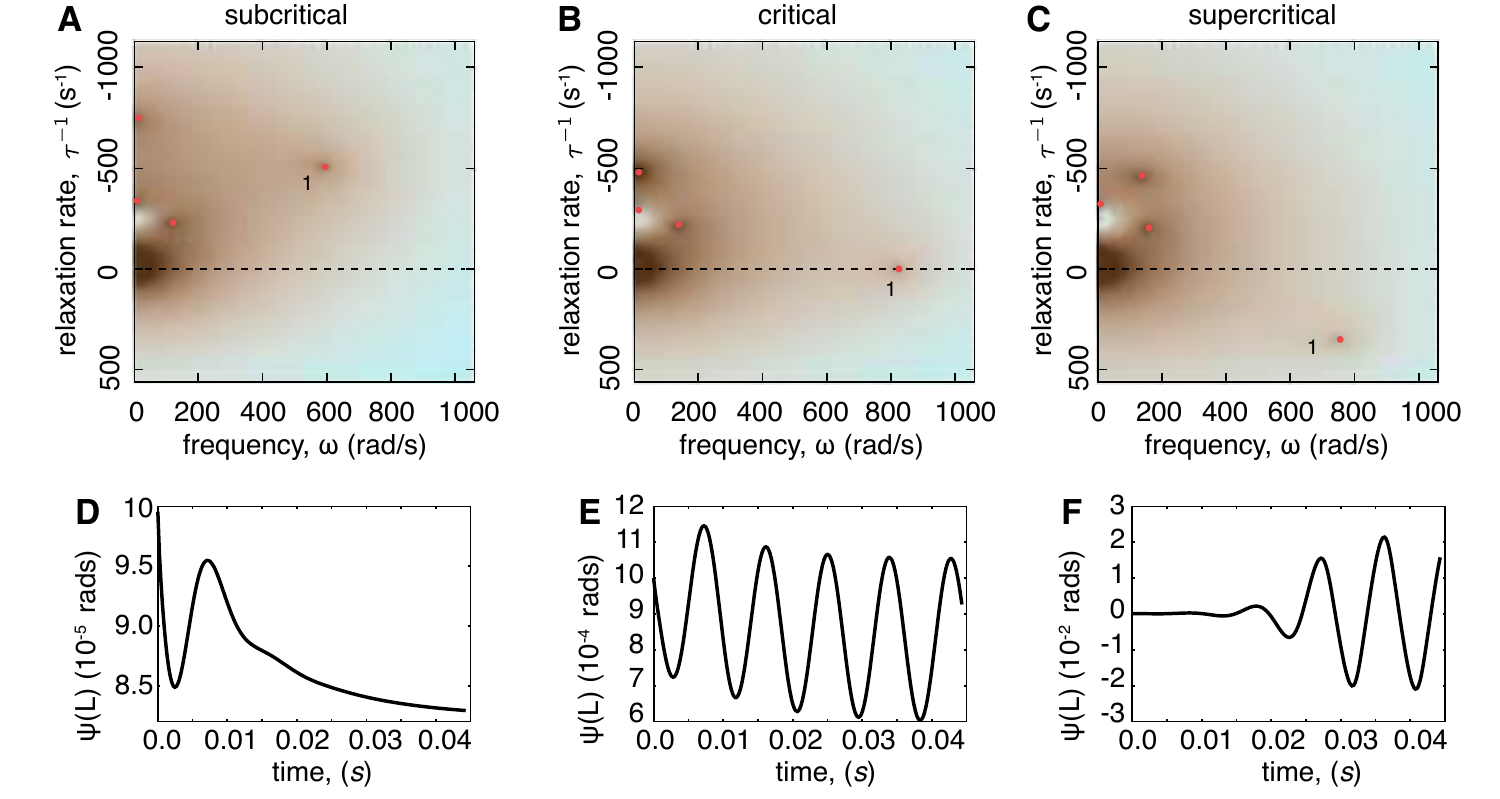}
\caption{\textbf{Onset of modes instability in cilia.} \textbf{A, B, C.} The value of $|\Gamma|^{-1}$ (with $\Gamma$ the boundary value determinant of a free cilium) is plotted as a function of the relaxation rate $\sigma'$ and  frequency $\sigma''$ using the motor model in Eq.~\ref{eq:moteff} (red dots correspond to divergencies, where modes lie). In the subcritical regime  all modes are stable, with negative relaxation rate. In the critical regime the mode denoted by 1 becomes unstable with a zero relaxation rate. In the supercritical regime mode 1 is unstable.  \textbf{D, E, F.} Time traces for the tip angle after solving the non-linear dynamic equations for the values used in A, B and C correspondingly. There is a clear transition from damped, to critical, and supercritical oscillations. Parameters used as in Fig.~\ref{fig:freesols}, and for the motor model $k=228\,\pN/\um^{2}$, $\xi_{\rm i}=0.1\,\pN\cdotp\s/\um^{2}$, $\alpha_3=2\cdotp10^{-4}\,\pN\cdotp \s^3/\um^4$, $\tau_0=0.0038\,\s$ and $\alpha_1=0.8\,\pN\cdotp \s/\um^2$ (for A and D), $\alpha_1=1.24\,\pN\cdotp \s/\um^2$ (for B and E) and $\alpha_1=1.52\,\pN\cdotp \s/\um^2$ (for C and F).}
\label{fig:instability}
\end{figure}

Consider now that at time $t=0$ a small complex sliding perturbation $\Delta(s,t)=\Delta(s)\exp(\sigma t)$ is turned on, where in general $\sigma=\tau^{-1}+i\omega$ with $\tau$ the relaxation time of the perturbation and $\omega$ its frequency. The effect of the sliding perturbation on the force can then be described by
\begin{align}
f(s,t)=-\chi(\sigma){\Delta}(s)\exp(\sigma t),
\end{align}
where $\chi(\sigma)$ is the sliding response coefficient. This coefficient  combines passive and active elements, and is given by 
\begin{align}
\chi(\sigma)=-\lambda(\sigma)+k+\sigma\xi_{\rm i}\quad,
\end{align}
with $k$ the sliding stiffness, $\xi_{\rm i}$ the internal viscosity, and $\lambda(\sigma)$ the response coefficient of the motor force. The response to small perturbations $\lambda(\sigma)$ can be calculated by inserting the perturbation {\it ansatz} in the non-linear response Eq.~\ref{eq:nonlinres}. Keeping the linear order results in $\lambda(\sigma)=\int_0^\infty\lambda(t-t')\e^{-\sigma(t-t')}\d t'$, which is the analytical continuation of the Fourier coefficient $\lambda(\omega)$ introduced before.

The tangent angle likewise evolves according to $\psi(s,t)=\psi(s)\e^{\sigma t}$. We can obtain an equation for the complex amplitude $\psi(s)$ of the angle by linearizing Eq.~\ref{eq:angnonlin}, which results in
\begin{align}
\sigma\xi_{\rm n}\psi(s)=-\kappa\ddddot{\psi}+a^2{\chi}(\sigma)\ddot{\psi}\quad.
\label{eq:sigma}
\end{align}
Provided the dependence on $\sigma$ of the response $\chi(\sigma)$  and a choice of boundary conditions, this boundary value problem can be solved. Using the boundary conditions a discrete set of solutions or modes can be obtained, each characterized by a relaxation time $\tau$ and an angular frequency $\omega$. 


In Fig.~\ref{fig:instability} we show the stability of the different modes of the motor model in \ref{sec:effmot}, for which $\lambda(\sigma)=\alpha_1\sigma/(1+\sigma\tau_0)$ with $\tau_0$ the relaxation time of the motors, and $\alpha_1$ the control parameter. For values $\alpha_1<\alpha_{\rm c}$ all modes (which appear as red dots in Fig.~\ref{fig:instability} A) have a negative relaxation rate $\tau<0$ and are thus stable. Their response is characterized by a damped oscillation of frequency $\omega$ (see Fig.~\ref{fig:instability} D). At the critical value $\alpha_1=\alpha_{\rm c}$ one mode, which we noted 1 in Fig.~\ref{fig:instability} B, becomes critical: it's characterized by $\tau\to\infty$. In this case the system shows small amplitude oscillations (see \ref{fig:instability} E). Finally, in the supercritical regime $\alpha_1>\alpha_{\rm c}$,  the system is stabilized by non-linearities and shows large amplitude oscillations (see Fig.~\ref{fig:instability} F, where the full non-linear problem was solved).


\subsection{Critical beats in frequency domain}
\label{sec:critbeat}
For a cilium that beats at its critical poin, a perturbation does not decay nor does it increase, as $\tau\to\infty$. Critical perturbations thus evolve as $\exp[\sigma t]=\exp[i\omega_{\rm c}t]$, producing sustained small amplitude oscillations of critical frequency $\omega_{\rm c}$. Since at the critical point only one harmonic is present, we have
\begin{align}
\psi(s,t) = {\psi}_1(s)\e^{i\omega_{\rm c}t}+{\psi}_1^*(s)\e^{-i\omega_{\rm c}t}\quad,
\label{eq:oneharm}
\end{align}
where we have used that the angle is a real quantity and thus ${\psi}_{-n}={\psi}^*_{n}$, and the sub-index 1 denotes the fundamental mode. For simplicity, we are considering that the oscillation has no static mode. Equations analogous to \ref{eq:oneharm} apply to the basal sliding and the motor force. Further, the relationship between the sliding force and the sliding displacement is given by
\begin{align}
{f}_1(s)=-\chi_{\rm c}{\Delta}_1(s)\quad,
\label{eq:critsli}
\end{align}
where $\chi_{\rm c}=\chi(\sigma=i\omega_{\rm c})$ with the control parameter at the critical point, $\Omega=\Omega_{\rm c}$. From now on we drop the sub-index $1$, since it applies to all geometrical and mechanical variables of the system. The relation between basal sliding $\Delta_0$ and basal force $F_0$ is
\begin{align}
F_0=\chi_0\Delta_0\quad,
\end{align}
with the critical value of the passive basal compliance $\chi_0=k_0+i\omega_{\rm c}\xi_0$.

Consistent with the presence of just one harmonic, at the critical point the amplitude of the oscillations is small. Critical oscillations are thus described by the linear equation \ref{eq:sigma}. Using $\sigma=i\omega_{\rm c}$ and  $\chi=\chi_{\rm c}$ results in
\begin{align}
\label{eq:sperm}
i\bar{\omega}_{\rm c}\psi=-\ddddot{\psi}+\bar{\chi}_{\rm c}\ddot{\psi}\quad,
\end{align}
where bars denote dimensionless quantities, and we have used the following transformations
\begin{align}
\label{eq:dimrules}
\bar{s}&=\frac sL \quad,&  \bar{t}&=t\omega\quad,&  \bar{\tau}&=\frac{L^2}{\kappa}\tau \quad,& \bar{\Delta}&=\frac{\Delta}{a_0}  \quad, &  \bar{f}&=\frac{a_0 L^2 f}{\kappa}\quad,\nonumber\\
 \bar{k}_{0}&=\frac{a_0^2L}{\kappa}k_{0}  \quad,& \bar{\xi}_{0}&=\frac{a_0^2L\omega}{\kappa }  \xi_{\rm i}\quad,&   \bar{k}&=\frac{a_0^2L^2}{\kappa }k \quad,& \bar{\xi}_{\rm i}&=\frac{a_0^2L^2\omega}{\kappa }  \quad,& \bar{\lambda}&=\frac{a_0^2L^2}{\kappa }\lambda\quad,\nonumber\\
\bar{\xi}&=\frac{\xi_{\rm n}}{\xi_{\rm t}}\quad,&\bar{\omega}&=\frac{ L^4\omega\xi_{\rm n}}{\kappa }\quad,&\bar{\chi}_0&=\bar{k}_0+i\bar{\xi}_0\quad,&\bar{\chi}_{\rm c}&=\bar{k}+i\bar{\xi}_{\rm i}-\bar{\lambda}\quad.
\end{align}

\begin{figure}[!ht]
\centering
\includegraphics[width=\textwidth] {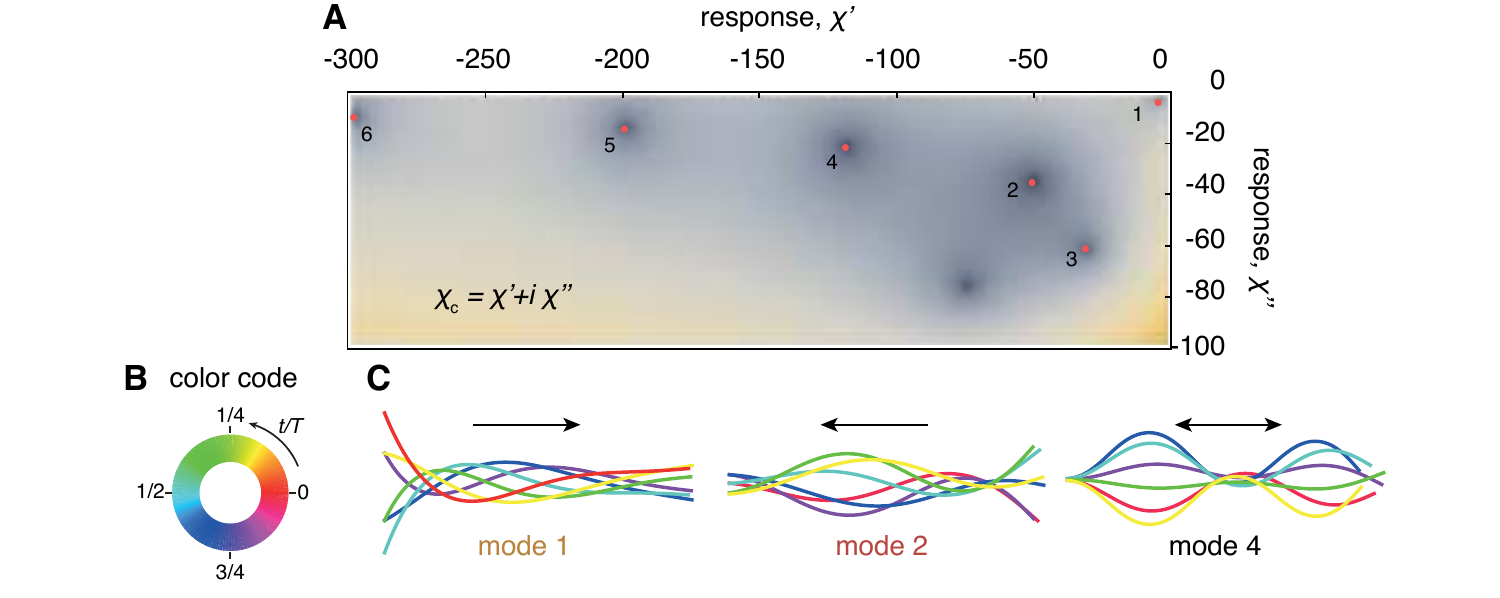}
\caption{\textbf{Critical beats of free cilia under sliding control.} \textbf{A.} Space of critical solutions for free ends. The value of $|\Gamma|^{-1}$ is plotted as a function of $\chi$, where $\Gamma$ is the determinant to the boundary value problem. Yellow to blue corresponds to increasing values, and divergences (red dots) correspond to critical modes. These are numbered by $|\chi_n|<|\chi_{n+1}|$ (one divergence corresponds to a trivial solution, and is not marked). \textbf{B.} Color code used throughout this thesis, as time progresses over one period the shade of the cilium changes color counter-clockwise. \textbf{C.} Three  critical beats for free cilia. While mode 1 shows a forward (base-to-tip) traveling wave which results in swimming towards the left, mode 2 has a backward traveling wave, and mode 4 shows no wave propagation, being a standing mode. Arrows denote direction of wave propagation. The parameters used were the typical ones of {\it Bull Sperm}: $\kappa=1730\,\pN\cdotp\um^2$, $\xi_{\rm n}=0.0034\,\pN\cdotp \s/\um^2$,  $L=58.3\,\um$, $\omega_{\rm c}=2\pi\cdotp 20\cdotp \Hz$, $a_0=0.185\,\um$ and $k_0=\infty$ (which ensures no basal sliding).}
\label{fig:freesols}
\end{figure}

Note that, since Eqs.~\ref{eq:normfor} and \ref{eq:teninc} are non-linear, tension and normal force vanish to first order, $\tau=0$ and $f_\perp=0$. We will see in the next chapter that this does not apply to asymmetric ciliary beats. Given the critical frequency $\omega_{\rm c}$, the  linear ordinary differential equation for $\psi(s)$ is an eigenvalue problem, which has a discrete set of solutions for the eigenvalues
\begin{align}
\chi_{\rm c}=\chi_n\quad {\rm with}\;\; n=1,2,\ldots
\end{align}
The set of eigenvalues $\chi_n$ is determined by the condition that the determinant $\Gamma(\bar{\chi}_{\rm c};\bar{\omega}_{\rm c},\bar{\chi}_0)$ of the homogeneous system of equations imposed by the boundary conditions is zero (see Appendix B for details). They are numbered according to the rule $|\chi_n|<|\chi_{n+1}|$.

In Fig.~\ref{fig:freesols} we show the value of $|\Gamma(\chi'+i\chi'')|^{-1}$ over the space of negative values of $\chi'$ and $\chi''$, which is where the active motor response $\lambda$ dominates over the passive response of $k+i\omega_{\rm c}\xi_{\rm i}$. The value of $\bar{\omega}_{\rm c}\sim10^{4}$ is a typical one for {\it Bull Sperm}, and the basal compliance has been taken very stiff $|\chi_0|\sim10^{5}$. Red dots correspond to divergences where a mode exists. A time-trace of some of the modes is also shown in  Fig.~\ref{fig:freesols}. The first mode for a free axoneme is forward swimming. The wave propagates forward from base to tip, thus allowing for propulsion towards the left. This mode is also the one that was numbered  1 in Fig.~\ref{fig:instability}. This is the first to become unstable with the motor model of \ref{sec:effmot}. The second mode is backwards swimming. Finally, there is an infinite number of standing modes,  of which an example is shown (see Appendix B for phase and amplitude profile of modes 1 and 2).

At the critical point the characteristics of the modes, such as direction of wave propagation or amplitude profile, depend on the boundary conditions, on the basal compliance $\bar{\chi}_0$, and on the dimensionless frequency $\bar{\omega}$. But besides this, they are generic and independent of the motor model dynamics. That is, they are independent of the functional dependence of $\bar{\chi}$ on $\bar{\omega}$, as long as $\chi(\omega_{\rm c})=\chi_n$ for a certain $n$. For now we restrict ourselves to the case of free cilia, and study the effect of $\bar{\chi}_0$ and $\bar{\omega}_{\rm c}$ on the modes. In particular, the dimensionless frequency $\bar{\omega}$ can be understood as a ratio of two length scales
\begin{align}
\bar{\omega} =\left( \frac{L}{\ell_0}\right)^4\quad,\quad\quad {\rm with}\quad\quad \ell_0=\left(\frac{\kappa}{\xi_{\rm n}\omega}\right)^{1/4}\quad.
\end{align}
The characteristic length $\ell_0$ is the typical wavelength of a cilium when driven from a  boundary. It is also the length beyond which hydrodynamic effects become negligible \cite{machin_wave_1958}. The length $L$ is simply the length of the flagellum.

\begin{figure}[!ht]
\centering
\includegraphics[width=\textwidth] {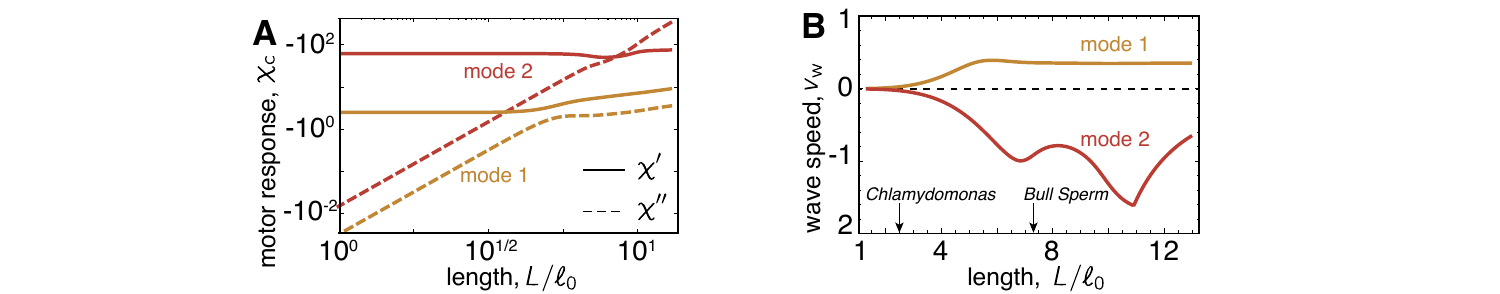}
\caption{\textbf{Length scaling of modes.} \textbf{A.} Scaling of the real ($\chi'$, solid lines) and imaginary ($\chi''$, dashed lines) parts of the eigenvalues for the first and second modes in Fig.~\ref{fig:freesols}. The real part vanishes (linearly) for short lengths. \textbf{B.} Scaling of the wave velocity (see Eq.~\ref{eq:velw}) with the length of the cilium, arrows mark typical values for  {\it Chlamydomonas} and {\it Bull Sperm}. Short cilia show no wave propagation, as all solutions become standing waves. Long cilia allow forward (in mode 1) and backward (in mode 2) wave propagation. The basal stiffness is $k_0\to\infty$, which prevents basal sliding.}
\label{fig:sliscaling}
\end{figure}

For short cilia $L\ll\ell_0$, we have $\bar{\omega}_{\rm c}\ll1$. In the limit of $\bar{\omega}_{\rm c}\to0$ the complex component of the response vanishes, $\chi''\to0$, as can be seen in Fig.~\ref{fig:sliscaling} A \cite{camalet_generic_2000}. In this regime viscous components (of the surrounding fluid and the sliding force) vanish and the phase profile of $\psi(s)$ becomes flat  for all modes. They become standing modes, which do not allow for wave propagation. This effect is better described using the wave propagation velocity 
\begin{align}
v_{\rm w}=-\int_0^L|\psi(s)|^2\partial_s\arg(\psi(s)) \d s\quad,
\label{eq:velw}
\end{align}
which is proportional to the swimming velocity \cite{camalet_generic_2000}. Note that, for waves propagating from base to tip, the wave velocity is positive; while for waves propagating from tip to base it is negative.  This can be seen in Fig.~\ref{fig:sliscaling} B, where $v_{\rm w}$ has been plot as a function of $L/\ell_0$, as cilia get shorter their wave velocity decreases, vanishing asymptotically as $L/\ell_0\to0$. It is important to note that, while in {\it Bull Sperm} $L/\ell_0$ is sufficiently large to allow for wave propagation, this is not the case for {\it Chlamydomonas}. This suggests that {\it Chlamydomonas} cilia are not regulated by sliding control.

\begin{figure}[!ht]
\centering
\includegraphics[width=\textwidth] {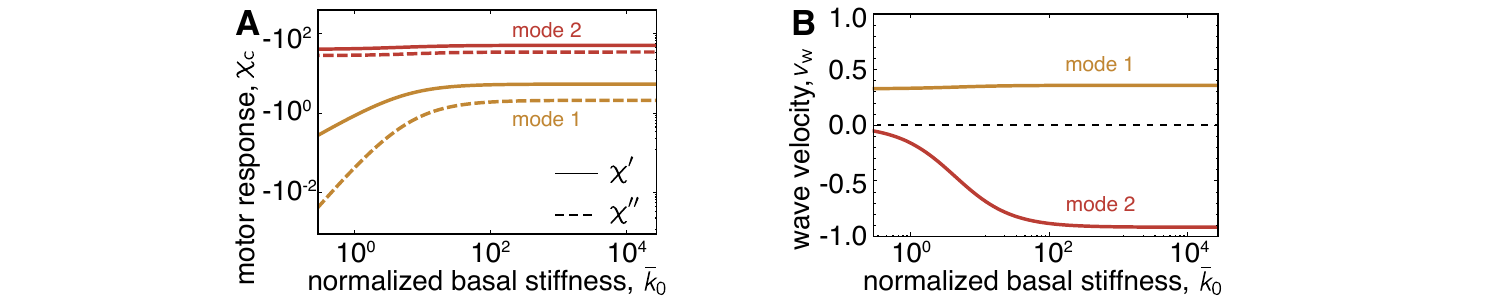}
\caption{\textbf{Scaling of modes with normalized basal stiffness.} \textbf{A.} Since mode 2 is a standing mode it does not vanish for small values of $\bar{k}_0=a_0^2Lk_{0}/\kappa$. Mode 1 however is a directional mode, and vanishes in the absence of basal compliance (as $k_0\to0$  we have $\chi'\propto k_0$ and $\chi''\propto k_0^{2}$). \textbf{B.} Mode 1, which is a directional mode, does not substantially alter its directionality as the basal stiffness changes. Mode 2 looses its directionality when the base decreases its stiffness. Parameters as those in Fig.~\ref{fig:freesols}, which correspond to $L/\ell_0=7.3$.}
\label{fig:basal}
\end{figure}

The basal compliance $\chi_0$ has a fundamental role in regulating the flagellar beat, particularly when motors are controlled by sliding. The reason is that Eq.~\ref{eq:sperm} is symmetric around the midpoint $s=1/2$, thus not reflecting the polar asymmetry of the axoneme (see section \ref{sec:polarasym}). The breaking of polar symmetry has to come from the boundary conditions or, for a free axoneme, from the presence of a basal compliance $\chi_0\neq0$. To better understand this we expand the eigenvalues in $k_0$: $\chi_n=\chi_n^{(0)}+k_0\chi_n^{(1)}+k_0^2\chi_n^{(2)}+\ldots\;$. Introducing this in the equation of the determinant and matching terms we can determine the coefficients $\chi_n^{(i)}$. We distinguish between {\it standing modes} (such as mode 2), for which $\chi_n^{(0)}\neq0$; and {\it traveling modes} (such as 1), for which $\chi_n^{(0)}=0$ (see Fig.~\ref{fig:basal} A).

Standing modes do not allow for wave propagation when $k_0=0$, in this limit they are standing waves. So  while static bending is impossible in the absence of a basal constrain, as  shown in section \ref{sec:minreq}, dynamic solutions do exist. These standing modes gain directionality as the basal compliance grows, resulting in an increase of wave velocity (see Fig.~\ref{fig:basal} B, mode 2). On the other hand, traveling modes vanish when $k_0=0$, but for an arbitrarily small basal compliance allow for strong wave propagation (see mode 1 in Fig.~\ref{fig:basal} A).

\subsection{Non-linear  beats in time-domain}

Although Fig.~\ref{fig:instability} B shows that mode 1 is the first to become unstable, in order to obtain finite amplitude beat patterns it is necessary to numerically integrate the set of Eqs.~\ref{eq:basaldyn}-\ref{eq:angnonlin} supplemented with Eq.~\ref{eq:moteff} for the motor force in the supercritical regime. This system of equations is a set of non-linear coupled partial differential equations of fourth order with an integral term for the basal sliding. The oscillatory transition corresponds to a Hopf bifurcation, and as the control parameter $\alpha_1$ moves away from the critical point $\alpha_{\rm c}$ the amplitude of the oscillation grows as $(\alpha_1-\alpha_{\rm c})^{1/2}$. Furthermore, the saturation term $\alpha_3$ of the motor model also controls the amplitude, in this case following the inverse scaling $\alpha_3^{-1/2}$ (see Appendix B, where both these scalings where verified). Thus by sufficiently decreasing  the saturation term and increasing the control parameter, finite amplitude beats will occur. We note here that, since sliding control only produces wave propagation for long cilia, we consider parameters such that $L/\ell_0\approx7$, as is the case in {\it Bull Sperm}.

\begin{figure}[t]
\centering
\includegraphics[width=\textwidth] {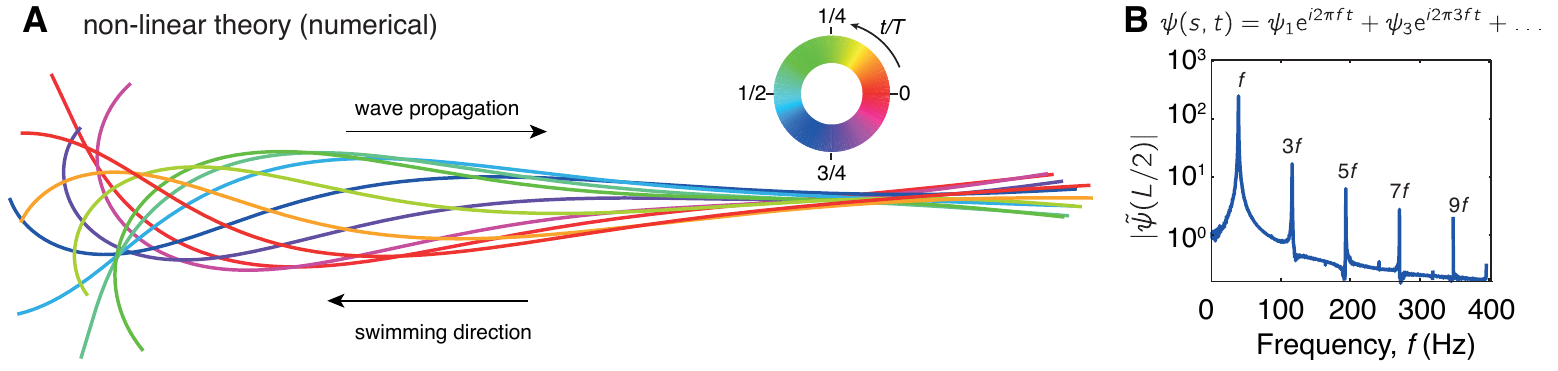}
\caption{\textbf{Non-linear beat pattern of a free cilium.} \textbf{A.} Time trace of the swimming cilium with mode 1 dominating (see Fig.~\ref{fig:freesols}). The base is to the left, and the wave propagates right towards the tip, making the cilium move towards the left. Parameters as those in Fig.~\ref{fig:instability}, with $\alpha_1=5.2\,\pN\cdotp \s/\um^2$ and $\alpha_3=2\cdotp10^{-3}\,\pN\cdotp \s^3/\um^4$.  \textbf{B.} The power spectrum of the angle at the mid-point of the cilium (obtained through a Discrete Fourier Transform of the numerical integration) shows peaks in the odd harmonics, due to the antisymmetry of the problem to the transformation $t\to-t$.}
\label{fig:detailedfree}
\end{figure}

The numerical integration of this dynamical system is far from trivial, and the custom-made numerical algorithm is described in Appendix B. In Fig.~\ref{fig:detailedfree} A we can see the time trajectory of a typical beat pattern in the supercritical regime. This beat corresponds to mode 1 in the non-linear regime, and it shows a forward traveling wave which makes the free cilia swim. To show that this beat is non-linear we  plot the power spectrum of the mid-angle, which exhibits higher odd harmonics. The lack of even harmonics is due to the odd parity of the equations of motion and the motor model.

\begin{figure}[!htb]
\centering
\includegraphics[width=\textwidth] {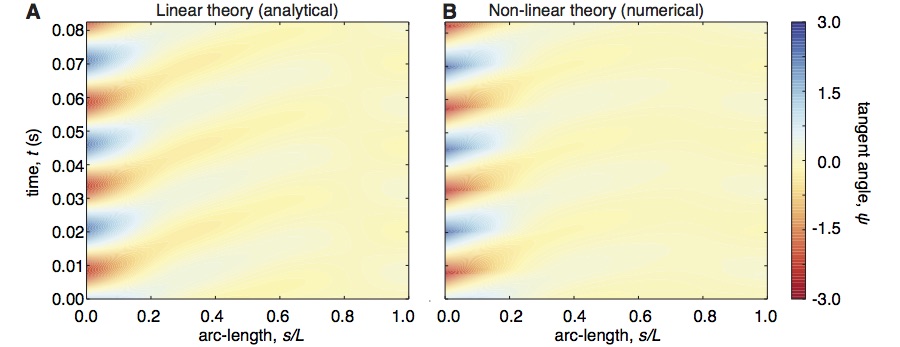}
\caption{\textbf{Tangent angle of linear and non-linear  solution.} \textbf{A.} Kymograph of the tangent angle of the analytical critical  solution over three periods. As one can see the waves propagate from base to tip. It is also clear that the amplitude decreases towards the tip. Height normalized to match that of the non-linear solution. \textbf{B.} Same as A but for the non-linear solution  obtained numerically. We can see that frequency, wave profile and amplitude are in good agreement with the analytical linear solution. Parameters as in Fig.~\ref{fig:detailedfree}.}
\label{fig:kymo}
\end{figure}

To better compare this non-linear beat pattern with the critical (thus, linear) beat of mode 1, we show in Fig.~\ref{fig:kymo}  a kymograph of the angle $\psi(s,t)$ by adjusting the amplitude of the linear analytical solution to that of the non-linear numerical solution. As one can readily see, the beat patterns are very similar. The frequency is roughly the same, and in both cases the wave clearly propagates towards the tip (forward) with a slowing down of the propagation towards the end. Furthermore, in both cases the amplitude of the beat decays over arc-length, with the decay being stronger in the non-linear theory than in the linear one.

\subsection{Role of boundary conditions}
\begin{figure}[!ht]
\centering
\includegraphics {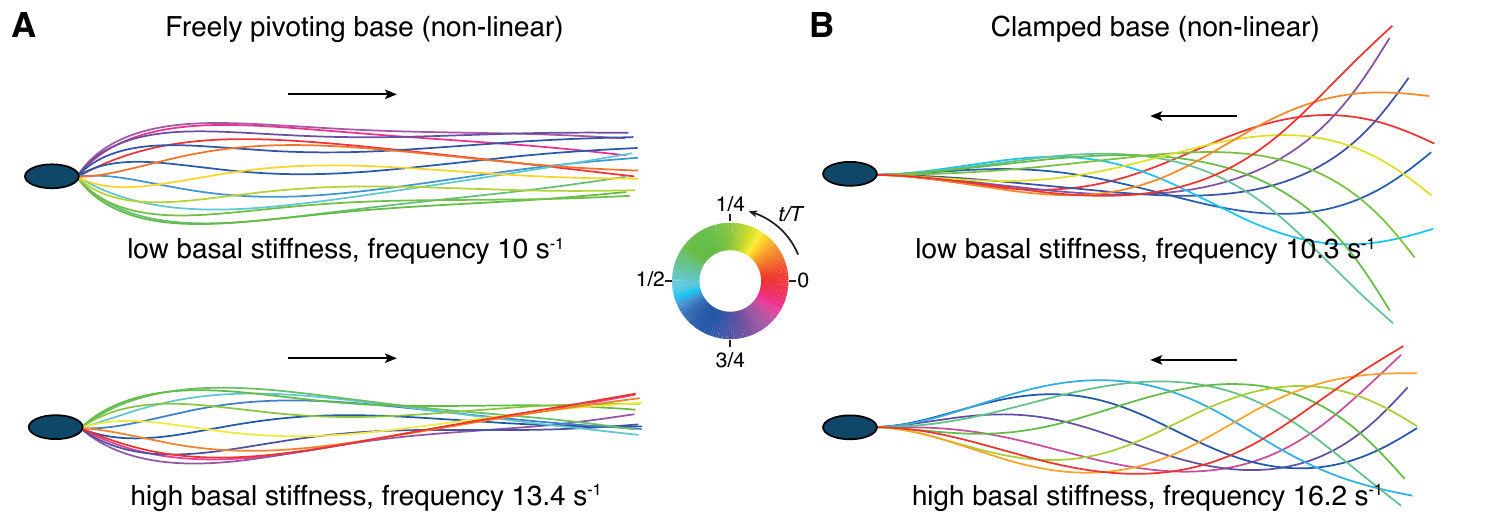}
\caption{\textbf{Beat patterns with different boundary conditions.}  \textbf{A.} Beat pattern for a freely pivoting base, the wave travels and the basal stiffness has little relevance. \textbf{B.}  Beat patterns for a clamped base show a backwards traveling wave. Changing the basal stiffness produces a significant change in the beat pattern. In A top $\xi_{0}=26.8\,\pN\cdotp s/\um$,  $k_0=10^2\,\pN/\um$, $\alpha_1=16.8\,\pN\cdotp \s/\um^2$, $\alpha_3=1.6\cdotp10^{-3}\,\pN\cdotp \s^3/\um^4$; bottom same besides $k_0=10^5\,\pN/\um$ and $\alpha_3=4\cdotp10^{-2}\,\pN\cdotp \s^3/\um^4$. In B $\alpha_1=18\,\pN\cdotp \s/\um^2$, $\alpha_3=4\cdotp10^{-3}\,\pN\cdotp \s^3/\um^4$, $\xi_0=268\,\pN\cdotp s/\um$, with top $k_0=10^3\,\pN/\um$ and bottom $k_0=5\cdotp10^5\,\pN/\um$.  Other parameters as in Fig.~\ref{fig:instability}, with $\xi_{\rm i}=9.3\,\pN\cdotp s/\um^2$ and $k=228\,\pN/\um^2$.}
\label{fig:pivclam}
\end{figure}

Boundary conditions have an important effect on the patterns of beating cilia \cite{camalet_self-organized_1999, camalet_generic_2000}. This can be readily seen in Fig.~\ref{fig:pivclam}, where we directly compare the non-linear beat patterns for freely pivoting and clamped bases (see Fig.~\ref{fig:boundaries}). These beat patterns differ qualitatively from those of a freely swimming cilium. Most importantly, while for a pivoting base the wave propagation is forward (Fig.~\ref{fig:pivclam} A, from base to tip), it is reversed in the case of a clamped base (Fig.~\ref{fig:pivclam} B, from tip to base).

This change in direction of wave propagation is not affected by changes in the basal stiffness. In fact, a decrease in basal stiffness has little effect in the beating patterns for a pivoting cilium (other than reducing the amplitude, which we compensated by also reducing $\alpha_3$, see Fig.~\ref{fig:pivclam} B). In the case of a clamped base a change in basal stiffness changes the beating mode (this is detailed in Appendix B). Yet, this change of mode does not affect the direction of wave propagation (see Fig.~\ref{fig:pivclam} A).

\begin{SCfigure}[\sidecaptionrelwidth][h!]
\includegraphics{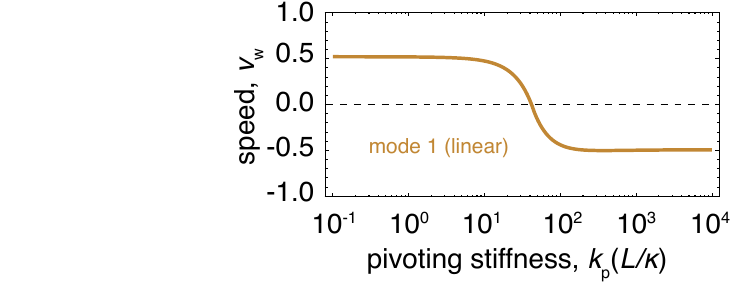}	
 	\caption{\textbf{Wave velocity vs pivoting stiffness.} As the pivoting stiffness increases, the wave velocity of mode 1 decreases. At a certain value the wave reverses and the wave propagates backwards. This is in agreement with Fig.~\ref{fig:pivclam}, which shows backwards waves for clamped ($k_{\rm p}\to\infty$) and forward traveling for pivoting ($k_{\rm p}\to0$). Parameters as in Fig.~\ref{fig:freesols}, with $k_0=8.7\cdotp10^3\,\pN/\um$.}
 	\label{fig:velpivot}
\end{SCfigure}

To better understand the change in direction of wave propagation as the base goes from pivoting to clamped, we analytically studied how the first mode is influenced by the pivoting stiffness $k_{\rm p}$. In Fig.~\ref{fig:velpivot} we plot the wave velocity of mode 1 as a function of the pivoting stiffness. For $k_{\rm p}=0$ the cilium is freely pivoting, and for $k_{\rm p}\to \infty$ we have that $\psi(0)=0$ and the cilium is clamped. As one can readily see, stiffening the basal pivot produces a reversal of wave direction, in agreement with the  non-linear beat patterns in Fig.~\ref{fig:pivclam}.

This dependence on the direction of wave propagation on the boundary conditions provides a simple and elegant mechanism of change in swimming direction for a micro-organism. Simply stiffening the connection between the head and the cilium can produce wave reversal. It can however also be seen as a limitation of  sliding control, not allowing forward wave propagation for clamped cilia. In the next section we show a way to circumvent this issue.

%
%
%

\subsection{Motor regulation with effective inertia and the beat of {\it Bull Sperm}}

\begin{figure}[!ht]
\centering
\includegraphics[width=\textwidth] {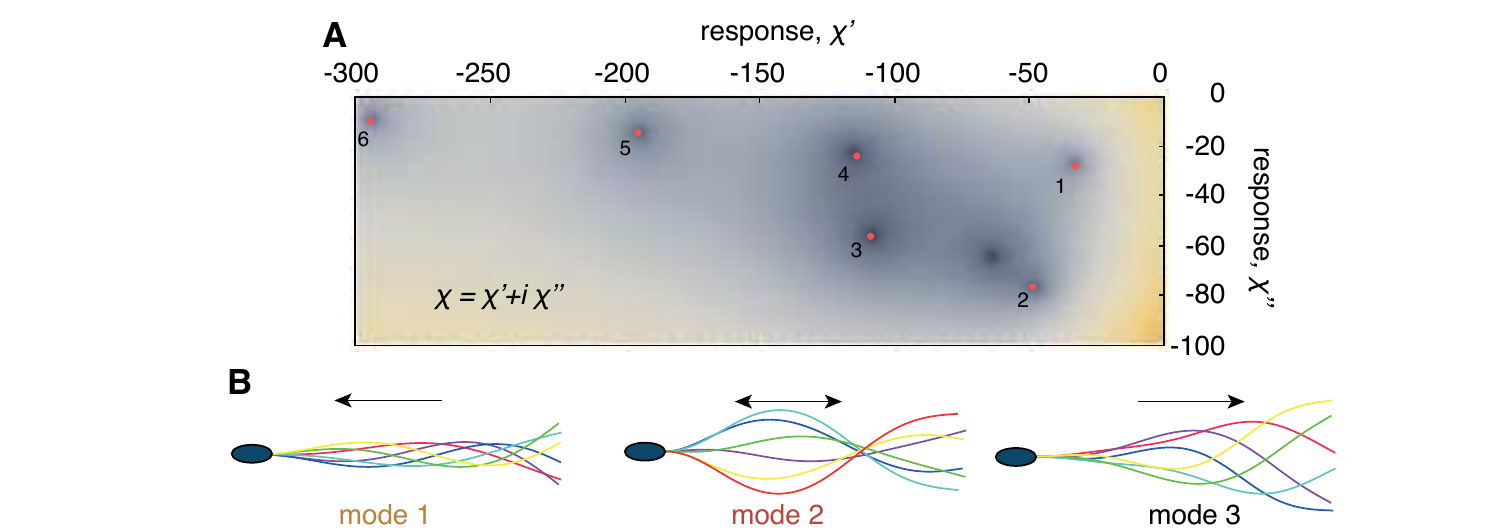}
\caption{\textbf{Linear beats of clamped cilia under sliding control.} \textbf{A.} Space of solutions for a clamped cilium, with red dots in the critical modes (again we see a trivial mode). \textbf{B.} Three critical beats. The first mode shows backwards wave propagation, as already discussed. The second mode is a standing mode. Mode 3 however shows forward propagation, and the beat patter is similar to that of  {\it Bull Sperm}. Parameters as in Fig.~\ref{fig:freesols}, but for a frequency of $14.7\Hz$.}
\label{fig:clampspace}
\end{figure}

The choice of motor model determines which mode gets selected, as shown in Fig.~\ref{fig:instability}. Since in Fig.~\ref{fig:freesols} we saw that different modes have different direction of wave propagation the relevance of the motor model is clear. In the case of a clamped cilium, we saw that the first mode is a backwards traveling one (see Fig.~\ref{fig:velpivot}), but there is a higher mode which is forward traveling (see Fig.~\ref{fig:clampspace}), in agreement with experimental observations for the beat of {\it Bull Sperm} \cite{riedelkruse_how_2007}.

\begin{figure}[!htb]
\centering
\includegraphics[width=\textwidth] {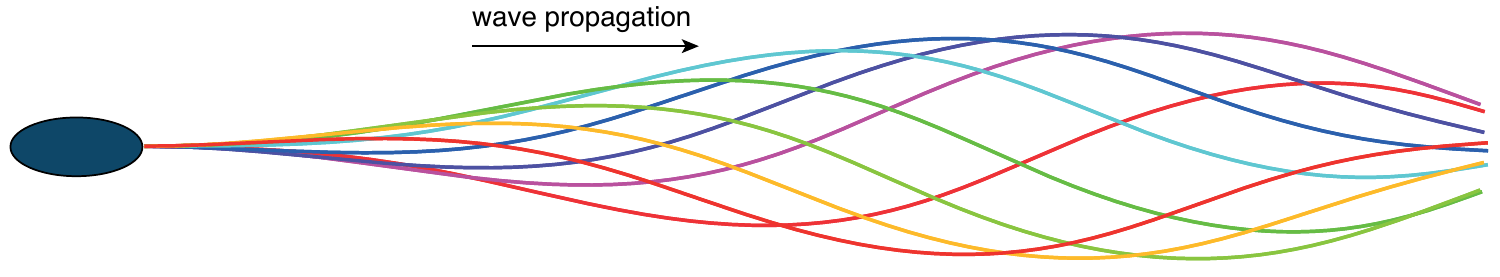}
\caption{\textbf{Forward wave in clamped cilium.} Using the motor model of Eq.~\ref{eq:motiner} allows to excite the second mode at a frequency of $14.7\,\Hz$. This beat is similar to that of {\it Bull Sperm}, although with lower amplitude. The parameters used are those in Fig.~\ref{fig:freesols}, with a basal stiffness $k_0=92352\,\pN/\um$, a basal compliance $\xi_{0}=268\,\pN\cdotp\s/\um$. The motor parameters are $\tau_0=10^{-3}\,\s$, $m=3.5\cdotp10^{-5}\,\s^2$, $\alpha_1=110\,\pN\cdotp\s/\um^2$, and $\alpha_3=0.93\,\pN\cdotp \s^3/\um^4$.}
\label{fig:hfsp}
\end{figure}

To excite this forward mode without activating the backwards one a more complex motor model than that of Eq.~\ref{eq:moteff} is necessary. A simple modification to that model is to include a higher order time-derivative for the motor force, that is 
\begin{align}
\alpha_0\partial_t^2f_{\rm m}+\tau_0\partial_t f_{\rm m} + f_{\rm m}=  \alpha_1\partial_t \Delta - \alpha_3(\partial_t\Delta)^3\quad ,
\label{eq:motiner}
\end{align} 
where $\alpha_0$ accounts for effective inertial effects (these need not come from mechanical inertia, they may be of chemical origin). Indeed, for an adequate choice of parameters this model permits to select the forward traveling mode. The resulting beat pattern of a numerical integration of the equations is given in Fig.~\ref{fig:hfsp}. As one can see, the wave propagation is very clearly backwards and the amplitude grows along the arc-length. 

The implication of this is that forward modes also exist for clamped cilia under sliding control, however they are modes of order higher than one. To be excited, a complex motor model which includes inertial effects is necessary. This suggests that, if indeed the beat of {\it Bull Sperm} is well explained by a sliding control mechanism, the microscopic details of such a mechanism must include some kind of chemical inertia.

\section{Bending waves under curvature control}
\label{sec:curcont}
So far we have studied the beat patterns produced when motors are controlled by sliding of the doublets. However motors can instead be controlled by changes in the curvature of the cilium. In this case the response of the sliding force to a sliding perturbation  $\Delta(s,t)=\Delta(s)\exp(\sigma t)$ is passive, characterized by the sliding stiffness $k$ and internal friction $\xi_{\rm i}$. However its response to curvature $\dot{\psi}=\dot{\Delta}/a$ can be active. We thus have
\begin{align}
f(s)=-\chi(\sigma)\Delta(s)-\beta(\sigma)\dot{\psi}(s)\quad,
\label{eq:slicur}
\end{align}
with $\chi(\sigma)= k+\xi_{\rm i}\sigma$ the passive viscoelastic response, and $\beta(\sigma)$ the linear response coefficient of the motors to curvature (see Eq.~\ref{eq:curres}).

\begin{figure}[!ht]
\centering
\includegraphics[width=\textwidth] {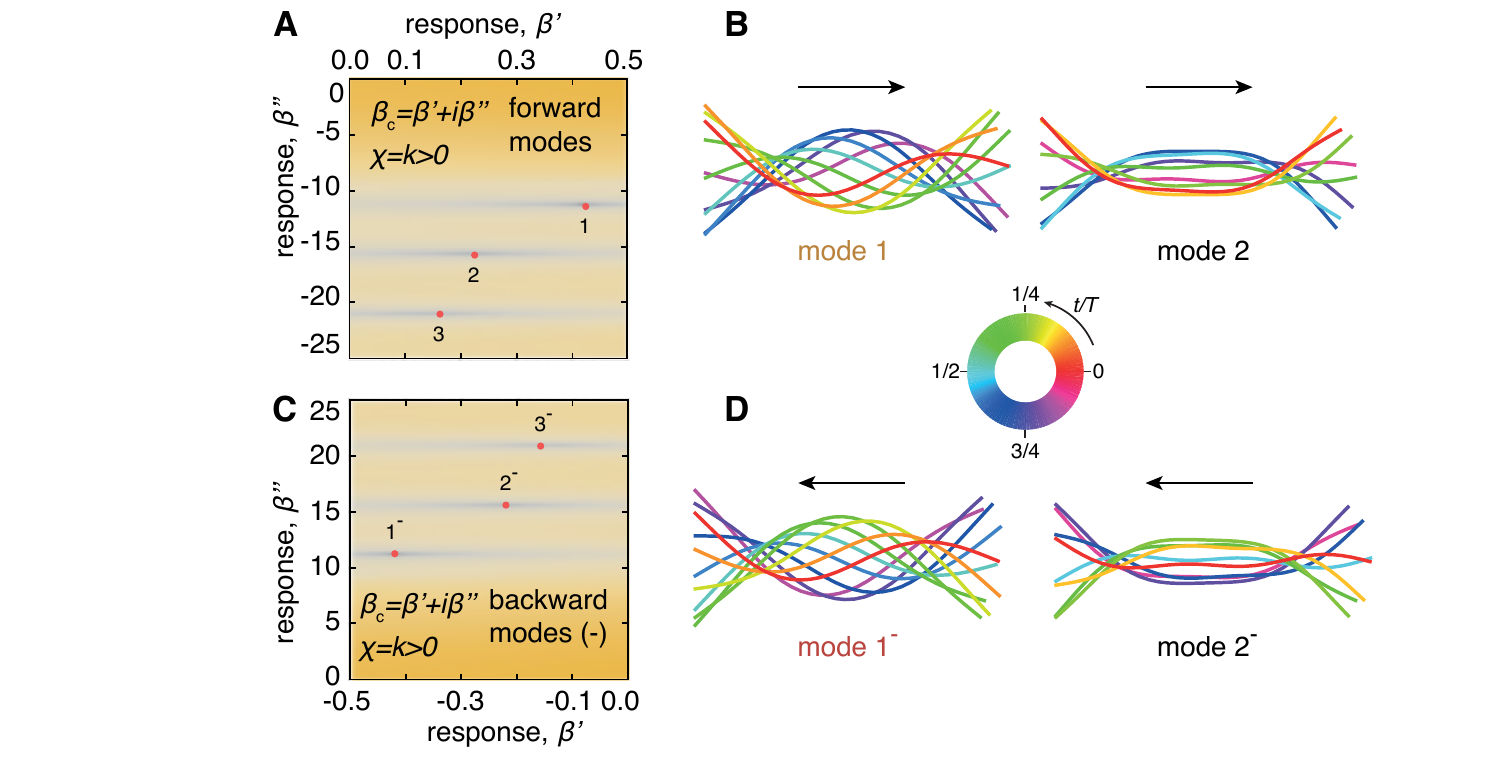}
\caption{\textbf{Critical beats of free cilia under curvature control without basal compliance.} \textbf{A,  C.} Space of critical solutions for free ends. The value of $|\Gamma|^{-1}$ is plotted as a function of $\beta$ for a fixed value of $\chi$. Divergences (red dots) correspond to critical modes, which are numbered by $|\beta_n|<|\beta_{n+1}|$. The region of orward modes is in A, and that of backward modes in C (these regions are symmetric to each other). \textbf{B ,  D.} Time traces of the first two modes, which in A show forward wave propagation (thus swimming to the left), and  backwards traveling waves in D. Arrows denote direction of wave propagation.The parameters used were the typical ones of {\it Chlamydomonas} cilium: $\kappa=400\,\pN\cdotp\um^2$, $k=15\cdotp10^3\pN/\um^2$, $\xi_{\rm n}=0.0034\,\pN\cdotp \s/\um^2$,  $L=12\,\um$, $\omega_{\rm c}=2\pi\cdotp 40\cdotp \Hz$, $a_0=0.06\,\um$ and $k_0=0$.}
\label{fig:freesolscc}
\end{figure}

\subsection{Critical beats in frequency domain}
At the critical point $\Omega=\Omega_{\rm c}$, the beat patterns exhibit only one harmonic with critical frequency $\omega_{\rm c}$. In this case the small amplitude beats are characterized by linearization of Eq.~\ref{eq:angnonlin} together with the linear response of  Eq.~\ref{eq:slicur}, which using $\sigma=i\omega_{\rm c}$ results in
\begin{align}
\label{eq:curchlam}
i\bar{\omega}\psi=-\ddddot{\psi}+\bar{\chi}\ddot{\psi}+\bar{\beta}_{\rm c}\dddot{\psi}\quad,
\end{align}
where we have defined the dimensionless curvature response at the critical point as $\bar{\beta}_{\rm c}=\frac{a_0L}{\kappa}\beta_{\rm c}$, and bars as before denote dimensionless quantities. Provided boundary conditions, obtaining the critical modes is an eigenvalue problem. In this case the discrete set of eigenvalues are given by the condition
\begin{align}
\beta_{\rm c}=\beta_n\quad {\rm with}\;\; n=1,2,\ldots
\end{align}
We now use the condition $|\beta_n|<|\beta_{n+1}|$ to label the modes, and drop bars as all quantities are dimensionless.

Note that there is a fundamental difference between Eq.~\ref{eq:sperm}, characteristic of sliding control; and Eq.~\ref{eq:curchlam}, characteristic of curvature control. The former is symmetric with respect to the change $s\to-s$, whereas the latter is not. Thus, motor regulation by curvature explicitly takes into account the polarity of the motor-doublet interaction. Importantly, Eq.~\ref{eq:curchlam} does remain unchanged under a change of $s\to-s$ and $\beta\to-\beta$. This means that there will be pairs of modes with opposite directions of wave propagation corresponding to values of $\beta_{\rm c}$ with opposite signs. In Fig.~\ref{fig:freesolscc} A and C we show two regions of the space of critical modes of a freely swimming cilium regulated by curvature, and see that this is indeed the case.  The first two modes corresponding to forward traveling waves appear in Fig.~\ref{fig:freesolscc} B, while those corresponding to backward  waves (which we note $1^{-} $ and $2^{-}$) are shown in Fig.~\ref{fig:freesolscc} D. These modes are identical aside from a different direction of wave propagation.

\begin{figure}[!ht]
\centering
\includegraphics[width=\textwidth] {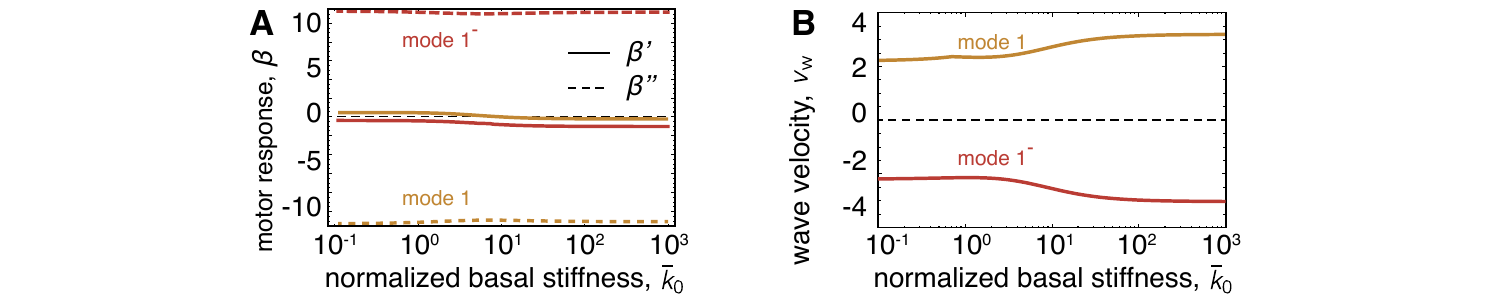}
\caption{\textbf{Depdence of modes with normalized basal stiffness.} \textbf{A.} The eigenvalues of first forward and backward modes show little dependence on the basal compliance $\bar{k}_0=a_0^2Lk_{0}/\kappa$. In particular, they are both directional. \textbf{B.} The wave velocity decreases as the basal stiffness decreases, but always maintaining its sign unaffected. Parameters as those in Fig.~\ref{fig:freesolscc}, which correspond to normalized length $L/\ell_0\approx2.4$ and normalized sliding stiffness $\bar{k}\approx22$}
\label{fig:basalcc}
\end{figure}

Since the polar symmetry is already broken in the motor-filament interaction, a basal compliance is not required to produce wave propagation under curvature control. Indeed, Fig.~\ref{fig:freesolscc} was obtained without basal compliance (i.e., $\chi_0=0$), which means that under curvature control there are no standing modes: all modes are traveling. In Fig.~\ref{fig:basalcc} we see how the forward and backward first modes ($1$ and $1^{-}$) depend on the basal compliance. The eigenvalues show little change. Furthermore, also the wave velocity remains little affected as the modes exhibit only a small change (see Appendix B). This is in  contrast with the role of the basal compliance under sliding control.

\begin{figure}[!ht]
\centering
\includegraphics[width=\textwidth] {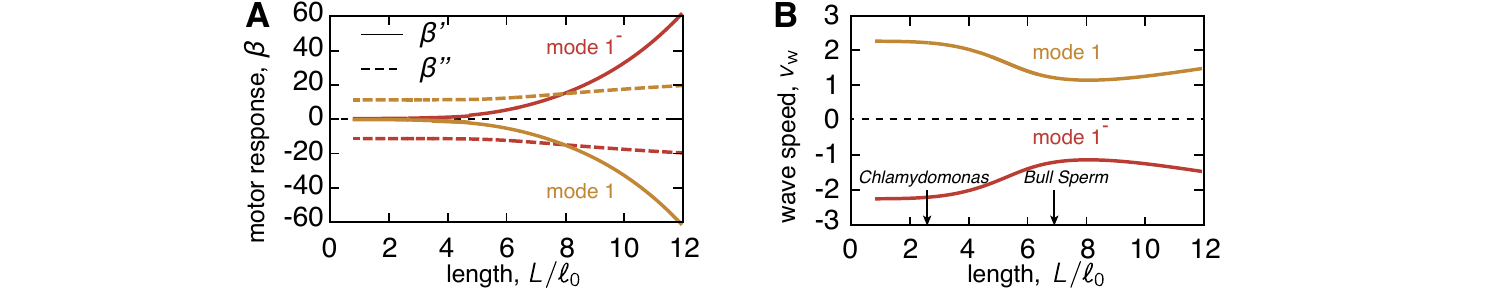}
\caption{\textbf{Scaling of modes with cilium length.} \textbf{A.} Scaling of the real ($\beta'$, solid lines) and imaginary ($\beta''$, dashed lines) parts of the eigenvalues for the first and second modes in Fig.~\ref{fig:freesolscc}. The real part vanishes for short lengths. \textbf{B.} Wave velocity  vs  length of the cilium, arrows note typical values for  {\it Chlamydomonas} and {\it Bull Sperm}. Unlike in sliding control, short and long cilia allow for strong wave propagation. Parameters as in Fig.~\ref{fig:freesolscc}.}
\label{fig:sliscalingcc}
\end{figure}

In section \ref{sec:minreq} we saw that not only a basal compliance was necessary to produce wave propagation in freely swimming cilia under sliding control, but also a minimum length (see Fig.~\ref{fig:sliscaling}). In Fig.~\ref{fig:sliscalingcc} we show that this is not the case under curvature control. While the eigenvalues of the modes depend strongly on the length of the cilium (see Fig.~\ref{fig:sliscalingcc} A), wave propagation can occur for short and long cilia, see Fig.~\ref{fig:sliscalingcc} B.

\subsection{Non-linear  beats in time-domain}

Integrating in time and space the dynamical system posed by Eqs.~\ref{eq:basaldyn}-\ref{eq:angnonlin} together with Eq.~\ref{eq:moteffcc} for the motor force we obtain non-linear beat  patterns under curvature control. The numerical scheme used was the same as for sliding control. The control parameter $\alpha_1$ was adjusted to the supercritical regime, and $\alpha_3$ was regulated so that the amplitude of the beats was large. Finally, since (unlike sliding control) curvature control can produce wave propagation for short cilia, we restricted ourselves to the regime relevant for {\it Chlamydomonas} in which $L/\ell_0\approx2$.

\begin{figure}[!ht]
\centering
\includegraphics[width=\textwidth] {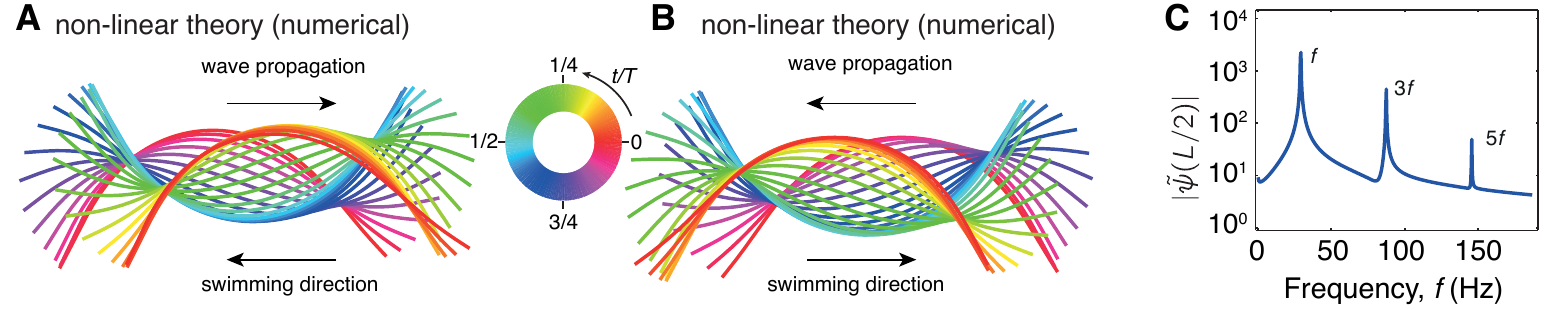}
\caption{\textbf{Non-linear beat pattern of a free cilium regulated by curvature.}  \textbf{A, B.}  Time traces of  swimming cilia with mode $1$ (in A) and mode $1^-$ in B dominating (see Fig.~\ref{fig:freesols}). In A the wave travels forward and the cilium to the left, vice-versa in B.   \textbf{C.} The power spectrum of the angle at the mid-point of the cilium for the trajectory in A has a fundamental frequency of $f=29.2\,\Hz$. Parameters in A as those in Fig.~\ref{fig:freesolscc}, with $\alpha_1=700 \,\nN$, $\alpha_3=5000\,\nN\cdotp\um^2$ and $\tau_0=0.3\,\s$. In B the same with change $\alpha_1\to-\alpha_1$ and $\alpha_3\to-\alpha_3$.}
\label{fig:detailedfreecc}
\end{figure}

In Fig.~\ref{fig:detailedfree} A and B we show two resulting beat patterns. As can be readily seen by comparison with Fig.~\ref{fig:freesolscc} B and D these beats correspond to the modes $1$ and $1^-$, and thus show wave propagation in the forward (for A) and backward (for B) directions. The difference in the motor model to activate one or the other is the following change of sign: $\alpha_1\to-\alpha_1$  and $\alpha_3\to-\alpha_3$. In each case there is  strong  propulsion in one beat, unlike for the case in Fig.~\ref{fig:detailedfree}. This is because the large amplitude of the beat is roughly constant along the arc-length.

\begin{figure}[!hb]
\centering
\includegraphics[width=\textwidth] {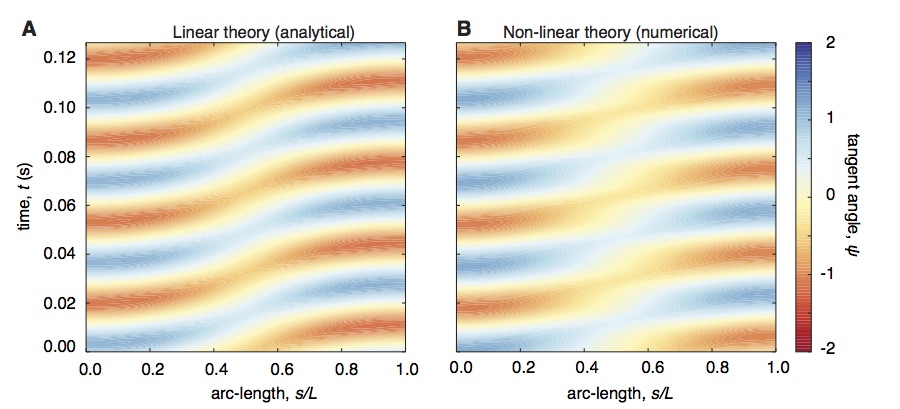}
\caption{\textbf{Tangent angle of linear and non-linear solution.} \textbf{A.} Kymograph of the tangent angle of the analytical critical  solution over three periods for  mode 1. Waves propagate forward (base to tip) and the amplitude is roughly constant along arc-length with a dip around the mid-point. Height normalized to match that of the non-linear solution. \textbf{B.} Same as A but for the non-linear solution which was obtained numerically. Frequency, wave profile and amplitude are in good agreement with the analytical linear. Parameters as in Fig.~\ref{fig:detailedfreecc}.}
\label{fig:kymocc}
\end{figure}

To compare this non-linear beat patterns with the critical modes we again use kymographs. Since mode $1^{-}$ is the mirror symmetric of mode 1, we just focus on the latter. In Fig.~\ref{fig:kymocc} the kymographs of the angle $\psi(s,t)$ show that the waves propagate from base to tip steadily. The amplitude is homogeneous along the length of the cilium, with a small dip around the midpoint in agreement with what was shown in Fig.~\ref{fig:wtchar} for the beat of {\it Chlamydomonas}.  There is good agreement between the numerical solution in Fig.~\ref{fig:kymocc} B corresponding to the non-linear beat patterns with the critical solution in Fig.~\ref{fig:kymocc} A obtained analytically in frequency domain.

\subsection{Role of boundary conditions}
We have seen that boundary conditions play a crucial role in the beat patterns produced by sliding control, being capable of reverting the direction of wave propagation. Under curvature control the direction of wave propagation is set by the motors, and preserved with different boundary conditions.

\begin{figure}[!ht]
\centering
\includegraphics{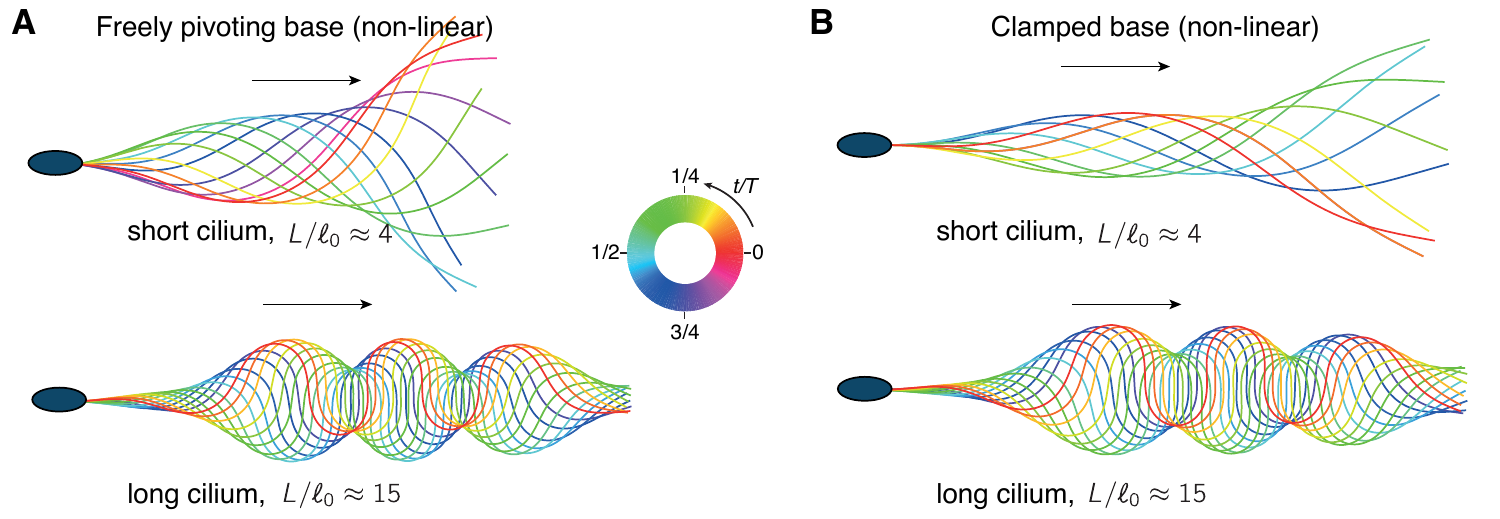}
\caption{\textbf{Beat patterns with different boundary conditions.}  \textbf{A.} Beat patterns of the forward and backward modes for a freely pivoting base. \textbf{B.}  Forward and backward beat patterns for a clamped base. Parameters as in Fig.~\ref{fig:freesolscc}, besides: in A and B top $\alpha_1=500\,\nN$, $\alpha_3 = 2000\,\nN\cdotp\um^2$ and $\tau_0=0.03\,\s$; in a and B bottom $\alpha_1=180\,\nN$, $\alpha_3 = 20\,\nN\cdotp\um^2$, and $\tau_0=0.03\,\s$. The emerging beat frequencies in A and B top to bottom are: $f=199.9\,\Hz$, $f=248.3\,\Hz$,  $f=132.3\,\Hz$ and $f=132.4\,\Hz$.}
\label{fig:pivclamcc}
\end{figure}

In the top panels of Fig.~\ref{fig:pivclamcc} A and B we have the non-linear beat patterns of a freely pivoting and clamped cilium with motors regulated by curvature. In both cases the wave propagates backwards, and the amplitude increases along the arc-length. The difference between the patterns is minor, thus unlike under sliding control changing the pivoting stiffness has a small effect on wave propagation. The same is true for the basal stiffness, which only has a minor effect on the beat patterns under curvature control (see Fig.~\ref{fig:curvaturebasal}). These beat patterns are remarkably similar to those of {\it Bull Sperm}, however they correspond to short cilia: while for the top panels of A and B cilia of $L=12.3\,\um$ were used, the typical length of the {\it Bull Sperm} cilium is $L=52.3\,\um$. We thus studied the beat patterns of long cilia under curvature control for clamped and freely pivoting boundary conditions, see bottom panels of Fig.~\ref{fig:pivclamcc} A and B. As in the case of short cilia, the wave propagation is forward for both boundary conditions. The wave number for these long cilia is around three,  which is larger than that observed for {\it Bull Sperm}. 

To conclude, the effect of boundary conditions on the beat pattern on a curvature control  mechanism are smaller than on a sliding control mechanism. To obtain wave-reversal under curvature control a change in the microscopic details of the motors is necessary, changes at the boundaries do not suffice. Loosely speaking, the reason is that the fundamental modes of curvature control are directional, while those of sliding control are standing and only become directional when a boundary asymmetry is included. 

\section{Conclusions}
\begin{itemize}

\item A cilium in which motors are controlled by sliding has the following requirements for producing wave propagation: a basal asymmetry, such as a basal compliance; and a length larger than the characteristic dynamic length $\ell_0$. In the absence of these conditions the cilium shows standing waves.

\item The direction of wave propagation of the first sliding control mode is determined by the boundary conditions. These conclusions are also valid for supercritical beats, analyzed numerically in the time domain.

\item Contrary to sliding control, when motors are regulated by curvature wave propagation exists in the absence of a basal asymmetry and for lengths of the cilium short compared to $\ell_0$. They are thus not subjected to the requirements of sliding control.

\item The direction of wave propagation in curvature control is determined by the motor response, and the main effect of changing the length of the cilium is changing the wave-number of the beat. These conclusions are valid for supercritical beats, analyzed numerically in the time domain. 

\end{itemize}


\chapter{Motor regulation in asymmetric beats}

{The beat of {\it Chlamydomonas} cilia is intrinsically asymmetric}. In this chapter we explore the consequences of this asymmetry in motor regulation. We develop a theory of  asymmetric beats, and show that due to the asymmetry normal forces can regulate the beat. We discuss the asymmetric beat patterns activated by sliding, curvature and normal force control. We end the chapter by comparing the beats patterns obtained from the asymmetric theory with those tracked of wild-type and mbo2 {\it Chlamydomonas} cilia.

\section{Dynamics of an asymmetrically beating cilium}
\begin{figure}[h]
\centerline{\includegraphics{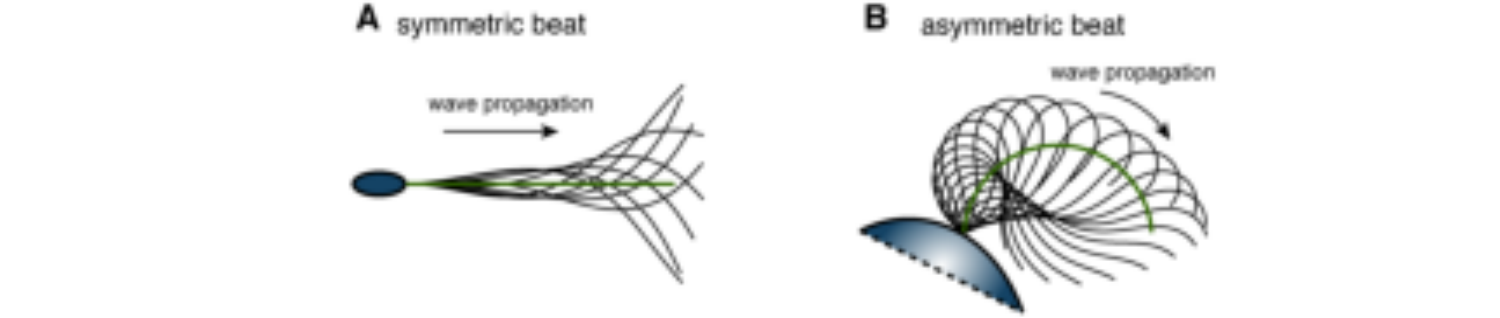}}
\caption{\textbf{Symmetric and asymmetric planar beats.} \textbf{A.} Symmetric beat patterns such as the one of {\it Bull Sperm} have a flat mean shape (green line).  \textbf{B.} Asymmetric beats such as that of the {\it Chlamydomonas} cilium show a curved mean shape (green line).}
\label{fig:chiralbeat}
\end{figure}

So far we have discussed beat patterns that are symmetric, as the one depicted in Fig.~\ref{fig:chiralbeat} A. Symmetric beats have a flat mean shape (green line). The beat of {\it Chlamydomonas} cilia however is asymmetric and thus has an average curvature, as  seen in Fig.~\ref{fig:chiralbeat} B.

If the beat is planar, the swimming path of an isolated asymmetric cilium is circular. This is indeed the case for the {\it Chlamydomonas} cilium swimming near a surface, as seen in Fig.~\ref{fig:tracktrace}. Furthermore, the same is also true for single flagellates such as {\it Bull Sperm} \cite{friedrich_stochastic_2008}. For non-planar asymmetric beats the path is a spiral \cite{friedrich_chemotaxis_2007}. Thus the asymmetry is of fundamental importance in determining the swimming path of a microswimmer.  We now provide a physical description for asymmetric ciliary beats.

We study the physics of asymmetrically beating cilia by describing the small amplitude dynamics of the cilium around a mean shape. To do this we perform an expansion of the dynamic equations of a beating cilium around a static shape. That is, consider the dynamics of the beating cilium described by 
\begin{align}
\label{eq:expan}
\psi(s,t) = \psi_0(s)+{\psi}_1(s)\e^{i\omega t}+{\psi}^{\rm *}_1(s)\e^{-i\omega t}\quad,
\end{align}
where $\psi_0$ is the static mode and $\psi_1$ the fundamental mode. Analogous expressions hold for other mechanical quantities (tension, sliding force, etc.). By introducing this into the dynamic equations of the beating cilium and expanding in powers of $\psi_1$ we obtain a hierarchy of equations. Using the static component of this hierarchy allows us to calculate the mean shape. With the static mode at hand, we can then solve the equation for the fundamental mode $\psi_1$.

The equations of the zeroth mode correspond to a static force balance, as derived in section \ref{sec:minreq}. These equations show that the average curvature of the cilium $\dot{\psi}_0$ is determined by the static contribution of the sliding force $f_0$. In the limit of small amplitude $\psi_1$ we can write the dynamics of the fundamental Fourier mode if the average shape is given:
\begin{align}
i{\omega}\xi_{\rm n}\psi_1&=-\ddddot{\psi_1} - a_0\ddot{f}_1+\dot{\psi}_0 \dot{\tau}_1 + \ddot{\psi}_0\tau_1+\frac{\xi_{\rm n}}{\xi_{\rm t}}\dot{\psi}_0(\kappa\dot{\psi}_0\ddot{\psi}_1 + a_0\dot{\psi}_0 f_1+\dot{\tau}_1)\quad,\nonumber\\
\frac{\xi_{\rm n}}{\xi_{\rm t}}\ddot{\tau}_1- \dot{\psi}_0^2\tau_1&=-\dot{\psi}_0(\kappa\dddot{\psi}_1+a_0\dot{f}_1) - \frac{\xi_{\rm n}}{\xi_{\rm t}}\partial_s[ \dot{\psi}_0 (\kappa\ddot{\psi}_1+a_0f_1)]\quad .
\label{eq:dynforbal}
\end{align}
This pair of equations is the generalization of the equations for the symmetric beat. In them, the fundamental mode is coupled to the static mode. Since the force balance of the static mode has been used, the coupling only appears through the mean curvature $\dot{\psi}_0$ and its derivatives. In the case in which the static mode vanishes ($\dot{\psi}_0=0$) we recover Eq.~\ref{eq:spermold}, and the tension also vanishes. However, for asymmetric beats all terms may be equally relevant. Solutions to the system above are obtained using the sliding force balance  $\chi_0\Delta_{0,1}=\int^1_s f_1(s')\d s'$ and the boundary conditions of Eq.~\ref{eq:bcdyn}. 

\subsection{Beats around a  circular arc}
\label{sec:circle}
The static mode of the {\it Chlamydomonas} beat corresponds within a good approximation to a circular arc (see Figs.~\ref{fig:wtchar} A and \ref{fig:takehome}). This motivates studying asymmetric beats in which the average curvature is constant. The balance of static forces states that $\kappa\dot{\psi}_0=a_0F_0$ (see section \ref{sec:minreq}). Thus to obtain a circular arc with constant curvature $\dot{\psi}_0=C_0$, the static force must be constant along the length of the cilium with magnitude $F_0=\kappa C_0/a_0$. This implies that the sliding force density is given by
\begin{align}
\label{eq:zerothdelta}
f_0=\delta(s-L)\kappa C_0/a_0\quad ,
\end{align}
which corresponds to an accumulation of the static sliding force at the distal end of the cilium, as already described in section \ref{sec:circles}. The static mode of the basal sliding corresponding to this shape with constant static curvature is $\Delta_{0,0} =\kappa C_0/(k_0 a_0)$, and the  sliding displacement along the cilium  is $\Delta_0(s)=\kappa C_0/(k_0 a_0)+aC_0 s$.  For beats in which the static shape corresponds to a circular arc, the linear dynamic equations become.
\begin{align}
\label{eq:asymsimple}
i{\omega}\xi_{\rm n}\psi_1&=-\kappa\ddddot{\psi}_1 -a_0 \ddot{f}_1+\left(1+\frac{\xi_{\rm n}}{\xi_{\rm t}}\right)C_0 \dot{\tau}_1 +\xi C_0^2(\kappa\ddot{\psi}_1 +a_0 f_1)\quad,\nonumber\\
\frac{\xi_{\rm n}}{\xi_{\rm t}}\ddot{\tau}_1 - C_0^2\tau_1&=-\left(1+\frac{\xi_{\rm n}}{\xi_{\rm t}}\right)C_0(\kappa\dddot{\psi}_1+a_0 \dot{f}_1) \quad .
\end{align}
These equations have constant coefficients, and as such can be analytically solved if boundary conditions and an additional equation for the motor force are provided.  

\section{Motor regulation in asymmetric beats}
\label{sec:asymmot}
Just like in the symmetric case, in asymmetric beats the motor force responds to the strains and stresses within the cilium (see chapter 4). There is, however, a fundamental difference. Due to the asymmetry, the normal force $f_\perp$ has a non-vanishing fundamental mode, which is given by
\begin{align}
f_{\perp,1} = \dot{\psi}_0(F_1+ \kappa \dot{\psi}_1/a_0)\quad.
\label{eq:fndyn}
\end{align}
Note that in the symmetric case $\dot{\psi}_0=0$  this mode vanishes, and the normal force only has modes of order two and higher.

This fundamental mode of the normal force in asymmetric beats allows that motors are dynamically regulated to linear order by normal stress. The most general form of the linear motor force response is thus 
\begin{align}
f_{{\rm m}, 1}=\lambda(\omega)\Delta_1 + \beta(\omega)\dot{\psi}_1 + \gamma(\omega)f_{\perp,1}\quad ,
\label{eq:fullmotmod}
\end{align}
which is a generalization of the models in section \ref{eq:nonlinfou} and allows for regulation by sliding, curvature, and normal forces. With this linear motor model, we obtain the following static mode of the sliding force
\begin{align}
f_0=(\lambda(0)-k)\Delta_0+ \beta(0)\dot{\psi}_0 + \gamma(0)f_{\perp,0}+\delta(s-L)\kappa C_0/a_0\quad,
\label{eq:circfor}
\end{align}
which also incorporates the role of cross-linkers of stiffness $k$, and where the last term corresponds to the static force at the tip responsible for creating the asymmetry. So far this motor description is completely general. We now discuss the conditions under which the static curvature is constant.

\subsection{Motor regulation around a  circular arc}
We have seen that when the static mode has a constant curvature, the static sliding force accumulates at the tip. In this scenario all terms besides the last one in Eq.~\ref{eq:circfor} vanish.  Since in sliding control motors are regulated by sliding velocity, the linear response $\lambda(\omega)$ vanishes at zero frequency, that is $\lambda(0)=0$. The same is true for curvature control.  Thus in mechanisms in which the motor force is regulated by the time-derivatives of the curvature, we will have  $\beta(0)=0$. 
Note that such a mechanism is different from what is usually referred to as curvature control, which is a delayed response of the motor force to curvature \cite{machin_wave_1958,brokaw_computer_2002,riedelkruse_how_2007}. The same applies to normal force control:  when the motor force depends on time-derivatives of the normal force, we have that $\gamma(0)=0$. Under these conditions the static force has a term corresponding to the cross-linkers, and one which accumulates at the tip.

The cross-linkers with stiffness $k$ will deviate the static mode away from a circular arc. To estimate how large their effect is, we compare the tip accumulated force $F_0=\kappa C_0 a_0$ with the total static sliding force $\int_0^L k\Delta_0$. Dividing these two terms we obtain
\begin{align}
\frac{\int_0^Lk\Delta_0(s)\d s}{F_0} = \frac{kL}{k_0}+\frac{L^2}{2\ell^2}\quad,
\end{align}
where $\ell$ is the  characteristic length associated to cross-linkers (see Eq.~\ref{eq:charlen}). In the case in which $\ell\gg L$ and for large values of the basal compliance $k_0\gg kL$ we have that the effect of cross-linkers can be neglected. In conclusion, for motor models in which the motor force depends on the time derivatives of sliding, curvature, or normal forces, and where the role of cross-linkers is small, we can have beats around a circular arc.

\section{Unstable modes of a circular cilium}
\subsection{Dispersion relation}
Before studying the critical modes of an asymmetric cilium, it is  instructive to study the stability of perturbations with a fixed wave-length $\lambda=1/q$. To do so we interpret the cilium as an infinite elastic medium which can become unstable due to the action of molecular motors. A perturbation of fixed wavelength will evolve as
\begin{align}
\psi(s,t)=\e^{\sigma t-iq s}\quad,
\label{eq:perturb}
\end{align}
where $\sigma=\tau^{-1}+i\omega$ with $\tau^{-1}$ the growth rate  and $\omega$ the frequency at which the perturbation evolves. Analogous equations hold for the tension, normal force and basal sliding.  Introducing this perturbation in the dynamic equations provides a dispersion relation $\sigma(q)$, which determines the stability of the different wave-lengths $1/q$.

We now focus on the case of constant static curvature $C_0$ with the general motor model introduced in the previous section. After replacing the perturbation in the dynamic equations, we obtain the following dispersion relation in implicit form:
\begin{align}
a_0^2\chi(\sigma)-ia_0q\beta(\sigma)-2i\kappa qC_0\gamma(\sigma) = -\kappa q^2 - \xi_{\rm n}\left(1-i\frac{C_0\gamma(\sigma)}{q}\right)A\sigma \quad,
\label{eq:dispersion}
\end{align}
which has units of a force balance. Here $\chi(\sigma)=k+\sigma\xi_{\rm i}-\lambda(\sigma)$ is the net sliding compliance, and $A$ is the function
\begin{align}
A(q,K)=\frac{q^2+\xi_{\rm t} C_0^2/\xi_{\rm n}}{(q^2-C_0^2)^2}\quad .
\end{align}
Provided the dependence of the motor response on $\sigma$ we can study the dispersion relation $\sigma(q)$.

\begin{figure}[!ht]
\centering
\includegraphics[width=\textwidth] {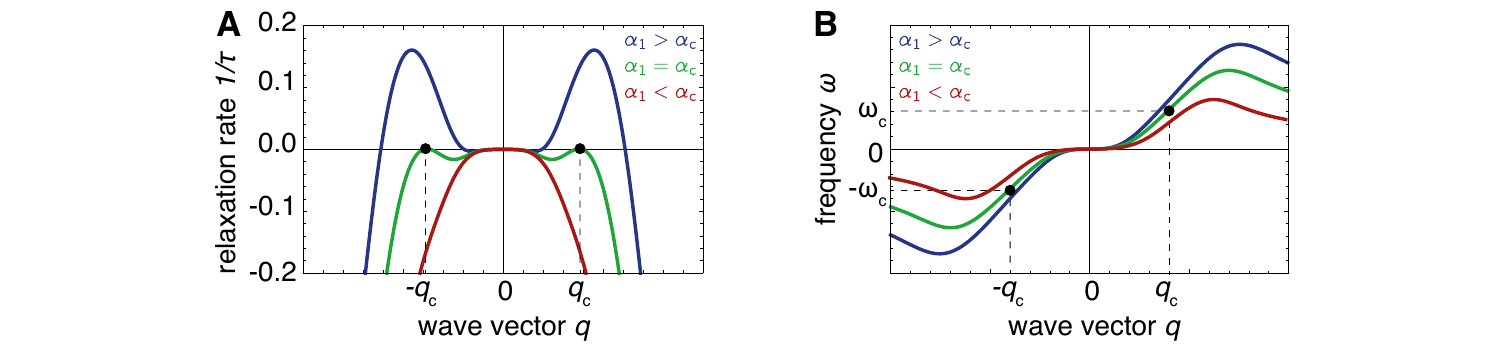}
\caption{\textbf{Dispersion relation for curvature control.} \textbf{A.} Relaxation rate of a fixed wave-length perturbation for three values of the control parameter. At the critical point two modes become unstable. \textbf{B.} Characteristic frequency of the modes. Since $q_{\rm c}$ has frequency $\omega_{\rm c}$ and $-q_{\rm c}$ has frequency $-\omega_{\rm c}$, both modes have the same direction of propagation. The values of the control parameter used are $\{\alpha_{\rm c}/2,\alpha_{\rm c}, 3\alpha_{\rm c}/2\}$ (in red, green and blue respectively), and time is in units of $\tau_0$.}
\label{fig:disper}
\end{figure}

As an example, consider the case in which motors are regulated by curvature and so $\gamma=0$ and $\lambda=0$. For simplicity we study the symmetric case $C_0=0$ with the motor model introduced in Eq.~\ref{eq:moteffcc}, which corresponds to $\beta(\sigma)=\alpha_1/(1+\tau_0\sigma)$. In Fig.~\ref{fig:disper} A we plot the dispersion relation for positive values of the control parameter $\Omega=\alpha_1$. As one can see at the critical value $\alpha_{\rm c}>0$ a  traveling mode with wave-vector $q_{\rm c}>0$ becomes unstable at the positive frequency $\omega_{\rm c}>0$, see also Fig.~\ref{fig:disper} B. Simultaneously, a mode $-q_{\rm c}$ becomes unstable at a negative frequency  $-\omega_{\rm c}$. Both these modes correspond to forward traveling waves. For negative values of the control parameter $\alpha_1$, modes can also become unstable. Indeed, at $\alpha_1=-\alpha_{\rm c}$ the mode $q_{\rm c}>0$ becomes unstable with a negative frequency $-\omega_{\rm c}$ and the mode  $q_{\rm c}<0$ is also critical with frequency $\omega_{\rm c}$. Both modes correspond in this case to backwards traveling waves. We can conclude that the instabilities result in traveling waves whose direction of propagation is determined by the value of $\alpha_1$, thus by motor model details.

This is different from the behavior for a sliding control mechanism, in which $\beta=0$ and $\gamma=0$. In sliding control Eq.~\ref{eq:dispersion} is symmetric with respect to $q$, which indicates that for positive and negative $q$ the frequencies will have the same sign. Thus at the critical point waves traveling in opposite directions appear, which result in a standing mode. These conclusions are in agreement with the conclusions of the previous chapter for the critical beats of a cilium of finite length with free ends as boundary conditions.

To analyze the dispersion relation the dependence of the motor model on $\sigma$ is necessary. However at the critical point $\Omega=\Omega_{\rm c}$ the behavior of the system is generic. When $\Omega=\Omega_{\rm c}$ the relaxation time of a critical wave-length diverges, and thus $\sigma=i\omega_{\rm c}$ with $\omega_{\rm c}$ the critical frequency. Using this condition in Eq.~\ref{eq:dispersion} we obtain that the critical values of the response coefficients satisfy certain conditions for each motor regulation mechanism. In particular, for sliding control  where $\beta=0$ and $\gamma=0$, we obtain that the real and imaginary part of $\chi$ have to be negative. For curvature  control we have that the real part of $\beta$ is positive and the imaginary part negative for forward traveling waves, the opposite is true for backward traveling waves. For normal force control the imaginary part of $\gamma$ has to be positive for forward and negative for backward traveling waves. The sign of the real part depends on motor parameters. Importantly, while for sliding and curvature control these signs are independent of the sign of the static curvature $C_0$, under normal force control the signs flip with the sign of $C_0$. This reveals that for the same set of motor parameters under normal force regulation the sign of the curvature regulates the direction of wave propagation. These results are summarized in table~\ref{tab:regimes}.

\begin{table}[!htb]
\begin{center}
\begin{tabular}{lllll}
\toprule
& wave direction & $\chi',\chi''$ & $\beta',\beta''$ & $\gamma',\gamma''$ \\
\toprule
sliding control & $\leftrightarrow$& $-,-$& $0,0$ & $0,0$ \\
\midrule
curvature control &  $\rightarrow$ & $+,+$ & $+,-$ & $0,0$ \\
\midrule
normal force control & $\rightarrow\,\,(C_0<0)$  & $+,+$ & $0,0$ &$\mp,+$ \\
\bottomrule
\end{tabular}
\label{tab:regimes}
\caption{\textbf{Response coefficients for different regulatory mechanisms.} The signs of the real part (single prime) and imaginary part (double prime) of the response coefficients to sliding ($\chi$), curvature ($\beta$) and normal force ($\gamma$) is given for three regulatory mechanisms: sliding control (with either direction of wave propagation), curvature control (for forward waves) and normal force control (for forward waves with a negative mean curvature). \label{tab:regimes}}
\end{center}
\end{table}

\begin{figure}[!hb]
\centering
\includegraphics[width=\textwidth] {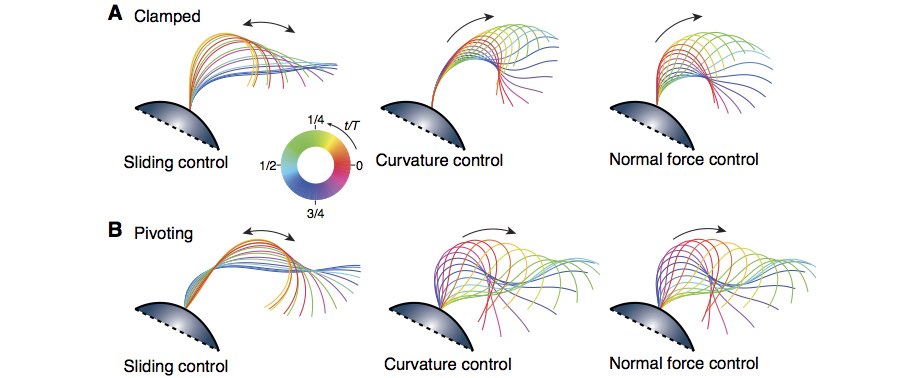}
\caption{\textbf{Examples of critical asymmetric beat patterns.} \textbf{A.} Beat patterns of a clamped cilium under sliding, curvature, and normal force control. Sliding control produces standing waves, curvature and normal force control produce forward traveling waves. \textbf{B.} Same as A but for a pivoting base boundary conditions, as corresponds to the  intact {\it Chlamydomonas} cilium. Arrows denote direction of wave propagation. The parameters used for the cilium were those typical of {\it Chlamydomonas} used in the previous chapter (as in Fig.~\ref{fig:freesolscc}) with a static curvature $C_0=-0.25\,\um^{-1}$. The resulting response coefficients preserved the sign convention of table \ref{tab:regimes}, with $\gamma'<0$.}
\label{fig:critasym}
\end{figure}

\subsection{Mode selection by boundary conditions}
To obtain critical beat patterns we need to consider a cilium of finite length, and provide boundary conditions. The procedure is  the same as the one outlined in Appendix B, but using this time equations~\ref{eq:asymsimple} for the dynamics and equation~\ref{eq:fullmotmod} for motor regulation. In Fig.~\ref{fig:critasym} we provide some examples of critical asymmetric beats of cilia with a clamped and pivoting base, with a free tip (free base beats are considered in the next section). Typical values of {\it Chlamydomonas} were used for the ciliary parameters (see caption).

Under sliding control there is no wave propagation, and no net swimming of the cell is possible. As already shown in the previous chapter, this is because wave propagation under sliding control only occurs for cilia long compared to the characteristic length $\ell_0$ (see section \ref{sec:critbeat}), but the {\it Chlamydomonas} cilium satisfies $L\sim\ell_0$. Curvature control produces strong wave propagation both for clamped and pivoting cilia. Finally, normal force control also produces strong wave propagation, revealing this new mechanism as a candidate for regulation of the beat of asymmetric cilia.

\section{Beat of freely swimming cilia: experimental comparison}
\begin{figure}[!ht]
\centering
\includegraphics[width=\textwidth] {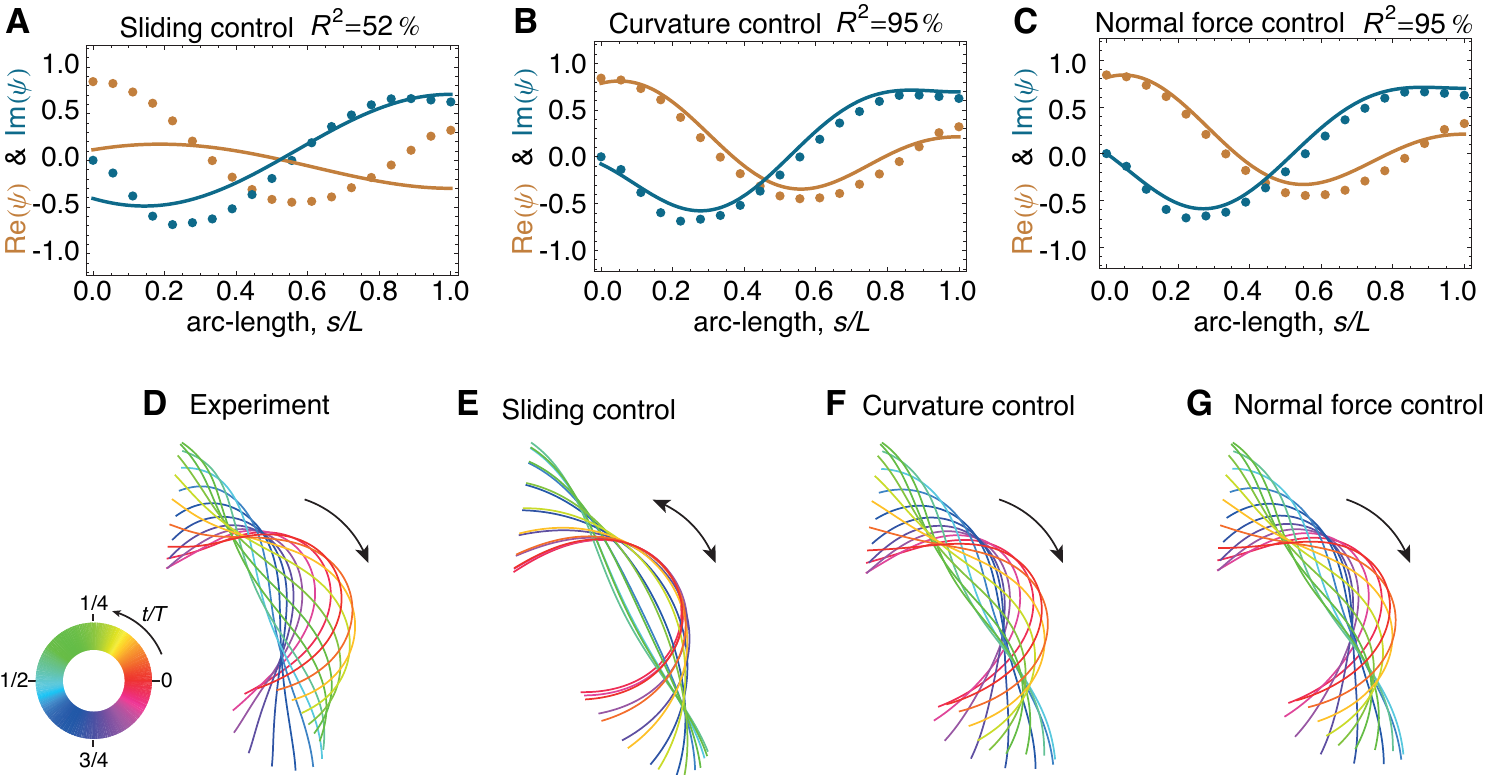}
\caption{\textbf{Fits of {\it Chlamydomonas} wild-type beat.} Best fits of the real (orange) and imaginary (blue) parts of the first Fourier mode of the tangent angle under sliding control (A), curvature control (B) and normal force control (C). Below, the corresponding position space tracked data for the first mode (D) followed by the  fits (E to G). Arrows denote direction of wave propagation. The data corresponds to a cilium of length $L=10.13\,\um$, frequency $f=73.05\,\Hz$, and mean curvature $C_0=-0.25\,\um^{-1}$ }
\label{fig:wtfits}
\end{figure}

Which motor regulatory mechanism is responsible for producing the bending waves observed in the beat of {\it Chlamydomonas} cilia? To answer this question we explored the three regions of parameter space  in table~\ref{tab:regimes}  for free end boundary conditions (no external torques or forces, see Fig.~\ref{fig:boundaries}). We defined the mean square displacement distance $R^2(\psi^{\rm the},\psi^{\rm exp})$ between the theoretically obtained critical mode and the experimental data, and found its minimum value in the corresponding region of parameter space. The result of a typical  fit is shown in Fig.~\ref{fig:wtfits}.

As we can see, both normal force control and curvature control provide good fits, while sliding control does not. Sliding control does not produce good fits since the length of the cilium is comparable to the characteristic dynamic length $\ell_0$, and thus does not allow for wave propagation (see section \ref{sec:critbeat}). Curvature control was already shown to produce traveling waves of constant amplitude for symmetric beats, this is now shown to also be true for asymmetric beats and to agree quantitatively with the {\it Chlamydomonas beat}. Normal force control  produces similar fits to curvature control since, according to  Eq.~\ref{eq:fndyn},  normal force is a probe for curvature sensing. Table \ref{tab:wtfits} collects average parameters resulting from the fits of 10 different cilia. The resulting basal stiffness is very large, since taking $\chi'$ to be due to cross-linkers, we have $k_0\gg\chi'L$. However, the characteristic length defined by $\chi'$ is short, $\ell\approx3\,\um$. With such strong cross-linkers it would be impossible to bend the {\it Chlamydomonas} cilium as much as observed, which suggests the origin of the sliding compliance may be motor response and not cross-linkers.

\begin{table}[!ht]
\begin{center}
\begin{tabular}{lllll}
\toprule
 & Sliding control & Curvature control & Normal Force control\\
\toprule
$R^2\;(\%)$ & $46\pm5$ &$94\pm1$&$95\pm1$\\
\midrule
$\chi',\chi''\; (\nN\cdotp\um^{-2})$ &$-16\pm3,-1.5\pm0.4$&
$ 15\pm5,2\pm1 $&
$14\pm2,1\pm1$\\
\midrule
$\beta',\beta''\;(\pN)$&$0$&$5\pm90,-5964\pm768$&$0$\\
\midrule
$\gamma',\gamma''$&$0$&$0$&$0.09\pm0.07,1.9\pm0.1$\\
\midrule
$\chi'_0,\chi''_0\; (\mN\cdotp\um^{-1})$ &$0.5\pm 1,0.6\pm 1$&
$0.04\pm0.1,6\cdotp10^{-3}\pm10^{-3}$&
$0.2\pm0.4,2\cdotp10^{-3}\pm3\cdotp10^{-3}$\\
\midrule
$\Delta_{0,0}\;(\nm)$&$-57\pm90$&$-37\pm45$&$-40\pm50$\\
\midrule
$|\Delta_{0,1}|\;(\nm)$&$16.6\pm7.4$&$10\pm10$&$8\pm8$\\
\bottomrule
\end{tabular}
\label{tab:wtfits}
\caption{\textbf{Average parameters extracted from fits of wild-type cilia.} Curvature controla and normal force control provide very good fits, unlike sliding control which is not capable of producing wave propagation due to the shortness of the axoneme (see Fig.~\ref{fig:wtfits}).\label{tab:wtfits}}
\end{center}
\end{table}

\begin{figure}[!hb]
\centering
\includegraphics[width=\textwidth] {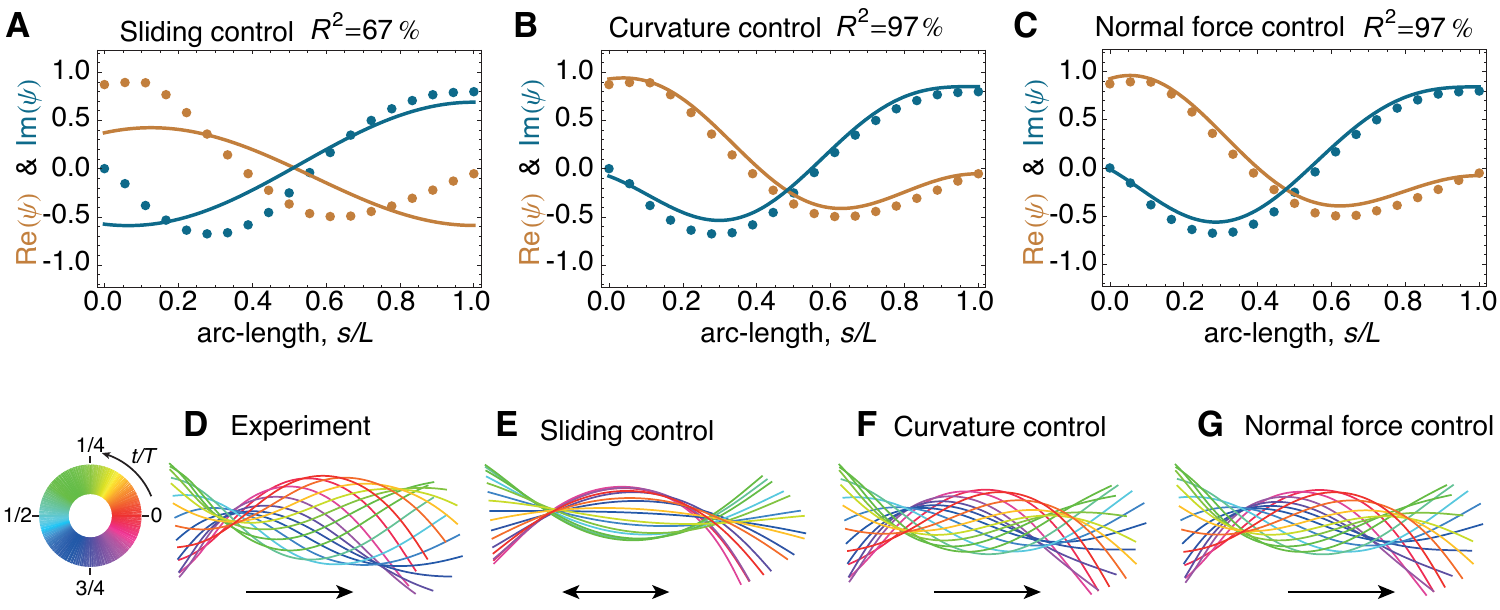}
\caption{\textbf{Fits of {\it Chlamydomonas} mbo2 beat.} Best fits of the real (orange) and imaginary (blue) parts of the first Fourier mode of the tangent angle under sliding control (A), curvature control (B) and normal force control (C). Below, the corresponding position space tracked data for the first mode (D) followed by the  fits (E to G).  The data corresponds to a cilium of length $L=9.43\,\um$, frequency $f=38.00\,\Hz$, and mean curvature $C_0=-0.027\,\um^{-1}$.}
\label{fig:mbo2fits}
\end{figure}

Normal force control and curvature control can both produce beats very similar to that of wild-type, and a priori there is no argument favoring one rather than the other in this asymmetric beat. However, for symmetric beats where $C_0=0$, the dynamic component of the normal force vanishes and normal force control as described in our planar model cannot be a regulatory mechanism.  We thus fitted the symmetric beating mutant mbo2, where the static curvature is reduced by one order magnitude as compared to wild-type (see section \ref{sec:experchlam}). In this case the results obtained were similar to the case of wild-type (see Fig.~\ref{fig:mbo2fits}): sliding control cannot produce wave propagation, while curvature and normal force control are in good agreement with the data. There is however an important difference between the fits of mbo2 and wild-type under curvature and normal force control. Under curvature control the value of the response coefficient $\beta$ is well preserved  from wild-type to mbo2, but this is not the case for $\gamma$ under normal force control (see table \ref{tab:mbo2fits}). The reason is that in mbo2 the small static curvature results in a small normal force, which requires a correspondingly larger response coefficient $\gamma$ to be sensed. While not impossible, such a change in sensitivity is hard to justify. Finally, we note that the inferred basal compliance for mbo2 is several orders of magnitude below the  value for wild-type.

To conclude: a length scale argument suggests that the beat of {\it Chlamydomonas} cilia can not be regulated by sliding, and the existence of wave propagation in symmetric beats suggests that it cannot be regulated by normal forces. This suggests that curvature is the likely mechanism through which the beat is regulated, in agreement with the good fits obtained.

\begin{table}
\begin{center}
\begin{tabular}{lllll}
\toprule
 & Sliding control & Curvature control & Normal Force control\\
\toprule
$R^2\;(\%)$ & $70\pm4$ &$96\pm1$&$96\pm1$\\
\midrule
$\chi',\chi''\; (\nN\cdotp\um^{-2})$ &$-19\pm2,-1.0\pm0.3$&
$ 16\pm5,5\pm5 $&
$13\pm4,2\pm1$\\
\midrule
$\beta',\beta''\;(\pN)$&$0$&$400\pm700,-6700\pm600$&$0$\\
\midrule
$\gamma',\gamma''$&$0$&$0$&$0.2\pm0.4,30\pm20$\\
\midrule
$\chi'_0,\chi''_0\; (\nN\cdotp\um^{-1})$ &$40\pm 20,0.67\pm 0.03$&
$14\pm7,10^2\pm10^2$&
$10\pm1,20\pm10$\\
\midrule
$\Delta_{0,0}\;(\nm)$&$-6\pm5$&
$-13\pm9$&
$-17\pm10$\\
\midrule
$|\Delta_{0,1}|\;(\nm)$&$40\pm10$&
$40\pm20$&
$50\pm20$\\
\midrule
\end{tabular}
\label{tab:mbo2fits}
\caption{\textbf{Average parameters extracted from fits of mbo2 cilia.} Curvature and normal force control provide very good fits, but the values of $\gamma$ in normal force control are very spread and different from those in the wild-type fits.\label{tab:mbo2fits}}
\end{center}
\end{table}

\section{Conclusions}
\begin{itemize}

\item We developed a theory for the asymmetric beat of cilia. In contrast to symmetric beats, the normal force has a linear dynamic component, which allows for normal force motor regulation.

\item For cilia short compared to the characteristic dynamic length $\ell_0$, curvature control and normal force control produce wave propagation for clamped, pivoting and free end boundary conditions. Due to the short length, in all these cases sliding control produces standing waves. 

\item Curvature and normal force control produce asymmetric beat patterns as those observed in isolated wild-type cilia. With similar motor parameters, curvature control also produces beats as those of the symmetric mbo2 mutant.

\end{itemize}

%
\chapter{Conclusions and future work}

{Cilia are highly conserved structures involved in many cellular processes.} Much is known about their structure, and their main constitutents are well identified. Yet, how these components self-organize in order to produce an orchestrated beat has remained a challenging question for the past fifty years. Today, it is widely believed that the response of dynein motors to the stresses and strains within the cilium are the key to unravel its beat pattern. Yet what the differences are between alternative regulatory mechanisms is poorly understood.  In this thesis we characterized the different beat patterns produced by three motor regulatory mechanisms, sliding, curvature, and normal force control;  and compared them to experimental data.

\section{Summary and conclusions}
\label{sec:table}
In this thesis we have used numerical and analytical tools in order to analyze different static and dynamic patterns of cilia, and compared them to experiments performed by our collaborators. Below we summarize the main findings.

\begin{itemize}
\item In chapter 3 we developed a planar theory for beating cilia by balancing hydrodynamic and mechanical forces. Importantly, we incorporated sliding and normal cross-linkers at the bulk of the cilium, as well as at the base. We then showed that basal cross-linkers are necessary to produce bending, and that those at the bulk define a fundamental length-scale beyond which the cilium is straight.

\item In chapter 4 we introduced the collective behavior of molecular motors, and showed that they are capable of producing dynamic instabilities such as oscillations. We then proposed a model by which motors can sense sliding, curvature and normal forces. In the static regime sensitivity to curvature and normal forces gives raise to circular arcs, in agreement with experimental evidence in disintegrated cilia. 

\item In chapter 5 we characterized the symmetric beat patterns corresponding to motor regulation by sliding and curvature. We showed that sliding control requires an asymmetry at the boundaries as well as a minimal length to produce wave propagation. This is in contrast to curvature control, which is capable of producing wave propagation in both directions irrespective of boundary conditions and length.  We also characterized the critical beats under several conditions, and obtained non-linear beat patterns by numerically solving the dynamic equations.

\item In chapter 6 we showed that asymmetrically beating cilia, such as those in {\it Chlamydomonas}, can be regulated via normal forces. We indeed showed that the resulting beat patterns obtained by curvature and normal force control are in good agreement with those measured for wild-type cilia. Furthermore, with similar parameters curvature control also produced beat patterns analogous to those of  the symmetric mbo2 {\it Chlamydomonas} mutant. Because these cilia are very short, sliding control does not produce  wave propagation.
\end{itemize}

To conclude, our work shows that  curvature, sliding, and normal forces can all regulate beat patterns. However, important differences exist on the conditions under which any of these mechanisms can operate: sliding requires long cilia and a constraint at one end, and normal force requires an asymmetry. Furthermore, the resulting beat patterns can also be very different. In the case of the wild-type {\it Chlamydomonas} cilia in particular, we have good evidence that curvature or normal forces can regulate its beat, while sliding can not. 

\section{Future work}
The results in this thesis suggest new lines of research to gain further insight into the regulatory mechanisms behind ciliary beat. Perhaps the most prominent unanswered question is the origin of the static asymmetry of the {\it Chlamydomonas} axoneme. While in chapter 4 we suggested a mechanism capable of producing circular arcs for disintegrated axonemes, the intact cilium is a complicated structure whose full understanding requires a three-dimensional treatment. Given the new light shed here about ciliary beat by the incorporation of normal forces and asymmetries, it is a necessary next step to develop a three-dimensional model of the cilium which includes radial and transverse elastic elements which may regulate the static bend.

Such a model could also point towards mechanisms by which motors can sense curvature or normal forces. For instance, radial stresses or the three-dimensional extension of the normal force here studied  could be the means by which curvature regulation arises. Furthermore, new imaging techniques could allow to indirectly observe the strains conjugate to such stresses. Together with a technique to observe the waves of dynein activity, this could provide direct evidence on the mechanism of motor regulation.

In this work we have used a coarse grained description of the motor force, which proved rich enough to allow for distinctive selection of beating modes, as well as amplitude and frequency selection. If different beating modes are experimentally characterized and then compared to the models here introduce, this could provide new information about the motor model. New patterns can be accessed changing experimental conditions as temperature, ATP, and viscosity; or characterizing mutants other than mbo2. Mutants with long flagella are particularly interesting, given the important role of the length in determining the beat pattern. 

\appendix

\chapter{Calculus of variations and test of resistive force theory}

{In this appendix we complete} the calculus of variations outlined in \ref{sec:stat} and \ref{sec:fluiddyn}. We also show how Resistive Force Theory is a valid model for axoneme-fluid interactions by comparing theoretical predictions of RFT to experimental measurements.

\section{Variations of mechanical and rayleigh functionals}
We begin by giving the full expressions for the curvature of filament A. First note that
\begin{align}
\dot{{\bf r}}_{\rm A} &=\left(1-\frac{a}{2}\dot{\psi}\right){\bf t} + \frac{\dot{a}}{2}{\bf n}\nonumber\\
\ddot{{\bf r}}_{\rm A} &=-\left(\dot{a}\dot{\psi} +\frac{a}{2}\ddot{\psi}\right){\bf t}+\left((1-\frac{a}{2}\dot{\psi})\dot{\psi} +\frac{\ddot{a}}{2} \right){\bf n}\quad.
\label{eq:appendixra}
\end{align}
The tangent vector of filament A is defined as
\begin{align}
{\bf t}_{\rm A} = \frac{\partial {\bf r}_{\rm A}}{\partial s_{\rm A}}=\frac{\dot{{\bf r}}_{\rm A}}{|\dot{{\bf r}}_{\rm A}|}\quad,
\end{align}
which can be readily calculated using Eqs.~\ref{eq:appendixra}. It is then straight-forward to calculate the normal vector ${\bf n}_{\rm A}$ which is normalized and perpendicular to ${\bf t}_{\rm A}$. Once normal and tangent vectors are known, the curvature is simply
\begin{align}
C_{\rm A}={\bf n}_{\rm A}\cdotp\frac{\partial {\bf t}_{\rm A}}{\partial s_{\rm A}}=\frac{{\bf n}_{\rm A}\cdotp\dot{ {\bf t}}_{\rm A} }{|\dot{{\bf r}}_{\rm A}|}\quad,
\end{align}
which again can be calculated using Eqs.~\ref{eq:appendixra}. Replacing $a\to-a$ we obtain the expression of $C_{\rm B}$, and expanding to leading (second) order we have Eq.~\ref{eq:curvAB}. 

To derive the variations of the energy functional, we begin by writing Eq.~\ref{eq:workfunc}  in the alternative form
\begin{align}
G &= \int_0^L\left[\frac{\kappa}{2}\dot{\psi}^2+\frac{\kappa}{2}\left(\frac{\ddot{a}}{2}\right)^2+ \frac{k}{2}\Delta^2 -f_{\rm m}(s)\Delta+ \frac{k_\perp}{2}(a-a_0)^2 +\frac\Lambda2({\bf \dot{r}}^2-1)\right]\d s+\frac{k_0}{2}\Delta_0^2
\end{align}
for which we have simply used Eq.~\ref{eq:curvAB}. We begin by calculating the variation with respect to the spacing $a$, which is:
\begin{align}
\int_0^L\d s\frac{\delta G}{\delta a}\delta a&=\int_0^L\left\{ \kappa\frac{\ddot{a}}{2}\delta \ddot{a}+k_\perp(a-a_0)\delta a-f \int_0^s\dot{\psi}\delta a\d s'  \right\}\d s\nonumber\\
&=\left[\kappa\frac{\ddot{a}}{2}\delta\dot{a}\right]^L_0-\int_0^Lf\d s\int_0^L\dot{\psi}\delta a\d s+\int_0^L\left\{ -\kappa\frac{\dddot{a}}{2}\delta \dot{a}+k_\perp(a-a_0)\delta a+\dot{\psi}\delta a\int_0^sf\d s'  \right\}\d s=\nonumber\\
&=\left[\kappa\frac{\ddot{a}}{2}\delta\dot{a}\right]^L_0-\left[\kappa\frac{\dot{a}}{2}\delta{a}\right]^L_0+\int_0^L\delta a\left\{ \kappa\frac{\ddddot{a}}{2}+k_\perp(a-a_0)-\dot{\psi}F  \right\}\d s
\end{align}
Where we have defined the net static sliding force density $f=f_{\rm m}-k\Delta$, the integrated force $F=\int_s^Lf\d s'$, we have used Eq.~\ref{eq:generalslide} for relating sliding to spacing, and in the last equality we have introduced a boundary term inside the integral.

We now calculate the variation with respect to the position of the center-line ${\bf r}$. Before proceeding\footnote{For completeness, we also note that $\delta\psi={\bf n}\cdotp\delta{\bf t}$}, we note that $\dot{\psi}={\bf n}\cdotp\dot{{\bf t}}$, so that $\delta\dot{\psi}={\bf n}\cdotp\delta\dot{{\bf t}}={\bf n}\cdotp\delta\ddot{{\bf r}}$. In the first equality we have used that $\dot{{\bf t}}\cdotp\delta{\bf n}=0$, which comes from imposing $|{\bf n}|=1$. After this preamble, we have
\begin{align}
\int_0^L\d s\frac{\delta G}{\delta {\bf r}}\delta {\bf r}&=\int_0^L\left\{ \kappa\dot{\psi}{\bf n}\cdotp\delta\ddot{\bf r}- f\int_0^sa{\bf n}\cdotp\delta\ddot{\bf r}\d s'+\Lambda{{\bf t}}\cdotp\delta{\dot{\bf r}} \right\}\d s=\int_0^L\left\{ (\kappa\dot{\psi}-aF){\bf n}\cdotp\delta\ddot{\bf r}+\Lambda{{\bf t}}\cdotp\delta{\dot{\bf r}} \right\}\d s\nonumber\\
&=\left[(\kappa\dot{\psi}-aF){\bf n}\cdotp\delta\dot{{\bf r}}\right]_0^L+\left[\left(-(\kappa\ddot{\psi}-\dot{a}F+af){\bf n} + \tau{\bf t}\right)\cdotp\delta{\bf r}\right]_0^L\nonumber\\
&+\int_0^L\partial_s\left\{ (\kappa\ddot{\psi}-\dot{a}F+af){\bf n} - \tau{\bf t} \right\} \cdotp\delta{{\bf r}} \d s
\end{align}
Where we have integrated by parts twice, and defined the tension $\tau=\Lambda+\kappa\dot{\psi}^2-aF\dot{\psi}$ following Eq.~\ref{eq:tensiondef}. The variation with respect to the basal sliding $\Delta_0$ is easy to calculate:
\begin{align}
\frac{\delta G}{\delta \Delta_0}\delta\Delta_0=(k_0\Delta_0-\int_0^Lf\d s)\delta\Delta_0\quad .
\end{align}

The variations of the Rayleigh dissipation function given in Eq.~\ref{eq:rayleigh} are calculated in much the same way, with all terms balancing those of the equilibrium variations. For the basal sliding velocity variations we have
\begin{align}
\frac{\delta R}{\delta\partial_t\Delta_0}\delta\partial_t\Delta_0 = (\xi_0\partial_t\Delta_0-F_{\rm i}(0))\delta\partial_t\Delta_0\quad ,
\end{align}
where we have defined the internal friction force $F_{\rm i}(s)=\int_s^Lf_{\rm i}\d s'$, with $f_{\rm i}=-\xi_{\rm i}\partial_t\Delta$ being the internal friction force density. For the spacing velocity the variation is
\begin{align}
\int_0^L\d s\frac{\delta R}{\delta\partial_t a}\delta\partial_t a =\int_0^L\left\{ \xi_\perp\partial_ta-\dot{\psi}F_i  \right\}\delta(\partial_t a)\d s\quad .
\end{align}
Finally, for the velocity of the filament the variation gives
\begin{align}
\int_0^L\d s\frac{\delta R}{\delta \partial_t{\bf r}}\delta \partial_t{\bf r}&=\left[-aF_{\rm i}{\bf n}\cdotp\delta\partial_t\dot{{\bf r}}\right]_0^L+\left[\left(-(-\dot{a}F_{\rm i}+af_{\rm i}){\bf n} + \tau_{\rm i}{\bf t}\right)\cdotp\delta\partial_t{\bf r}\right]_0^L\nonumber\\
&+\int_0^L\left((\xi_{\rm n}{\bf n}{\bf n}+\xi_{\rm t}{\bf t}{\bf t})\cdotp\partial_t{\bf r} +\partial_s\left\{  (-\dot{a}F_{\rm i}+af_{\rm i}){\bf n} - \tau_{\rm i}{\bf t} \right\} \right)\cdotp\delta\partial_t{{\bf r}} \d s
\end{align}
where we have defined $\tau_{\rm i}=-aF_{\rm i}\dot{\psi}$ as the dissipative component of the tension.

\section{Plane wave and hydrodynamic test}
In section \ref{sec:experchlam} it was stated that the axonemal beat of the {\it Chlamydomonas} axoneme can be understood as an angular  plane wave (see Fig.~\ref{fig:takehome}), that is
\begin{align}
\label{eq:simpshape}
\psi(s,t)=f_{\rm rot}t+C_0s +A\cos\left(2\pi(\nu t -s/\lambda)\right)
\end{align}
where $C_0$ is the mean curvature, $f_{\rm rot}$ the global rotational frequency , $\lambda$ is the wavelength, $A$ is the amplitude and $\nu$ the frequency. We now use this simple description to verify the model of  axoneme-fluid interaction introduced in \ref{sec:fluiddyn} (Resistive Force Theory). In RFT the fluid force is given by
\begin{align}
\label{eq:rftappen}
{\bf f}^{\rm fl} = -( {\bf nn}\xi_{\rm n} + {\bf tt}\xi_{\rm t})\cdotp\partial_t{\bf r}\quad ,
\end{align}
with ${\bf r}(s,t)$ the position of the axoneme over time which relates to the shape $\psi(s,t)$ via 
\begin{align}
{\bf r}(s,t) = {\bf r}_0(t) +{\bf R}(s,t)={\bf r}_0(t) +\int_0^s\begin{pmatrix}\cos(\psi(s',t))\\\sin(\psi(s',t))\end{pmatrix}\d s'\quad ,
\end{align}
with ${\bf r}_0(t)$ being the trajectory of the basal end, and ${\bf R}(s,t)$ the shape of the axoneme. Note that since $\dot{{\bf r}}=\dot{{\bf R}}$, the shape alone defines the tangent and normal vector fields, independent of the position of the base ${\bf r}_0$.

Since in RFT the inertia of the fluid is neglected, the sum of all forces exerted on it must vanish, and we thus have
\begin{align}
\int_0^L{\bf f}^{\rm fl}(s,t)\d s=-\int_0^L\left({\bf nn}\xi_{\rm n} + {\bf tt}\xi_{\rm t}\right)\cdotp(\partial_t{\bf r}_0(t)+\partial_t{\bf R}_0(s,t))\d s=0\quad .
\end{align}
Taking the tangent angle $\psi(s,t)$ given by Eq.~\ref{eq:simpshape} the only unknowns on this equation are the components of the basal velocity $\partial_t{\bf r}_0(t)$. We can thus numerically solve this equation in order to obtain the velocity of the base at each time-step, and reconstruct the axonemal trajectory.

\begin{figure}[!ht]
\centering
\includegraphics[width=\textwidth] {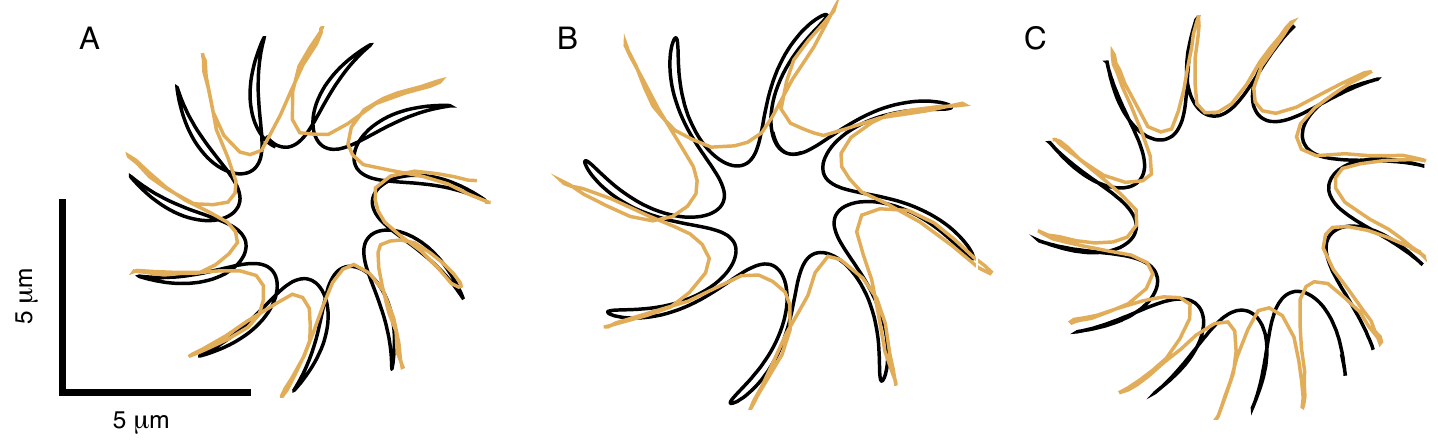}
\caption{\textbf{Reconstructed vs tracked trajectories.}. Tracked trajectories (in yellow) and reconstructed trajectories (in black) using the plane wave approximation of the beat (in Eq.~\ref{eq:simpshape}) and a RFT model of the fluid (Eq.~\ref{eq:rftappen}). The values used in the panels $\{A,B,C\}$ are: $C_0=\{0.26,0.22,0.28\}\um ^{-1}$, $\lambda=\{11.2,11.0,10.1\}\um$, $L=\{12 ,11,11\}\um$, $f_{\rm rot}=\{7.7,5.0,5.9\}\Hz$, $\nu=\{62.5,62.7,65.1\}\Hz$, and $A=\{0.67,0.54,0.66\}$. Figure adapted from \cite{geyer_characterization_2013}.}
\label{fig:hydrover}
\end{figure}

In Fig.~\ref{fig:hydrover} three comparisons of the reconstructed trajectories (in red) to the actual trajectories (in black) are given. Notice the very good agreement, given that the shape information used corresponds to a plane wave approximation of the beat. This neglects amplitude profile, higher harmonics, and fluctuations. Fig.~\ref{fig:hydrover} thus confirms not only that RFT applies to {\it Chlamydomonas} axonemes,  but that the plane wave approximation on the axonemal beat discussed in \ref{sec:experchlam} is accurate.

%

\chapter{Critical modes and numerical methods}

{In this appendix detail different beating modes referred to in chapter 5.} We show the beat patterns, amplitude and phase profiles of several critical beats, as well as non-linear beats obtained numerically. We also describe the numerical method used to solve the dynamical system, and  characterize the dynamical transition. 

\section{Characterization of critical modes}
The equation which characterizes the critical beats of symetric cilia regulated via sliding or curvature is
\begin{align}
i{\omega}\psi=-\ddddot{\psi}+{\chi}_{\rm c}\ddot{\psi}+\bar{\beta}_{\rm c}\dddot{\psi}\quad,
\end{align}
where ${\chi}_{\rm c}$ and $\beta_{\rm c}$ are the rlinear response coefficients to sliding and curvature respectively, and we are using the normalization rules in Eqs.~\ref{eq:dimrules}  to make all quantities dimensionless. The general solution to this linear equation is
\begin{align}
\label{eq:homo}
\tilde{\psi}_1(s) = \sum_{\alpha=1}^4A_{\alpha}\exp(k_{\alpha} s) 
\end{align}
where $\{A_{\alpha}\}$ are the amplitudes determined by the boundary conditions, and $\{k_{\alpha}\}$ are the solutions to the characteristic equation, which is:
\begin{align}
i\eta=-k^4_\alpha+\chi_{\rm c}k^2_\alpha+\beta_{\rm c}k^3_\alpha\quad.
\end{align}

To obtain the coefficients $\{A_\alpha\}$ we insert Eq.~\ref{eq:homo} into the boundary conditions, which gives a linear set of equations in $\{A_{\alpha}\}$. This linear system will  depend on $\omega$, $\chi_{\rm c}$ and $\beta_{\rm c}$ through the solutions of the characteristic polynomial $k_\alpha(\omega,\chi_{\rm c},\beta_{\rm c})$. It will also depend on the basal sliding $\Delta_0$, which makes the system appear homogeneous. However, using the sliding force balance
\begin{align}
\chi_0\Delta_0=\int_0^1\left[\chi_{\rm c}\Delta(s)+\beta\dot{\psi}(s)\right]\d s
\end{align}
together with $\Delta(s)=\Delta_0+\psi(s)-\psi(0)$ we can eliminate the basal sliding dependence, so that we have a homogeneous system of four equations. The determinant of the coefficient matrix of this system $\Gamma$ is a function of the set of seven parameters $\{\omega,\chi',\chi'',\beta',\beta'',\chi_0',\chi_0''\}$ (where primes and double primes denote real and imaginary part), and it has to be null to obtain non-trivial solutions. Given five of these parameters, the condition $\Gamma=0$ is a complex equation that gives a discrete spectrum of solutions for the other two. For example, in sliding control we typically fix all parameters besides $\chi'$ and $\chi''$, which we determine from the condition $\Gamma=0$. In this case we  order the modes according to $|\chi_n|<|\chi_{n+1}|$. In curvature control it is $\beta'$ and $\beta''$ which are determined from the determinant condition, and the ordering of the modes follows $|\beta_n|<|\beta_{n+1}|$.

\begin{figure}[!ht]
\centering
\includegraphics{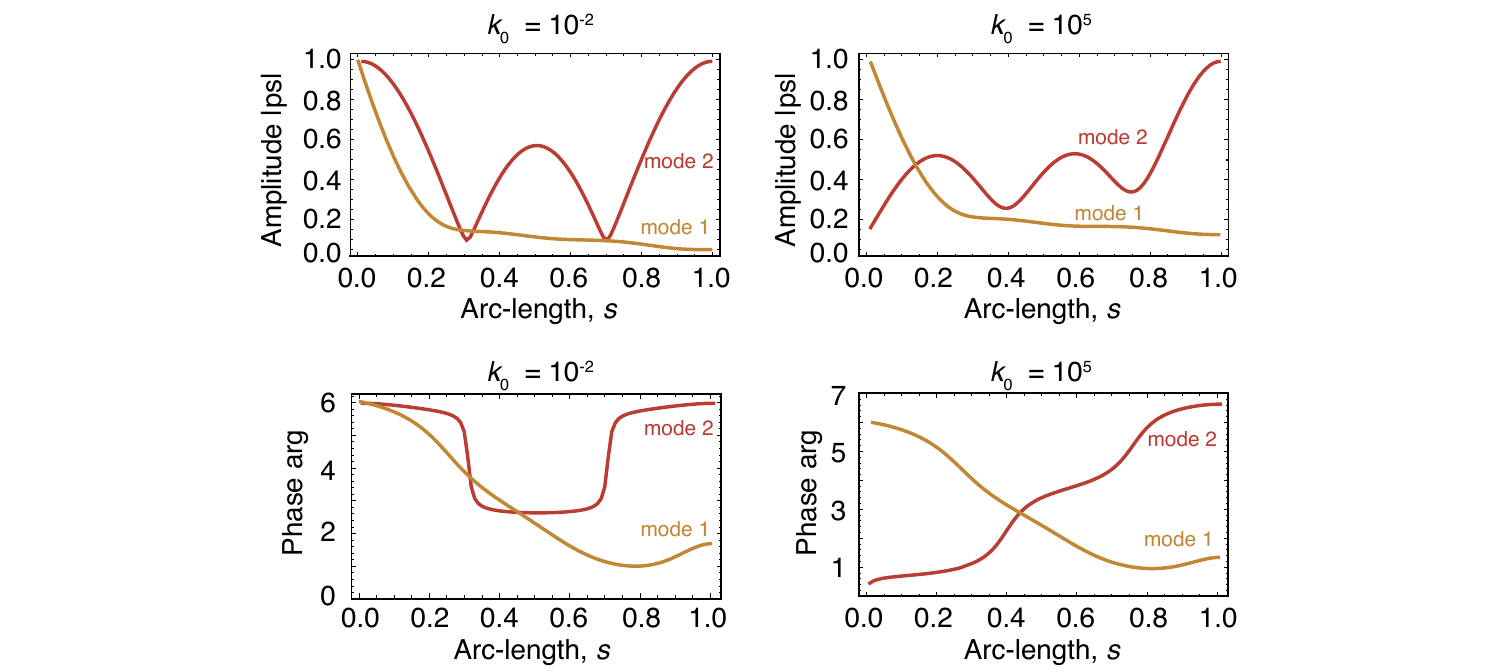}
\caption{\textbf{Modes 1 and 2 of sliding control for free ends.} As one can readily see, for a large basal compliance (panels to the right) both modes are directional and have opposing directions of wave propagation. As the basal compliance vanishes mode 2 looses its directionality and becomes a standing mode, while mode 1 remains directional. Parameters as in Fig.~\ref{fig:freesols}. }
\label{fig:slidingbasal}
\end{figure}

In Fig.~\ref{fig:slidingbasal} we show the amplitude and phase profile of the first two modes of a freely swimming cilium with a small and a high basal compliance. As one can see mode 1 is a directional mode, which shows strong wave propagation from base to tip for high and low basal compliance alike (yellow curves, panes A and C for small basal stiffness and B and D for large). This is contrast with mode 2 (red curves), which shows a dramatic change as the basal compliance becomes smaller: while for a  large value of the basal compliance mode 2 show wave propagation from tip to base (panels B and D), when the basal compliance is small it becomes a standing mode with no wave propagation (panels A and C). The length in all cases is  $L/\ell_0\approx10$ (typical of {\it Bull Sperm}), guaranteeing that wave propagation is possible.

\begin{figure}[!ht]
\centering
\includegraphics{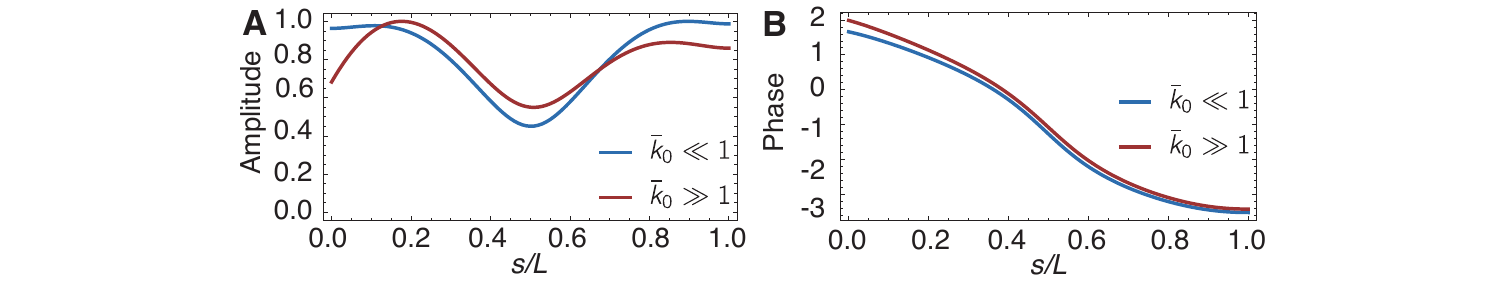}
\caption{\textbf{Mode 1 of curvature control for free ends.}  A large change in the basal stiffness has a minimal effect on the beat pattern of this mode. Parameters as in Fig.~\ref{fig:freesolscc}, with low and high basal compliances corresponding to  $k_0=0$ and $k_0=10^4\,\pN\cdotp\um^{-1}$.}
\label{fig:curvaturebasal}
\end{figure}

The basal compliance has a less dramatic effect in beats controlled by curvature. In Fig.~\ref{fig:curvaturebasal} we compare the amplitude and phase profiles of the first unstable mode under curvature control for a high (in red) and low (in blue) basal compliance. As one can see, while the increase in basal compliance makes the amplitude decrease near the base,  the effect is much smaller than for sliding controlled beat patterns. 

\begin{figure}[!ht]
\centering
\includegraphics{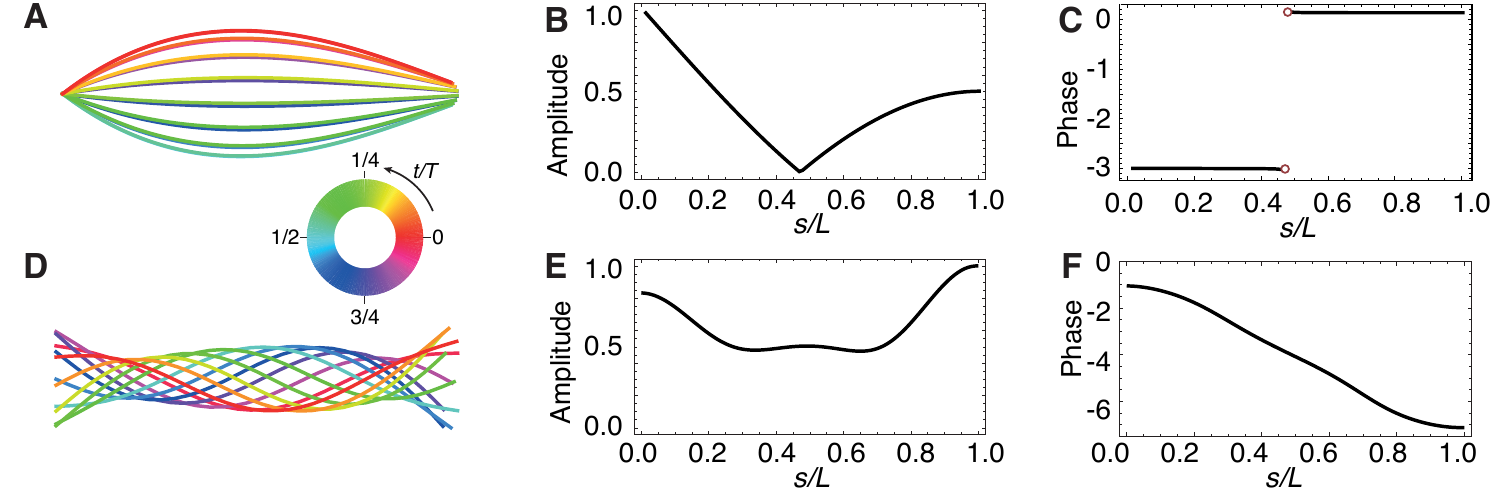}
\caption{\textbf{Beat patterns of long and short cilia.} \textbf{A, B \& C.} Beat pattern of a short ($L/\ell_0\approx3$) freely swimming cilium regulated by sliding. In agreement with Fig.~\ref{fig:sliscaling} no wave propagation occurs, unlike the beat of long cilia (see Fig.~\ref{fig:detailedfree}). \textbf{D, E \& F.} Beat pattern of a long ($L/\ell_0\approx15$) freely swimming cilium regulated by curvature. Wave propagation is strong, just as for short cilia (see Fig.~\ref{fig:sliscalingcc})}
\label{fig:lengths}
\end{figure}

As stated in the main text, all sliding control modes loose directionality when the length of the cilium $L$ becomes comparable to the characteristic length $\ell_0$. In Fig.~\ref{fig:lengths} A, B and C we have a time trace as well as amplitude and phase profiles of mode 1 of a freely beating cilium under sliding control. Although the basal compliance is high such that all modes can be directional, due to the small length of the cilium $L/\ell_0\approx3$ they show no wave propagation (see A, and the flat phase profile in C). Cilia regulated via curvature however can exhibit wave propagation for short (see Fig.~\ref{fig:freesolscc}) as well as long lengths (see Fig.~\ref{fig:lengths} D, E and F).

\section{Non-linear dynamics of beating cilium}

\begin{figure}[!ht]
\centering
\includegraphics[width=\textwidth] {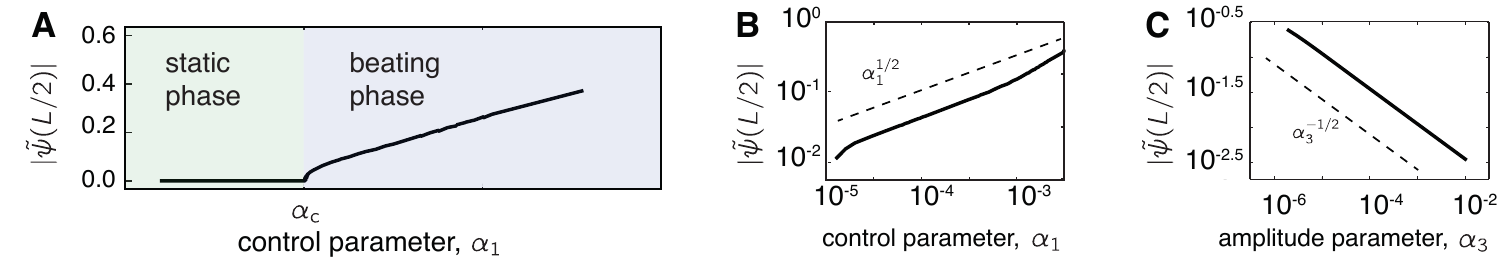}
\caption{\textbf{Numerical characterization of bifurcation.}  \textbf{A.} For values of the control parameter smaller than the critical one, the system is stable and there are no oscillations (green area). For values higher than the critical, finite amplitude oscillations  appear. \textbf{B.} The amplitude increases as the critical parameter grows as a power law. \textbf{C.} The saturation parameter also scales with the amplitude as a power law. Results obtained by performing a series of simulations for a pivoting cilium regulated via sliding.}
\label{fig:hopf}
\end{figure}

The onset of the oscillatory regime in our description of cilia occurs via a Hopf bifurcation. That is, as the control parameter $\Omega$ crosses some critical value $\Omega_{\rm c}$, the system starts oscillating with small amplitude at a critical frequency $\omega_{\rm c}$. We characterized the properties of this Hopf bifurcation by numerically integrating the dynamical system for a range of values of the control parameter $\Omega=\alpha_1$. As shown in Fig.~\ref{fig:hopf} A the oscillatory behavior can be understood as a dynamic phase transition. There are two associated scaling laws near the critical point: the amplitude of the beat increases as $(\alpha_1-\alpha_{\rm c})^{1/2}$, in Fig.~\ref{fig:hopf} B; and it decreases as $\alpha_3^{-1/2}$, in panel Fig.~\ref{fig:hopf} C

\begin{figure}[!ht]
\centering
\includegraphics[width=\textwidth] {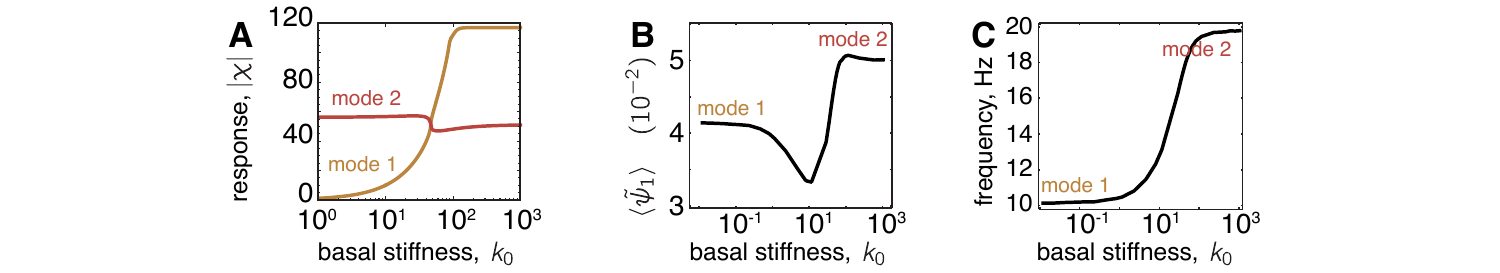}
\caption{\textbf{Mode swapping as a function of the basal stiffness.} \textbf{A.} As the basal stiffness changes the two modes swap their order: for high basal compliance $|\chi_1|>|\chi_2|$, for low $|\chi_1|<|\chi_2|$. \textbf{B.} As one or the other modes are gets activated there is a jump in the amplitude of the beat. While this jump is  discrete, the two modes are very similar near the cross-over point so the jump appears continuous. \textbf{C.} As the modes change, so does their characteristic frequency.}
\label{fig:flip}
\end{figure}

It was seen in chapter X that modifying the basal stiffness could produce a significant change in the beat pattern of a clamped cilium. In Fig.~\ref{fig:flip} A  we analyze how the motor response of the first two modes changes as a function of the basal stiffness $k_0$. The crucial point is that at some critical value of $k_0$ their order flips. Thus the mode which is second at a low basal stiffness, becomes first at a high basal stiffness. As a consequence the mode gets excited. By numerically computing a series of simulations of the non-linear theory we indeed observed a change in the beat patterns of the cilium, which we represent in Fig.~\ref{fig:flip} B and C by the amplitude and frequency of the beat respectively. Interestingly, this change is not abrupt as should be expected. The reason is that near the critical point both modes become very similar, although not equal, thus the discrete jump to be expected in all quantities is very much smoothed out.

\section{Numerical methods}
To solve the dynamical system defined by the ciliary mechanics and the motor dynamics in the time domain, we used a custom made numerical algorithm. It consists of an adaptation of the IMEX finite differences algorithm in \cite{tornberg_simulating_2004}, but including a predictor-corrector iteration loop \cite{_numerical_2007}. We now outline the key steps of the algorithm.

For concreteness, we write down the equations that were solved for the case of a clamped cilium where motors were regulated via sliding velocity. The dynamic equations are in this case:
\begin{align}
\green{\partial_t\psi}&=\xi_{\rm n}^{-1}(-\kappa\green{\ddddot{\psi}} -a_0 \ddot{f} +   \blue{\dot{\psi}\dot{\tau}} + \blue{\tau\ddot{\psi}}) +\xi_{\rm t}^{-1} \blue{\dot{\psi}}(\kappa\blue{\dot{\psi}\ddot{\psi}}+a_0f\blue{\dot{\psi}}+\blue{\dot{\tau}})\label{eq:dynangapp}\\
\frac{\xi_{\rm n}}{\xi_{\rm t}}\green{\ddot{\tau}}-\green{\dot{\psi}}^2\green{\tau} &= -\green{\dot{\psi}}(\kappa\green{\dddot{\psi}}+ a_0 \green{\dot{f}})-\frac{\xi_{\rm n}}{\xi_{\rm t}}\partial_s[\green{\dot{\psi}}(\kappa\green{\ddot{\psi}} + a_0 \green{f})]\label{eq:tensapp}\\
f&=\blue{f_{\rm m}}-k\blue{\Delta}-\xi_{\rm i}\green{\partial_t\Delta}\label{eq:sliforapp}\\
\green{\partial_t f_{\rm m}} &= -\frac{1}{\tau_0}(\blue{f_{\rm m}} - \alpha_1\blue{\partial_t \Delta} + \alpha_3(\blue{\partial_t\Delta})^3)\\
\xi_0\green{\partial_t\Delta_0}&=\int_0^L\blue{f(s)}\d s-k_0\blue{\Delta_0}\label{eq:basalapp}
\end{align}
with $\Delta(s)=\Delta_0+a_0(\psi(s)-\psi(0))$, and boundary conditions
\begin{align}
\kappa\green{\dddot{\psi}(0)}+a_0\green{\dot{f}(0)}-\green{\dot{\psi}(0)\tau(0)}=0\quad&;\quad
\kappa\green{\ddot{\psi}(L)}+ a \green{f(L)}=0\nonumber\\
\green{\psi(0)}=0\quad&;\quad
\green{\dot{\psi}(L)}=0\nonumber\\
\kappa\green{\dot{\psi}(0)\ddot{\psi}(0)}+a_0\green{\dot{\psi}(0)f(0)}+\green{\dot{\tau}(0)}=0\quad&;\quad
\green{\tau(L)}=0\quad.
\end{align}
The meaning of the color scheme will be explained below. The main challenges to numerically integrate these equations are the following: they have a  high order in space, they are non-linear, and they have coupled time-derivatives.

To proceed, we first discretize the angle, tension, and motor force along the arc-length. We also discretized all the differential operators using a second order finite differences scheme, with side derivatives at the boundaries \cite{tornberg_simulating_2004}. To deal with the fourth order spatial derivative we use an IMEX scheme. That is, we treat the fourth order term  implicitly (that is, in the new time step, noted by green) and all the others explicitly (in the old time step, noted by blue). Finally, in order to evolve the system in time, we discretize the time-derivatives with a first order Euler scheme, thus $ \green{\partial_t\psi} = (\green{\psi} -\blue{\psi}) /\d t$ with $\d t$ the discretization time.

There is a draw-back with the previous iteration scheme. In the boundary conditions and tension equations all terms must be evaluated in the new time step. The boundary conditions, which are required to obtain the updated $\green{\psi}$, involve the updated tension $\green{\tau}$. But according to Eq.~\ref{eq:tensapp} to obtain $\green{\tau}$  we need $\green{\psi}$. To solve this issue we perform a loop within each time step, such that initially we use the old tension $\blue{\tau}$ in the boundary conditions, and with it obtain an estimate for the updated angle. We then solve the tension equation, and obtain an estimate for the updated tension. We then use this estimated tension in the boundary conditions to re-estimate the angle, and also re-estimate the tension. This process is iterated until the boundary conditions are satisfied with enough accuracy, and then we move to the next time step. 

The accuracy of the method was ensured by calculating the residual sliding forces (using Eq.~\ref{eq:basalapp}, all quantities evaluated in the same time step) and the differential equation residue (using Eq.~ \ref{eq:dynangapp}, also quantities in the same time step). In all cases the relative errors were kept below $10^{-2}$, with often much smaller errors. The stability of the algorithm changed depending on the region of the parameter space, mainly determined by the length of the cilium which strongly affects the sperm number $(L/\ell_0)^4$. This parameter was changed up to three orders of magnitude. The different types of boundary conditions, and the ratio of time-scales defined by the different viscosities was also a limiting factor in the numerical stability of the algorithm. While several simple calibration methods were used for debugging the code, the most reliable is the direct comparison of the obtained numerical solutions with the analytical ones at the bifurcation point (see for example Figs.~\ref{fig:detailedfree} and \ref{fig:detailedfreecc}).


\clearpage
\addcontentsline{toc}{chapter}{References}
\bibliographystyle{plain}
\bibliography{axonemezotero}

\begin{thebibliography}{100}

\bibitem{CNRS}
Cnrs phototeque.
\newblock \url{http://phototheque.cnrs.fr/}.
\newblock Accessed: 2015-01-05.

\bibitem{_numerical_2007}
{\em Numerical Recipes 3rd Edition: The Art of Scientific Computing}.
\newblock Cambridge University Press, September 2007.

\bibitem{afzelius_human_1976}
B.~A. Afzelius.
\newblock A human syndrome caused by immotile cilia.
\newblock {\em Science}, 193(4250):317--319, July 1976.

\bibitem{aoyama_cyclical_2005}
Susumu Aoyama and Ritsu Kamiya.
\newblock Cyclical interactions between two outer doublet microtubules in split
  flagellar axonemes.
\newblock {\em Biophysical Journal}, 89(5):3261--3268, November 2005.

\bibitem{barber_three-dimensional_2012}
Cynthia~F. Barber, Thomas Heuser, Blanca~I. Carbajal-Gonz{\'a}lez, Vladimir~V.
  Botchkarev, and Daniela Nicastro.
\newblock Three-dimensional structure of the radial spokes reveals
  heterogeneity and interactions with dyneins in chlamydomonas flagella.
\newblock {\em Molecular Biology of the Cell}, 23(1):111--120, January 2012.

\bibitem{bell_models_1978}
G.~I. Bell.
\newblock Models for the specific adhesion of cells to cells.
\newblock {\em Science}, 200(4342):618--627, May 1978.

\bibitem{bessen_calcium_1980}
M.~Bessen, R.~B. Fay, and G.~B. Witman.
\newblock Calcium control of waveform in isolated flagellar axonemes of
  chlamydomonas.
\newblock {\em The Journal of Cell Biology}, 86(2):446--455, August 1980.

\bibitem{bower_n-drc_2013}
Raqual Bower, Douglas Tritschler, Kristyn VanderWaal, Catherine~A. Perrone,
  Joshua Mueller, Laura Fox, Winfield~S. Sale, and M.~E. Porter.
\newblock The n-{DRC} forms a conserved biochemical complex that maintains
  outer doublet alignment and limits microtubule sliding in motile axonemes.
\newblock {\em Molecular Biology of the Cell}, 24(8):1134--1152, April 2013.

\bibitem{brokaw_bend_1971}
C.~J. Brokaw.
\newblock Bend propagation by a sliding filament model for flagella.
\newblock {\em Journal of Experimental Biology}, 55(2):289--304, October 1971.

\bibitem{brokaw_molecular_1975}
C.~J. Brokaw.
\newblock Molecular mechanism for oscillation in flagella and muscle.
\newblock {\em Proceedings of the National Academy of Sciences},
  72(8):3102--3106, August 1975.

\bibitem{brokaw_direct_1989}
C.~J. Brokaw.
\newblock Direct measurements of sliding between outer doublet microtubules in
  swimming sperm flagella.
\newblock {\em Science}, 243(4898):1593--1596, March 1989.

\bibitem{brokaw_bending_1983}
C.~J. Brokaw and D.~J.~L. Luck.
\newblock Bending patterns of chlamydomonas flagella i. wild-type bending
  patterns.
\newblock {\em Cell Motility}, 3(2):131--150, January 1983.

\bibitem{brokaw_flagellar_1972}
Charles~J. Brokaw.
\newblock Flagellar movement: A sliding filament model an explanation is
  suggested for the spontaneous propagation of bending waves by flagella.
\newblock {\em Science}, 178(4060):455--462, November 1972.

\bibitem{brokaw_computer_2002}
Charles~J. Brokaw.
\newblock Computer simulation of flagellar movement {VIII}: Coordination of
  dynein by local curvature control can generate helical bending waves.
\newblock {\em Cell Motility and the Cytoskeleton}, 53(2):103--124, October
  2002.

\bibitem{brokaw_thinking_2009}
Charles~J. Brokaw.
\newblock Thinking about flagellar oscillation.
\newblock {\em Cell Motility and the Cytoskeleton}, 66(8):425--436, August
  2009.

\bibitem{bui_polarity_2012}
Khanh~Huy Bui, Toshiki Yagi, Ryosuke Yamamoto, Ritsu Kamiya, and Takashi
  Ishikawa.
\newblock Polarity and asymmetry in the arrangement of dynein and related
  structures in the chlamydomonas axoneme.
\newblock {\em The Journal of Cell Biology}, 198(5):913--925, September 2012.

\bibitem{burgess_dynein_2003}
Stan~A. Burgess, Matt~L. Walker, Hitoshi Sakakibara, Peter~J. Knight, and
  Kazuhiro Oiwa.
\newblock Dynein structure and power stroke.
\newblock {\em Nature}, 421(6924):715--718, February 2003.

\bibitem{camalet_generic_2000}
S{\'e}bastien Camalet and Frank J{\"u}licher.
\newblock Generic aspects of axonemal beating.
\newblock {\em New Journal of Physics}, 2(1):24, October 2000.

\bibitem{camalet_self-organized_1999}
S{\'e}bastien Camalet, Frank J{\"u}licher, and Jacques Prost.
\newblock Self-organized beating and swimming of internally driven filaments.
\newblock {\em Physical Review Letters}, 82(7):1590--1593, February 1999.

\bibitem{carter_mechanics_2005}
N.~J. Carter and R.~A. Cross.
\newblock Mechanics of the kinesin step.
\newblock {\em Nature}, 435(7040):308--312, May 2005.

\bibitem{carvalho-santos_tracing_2011}
Zita Carvalho-Santos, Juliette Azimzadeh, Jos{\'e}~B. Pereira-Leal, and
  M{\'o}nica Bettencourt-Dias.
\newblock Tracing the origins of centrioles, cilia, and flagella.
\newblock {\em The Journal of Cell Biology}, 194(2):165--175, July 2011.

\bibitem{feistel_three_2006}
Kerstin Feistel and Martin Blum.
\newblock Three types of cilia including a novel 9+4 axoneme on the notochordal
  plate of the rabbit embryo.
\newblock {\em Developmental Dynamics}, 235(12):3348--3358, December 2006.

\bibitem{friedrich_stochastic_2008}
B.~M. Friedrich and F.~J{\"u}licher.
\newblock The stochastic dance of circling sperm cells: sperm chemotaxis in the
  plane.
\newblock {\em New Journal of Physics}, 10(12):123025, December 2008.

\bibitem{friedrich_high-precision_2010}
B.~M. Friedrich, I.~H. Riedel-Kruse, J.~Howard, and F.~J{\"u}licher.
\newblock High-precision tracking of sperm swimming fine structure provides
  strong test of resistive force theory.
\newblock {\em The Journal of Experimental Biology}, 213(8):1226--1234, April
  2010.

\bibitem{friedrich_chemotaxis_2007}
Benjamin~M. Friedrich and Frank J{\"u}licher.
\newblock Chemotaxis of sperm cells.
\newblock {\em Proceedings of the National Academy of Sciences},
  104(33):13256--13261, August 2007.

\bibitem{fujimura_requirement_2006}
Miki Fujimura and Makoto Okuno.
\newblock Requirement of the fixed end for spontaneous beating in flagella.
\newblock {\em Journal of Experimental Biology}, 209(7):1336--1343, April 2006.

\bibitem{gennerich_force-induced_2007}
Arne Gennerich, Andrew~P. Carter, Samara~L. Reck-Peterson, and Ronald~D. Vale.
\newblock Force-induced bidirectional stepping of cytoplasmic dynein.
\newblock {\em Cell}, 131(5):952--965, November 2007.

\bibitem{geyer_characterization_2013}
Veikko Geyer.
\newblock Characterization of the flagellar beat of the single cell green alga
  chlamydomonas reinhardtii.
\newblock December 2013.

\bibitem{goldstein_motility_1981}
S.~F. Goldstein.
\newblock Motility of basal fragments of sea urchin sperm flagella.
\newblock {\em Journal of Cell Science}, 50(1):65--77, August 1981.

\bibitem{gray_propulsion_1955}
J.~Gray and G.~J. Hancock.
\newblock The propulsion of sea-urchin spermatozoa.
\newblock {\em Journal of Experimental Biology}, 32(4):802--814, December 1955.

\bibitem{grill_theory_2005}
Stephan~W. Grill, Karsten Kruse, and Frank J{\"u}licher.
\newblock Theory of mitotic spindle oscillations.
\newblock {\em Physical Review Letters}, 94(10):108104, March 2005.

\bibitem{hakan_bozkurt_morphology_1993}
H.~Hakan~Bozkurt and David~M. Woolley.
\newblock Morphology of nexin links in relation to interdoublet sliding in the
  sperm flagellum.
\newblock {\em Cell Motility and the Cytoskeleton}, 24(2):109--118, January
  1993.

\bibitem{hancock_self-propulsion_1953}
G.~J. Hancock.
\newblock The self-propulsion of microscopic organisms through liquids.
\newblock {\em Proceedings of the Royal Society of London. Series A.
  Mathematical and Physical Sciences}, 217(1128):96--121, March 1953.

\bibitem{heuser_dynein_2009}
Thomas Heuser, Milen Raytchev, Jeremy Krell, Mary~E. Porter, and Daniela
  Nicastro.
\newblock The dynein regulatory complex is the nexin link and a major
  regulatory node in cilia and flagella.
\newblock {\em The Journal of Cell Biology}, 187(6):921--933, December 2009.

\bibitem{hilfinger_chirality_2008}
A.~Hilfinger and F.~J{\"u}licher.
\newblock The chirality of ciliary beats.
\newblock {\em Physical Biology}, 5(1):016003, March 2008.

\bibitem{hoops_outer_1983}
H.~J. Hoops and G.~B. Witman.
\newblock Outer doublet heterogeneity reveals structural polarity related to
  beat direction in chlamydomonas flagella.
\newblock {\em The Journal of Cell Biology}, 97(3):902--908, September 1983.

\bibitem{howard_mechanics_2001}
Jonathon Howard.
\newblock {\em Mechanics of Motor Proteins \& the Cytoskeleton}.
\newblock Sinauer Associates, February 2001.

\bibitem{howard_mechanical_2009}
Jonathon Howard.
\newblock Mechanical signaling in networks of motor and cytoskeletal proteins.
\newblock {\em Annual Review of Biophysics}, 38(1):217--234, 2009.

\bibitem{johnson_pathway_1983}
K.~A. Johnson.
\newblock The pathway of {ATP} hydrolysis by dynein. kinetics of a presteady
  state phosphate burst.
\newblock {\em Journal of Biological Chemistry}, 258(22):13825--13832, November
  1983.

\bibitem{johnson_structure_1983}
K.~A. Johnson and J.~S. Wall.
\newblock Structure and molecular weight of the dynein {ATPase}.
\newblock {\em The Journal of Cell Biology}, 96(3):669--678, March 1983.

\bibitem{johnson_flagellar_1979}
R.~E. Johnson and C.~J. Brokaw.
\newblock Flagellar hydrodynamics. a comparison between resistive-force theory
  and slender-body theory.
\newblock {\em Biophysical Journal}, 25(1):113--127, January 1979.

\bibitem{julicher_cooperative_1995}
Frank J{\"u}licher and Jacques Prost.
\newblock Cooperative molecular motors.
\newblock {\em Physical Review Letters}, 75(13):2618--2621, September 1995.

\bibitem{julicher_spontaneous_1997}
Frank J{\"u}licher and Jacques Prost.
\newblock Spontaneous oscillations of collective molecular motors.
\newblock {\em Physical Review Letters}, 78(23):4510--4513, June 1997.

\bibitem{kagami_translocation_1992}
O.~Kagami and R.~Kamiya.
\newblock Translocation and rotation of microtubules caused by multiple species
  of chlamydomonas inner-arm dynein.
\newblock {\em Journal of Cell Science}, 103(3):653--664, November 1992.

\bibitem{kamimura_high-frequency_1992}
S.~Kamimura and R.~Kamiya.
\newblock High-frequency vibration in flagellar axonemes with amplitudes
  reflecting the size of tubulin.
\newblock {\em The Journal of Cell Biology}, 116(6):1443--1454, March 1992.

\bibitem{kamiya_extrusion_1982}
Ritsu Kamiya.
\newblock Extrusion and rotation of the central-pair microtubules in
  detergent-treated chlamydomonas flagella.
\newblock {\em Cell Motility}, 2(S1):169--173, January 1982.

\bibitem{king_dyneins:_2011}
Stephen~M. King and King~M. Stephen.
\newblock {\em Dyneins: Structure, Biology and Disease}.
\newblock Academic Press, 2011.

\bibitem{lin_structural_2014}
Jianfeng Lin, Kyoko Okada, Milen Raytchev, Maria~C. Smith, and Daniela
  Nicastro.
\newblock Structural mechanism of the dynein power stroke.
\newblock {\em Nature Cell Biology}, advance online publication, April 2014.

\bibitem{lin_building_2011}
Jianfeng Lin, Douglas Tritschler, Kangkang Song, Cynthia~F. Barber, Jennifer~S.
  Cobb, Mary~E. Porter, and Daniela Nicastro.
\newblock Building blocks of the nexin-dynein regulatory complex in
  chlamydomonas flagella.
\newblock {\em Journal of Biological Chemistry}, 286(33):29175--29191, August
  2011.

\bibitem{lindemann_geometric_1994}
Charles~B. Lindemann.
\newblock A "geometric clutch" hypothesis to explain oscillations of the
  axoneme of cilia and flagella.
\newblock {\em Journal of Theoretical Biology}, 168(2):175--189, May 1994.

\bibitem{lindemann_geometric_2002}
Charles~B. Lindemann.
\newblock Geometric clutch model version 3: The role of the inner and outer arm
  dyneins in the ciliary beat.
\newblock {\em Cell Motility and the Cytoskeleton}, 52(4):242--254, August
  2002.

\bibitem{lindemann_structural-functional_2003}
Charles~B. Lindemann.
\newblock Structural-functional relationships of the dynein, spokes, and
  central-pair projections predicted from an analysis of the forces acting
  within a flagellum.
\newblock {\em Biophysical Journal}, 84(6):4115--4126, June 2003.

\bibitem{lindemann_geometric_2007}
Charles~B. Lindemann.
\newblock The geometric clutch as a working hypothesis for future research on
  cilia and flagella.
\newblock {\em Annals of the New York Academy of Sciences}, 1101(1):477--493,
  April 2007.

\bibitem{lindemann_flagellar_2010}
Charles~B. Lindemann and Kathleen~A. Lesich.
\newblock Flagellar and ciliary beating: the proven and the possible.
\newblock {\em Journal of Cell Science}, 123(4):519--528, February 2010.

\bibitem{lindemann_counterbend_2005}
Charles~B. Lindemann, Lisa~J. Macauley, and Kathleen~A. Lesich.
\newblock The counterbend phenomenon in dynein-disabled rat sperm flagella and
  what it reveals about the interdoublet elasticity.
\newblock {\em Biophysical Journal}, 89(2):1165--1174, August 2005.

\bibitem{lindemann_evidence_2007}
Charles~B. Lindemann and David~R. Mitchell.
\newblock Evidence for axonemal distortion during the flagellar beat of
  chlamydomonas.
\newblock {\em Cell Motility and the Cytoskeleton}, 64(8):580--589, August
  2007.

\bibitem{machin_wave_1958}
K.~E. Machin.
\newblock Wave propagation along flagella.
\newblock {\em Journal of Experimental Biology}, 35(4):796--806, December 1958.

\bibitem{minoura_direct_1999}
Itsushi Minoura, Toshiki Yagi, and Ritsu Kamiya.
\newblock Direct measurement of inter-doublet elasticity in flagellar axonemes.
\newblock {\em Cell Structure and Function}, 24(1):27--33, 1999.

\bibitem{mitchell_bend_2004}
David~R. Mitchell and Masako Nakatsugawa.
\newblock Bend propagation drives central pair rotation in chlamydomonas
  reinhardtii flagella.
\newblock {\em The Journal of Cell Biology}, 166(5):709--715, August 2004.

\bibitem{mukundan_motor_2014}
V.~Mukundan, P.~Sartori, V.~F. Geyer, F.~J{\"u}licher, and J.~Howard.
\newblock Motor regulation results in distal forces that bend partially
  disintegrated chlamydomonas axonemes into circular arcs.
\newblock {\em Biophysical Journal}, 106(11):2434--2442, June 2014.

\bibitem{nicastro_molecular_2006}
Daniela Nicastro, Cindi Schwartz, Jason Pierson, Richard Gaudette, Mary~E.
  Porter, and J.~Richard McIntosh.
\newblock The molecular architecture of axonemes revealed by cryoelectron
  tomography.
\newblock {\em Science}, 313(5789):944--948, August 2006.

\bibitem{nonaka_randomization_1998}
Shigenori Nonaka, Yosuke Tanaka, Yasushi Okada, Sen Takeda, Akihiro Harada,
  Yoshimitsu Kanai, Mizuho Kido, and Nobutaka Hirokawa.
\newblock Randomization of left{\textendash}right asymmetry due to loss of
  nodal cilia generating leftward flow of extraembryonic fluid in mice lacking
  {KIF}3b motor protein.
\newblock {\em Cell}, 95(6):829--837, December 1998.

\bibitem{pelle_mechanical_2009}
Dominic~W. Pelle, Charles~J. Brokaw, Kathleen~A. Lesich, and Charles~B.
  Lindemann.
\newblock Mechanical properties of the passive sea urchin sperm flagellum.
\newblock {\em Cell Motility and the Cytoskeleton}, 66(9):721--735, 2009.

\bibitem{phillips_exceptions_1969}
David~M. Phillips.
\newblock Exceptions to the prevailing pattern of tubules (9 + 9 + 2) in the
  sperm flagella of certain insect species.
\newblock {\em The Journal of Cell Biology}, 40(1):28--43, January 1969.

\bibitem{pigino_axonemal_2012}
Gaia Pigino and Takashi Ishikawa.
\newblock Axonemal radial spokes.
\newblock {\em Bioarchitecture}, 2(2):50--58, February 2012.

\bibitem{pigino_comparative_2012}
Gaia Pigino, Aditi Maheshwari, Khanh~Huy Bui, Chikako Shingyoji, Shinji
  Kamimura, and Takashi Ishikawa.
\newblock Comparative structural analysis of eukaryotic flagella and cilia from
  chlamydomonas, tetrahymena, and sea urchins.
\newblock {\em Journal of Structural Biology}, 178(2):199--206, May 2012.

\bibitem{placais_spontaneous_2009}
P.-Y. Pla{\c c}ais, M.~Balland, T.~Gu{\'e}rin, J.-F. Joanny, and P.~Martin.
\newblock Spontaneous oscillations of a minimal actomyosin system under elastic
  loading.
\newblock {\em Physical Review Letters}, 103(15):158102, October 2009.

\bibitem{porter_transient_1983}
M.~E. Porter and K.~A. Johnson.
\newblock Transient state kinetic analysis of the {ATP}-induced dissociation of
  the dynein-microtubule complex.
\newblock {\em Journal of Biological Chemistry}, 258(10):6582--6587, May 1983.

\bibitem{purcell_life_1976}
E.~M. Purcell.
\newblock Life at low reynolds number.
\newblock {\em {AIP} Conference Proceedings}, 28(1):49--64, December 1976.

\bibitem{ravelli_insight_2004}
Raimond B.~G. Ravelli, Beno{\^i}t Gigant, Patrick~A. Curmi, Isabelle Jourdain,
  Sylvie Lachkar, Andr{\'e} Sobel, and Marcel Knossow.
\newblock Insight into tubulin regulation from a complex with colchicine and a
  stathmin-like domain.
\newblock {\em Nature}, 428(6979):198--202, March 2004.

\bibitem{riedelkruse_how_2007}
Ingmar~H. Riedel-Kruse, Andreas Hilfinger, Jonathon Howard, and Frank
  J{\"u}licher.
\newblock How molecular motors shape the flagellar beat.
\newblock {\em {HFSP} Journal}, 1(3):192--208, 2007.

\bibitem{ringo_flagellar_1967}
David~L. Ringo.
\newblock Flagellar motion and fine structure of the flagellar apparatus in
  chlamydomonas.
\newblock {\em The Journal of Cell Biology}, 33(3):543--571, June 1967.

\bibitem{ruhnow_tracking_2011}
Felix Ruhnow, David Zwicker, and Stefan Diez.
\newblock Tracking single particles and elongated filaments with nanometer
  precision.
\newblock {\em Biophysical Journal}, 100(11):2820--2828, June 2011.

\bibitem{s_flexural_1994}
Ishijima S and Hiramoto Y.
\newblock Flexural rigidity of echinoderm sperm flagella.
\newblock {\em Cell structure and function}, 19(6):349--362, December 1994.

\bibitem{sale_substructure_1985}
W.~S. Sale, U.~W. Goodenough, and J.~E. Heuser.
\newblock The substructure of isolated and in situ outer dynein arms of sea
  urchin sperm flagella.
\newblock {\em The Journal of Cell Biology}, 101(4):1400--1412, October 1985.

\bibitem{sanchez_cilia-like_2011}
Timothy Sanchez, David Welch, Daniela Nicastro, and Zvonimir Dogic.
\newblock Cilia-like beating of active microtubule bundles.
\newblock {\em Science}, 333(6041):456--459, July 2011.

\bibitem{sanders_centrin-mediated_1989}
M.~A. Sanders and J.~L. Salisbury.
\newblock Centrin-mediated microtubule severing during flagellar excision in
  chlamydomonas reinhardtii.
\newblock {\em The Journal of Cell Biology}, 108(5):1751--1760, May 1989.

\bibitem{schmidt_insights_2012}
Helgo Schmidt, Emma~S. Gleave, and Andrew~P. Carter.
\newblock Insights into dynein motor domain function from a 3.3-{\r a} crystal
  structure.
\newblock {\em Nature Structural \& Molecular Biology}, 19(5):492--497, May
  2012.

\bibitem{schmitz-lesich_direct_2004}
Kathleen~A. Schmitz-Lesich and Charles~B. Lindemann.
\newblock Direct measurement of the passive stiffness of rat sperm and
  implications to the mechanism of the calcium response.
\newblock {\em Cell Motility and the Cytoskeleton}, 59(3):169--179, November
  2004.

\bibitem{segal_mutant_1984}
R.~A. Segal, B.~Huang, Z.~Ramanis, and D.~J. Luck.
\newblock Mutant strains of chlamydomonas reinhardtii that move backwards only.
\newblock {\em The Journal of Cell Biology}, 98(6):2026--2034, June 1984.

\bibitem{shingyoji_dynein_1998}
Chikako Shingyoji, Hideo Higuchi, Misako Yoshimura, Eisaku Katayama, and Toshio
  Yanagida.
\newblock Dynein arms are oscillating force generators.
\newblock {\em Nature}, 393(6686):711--714, June 1998.

\bibitem{sleigh_propulsion_1988}
Michael~A. Sleigh, John~R. Blake, and Nadav Liron.
\newblock The propulsion of mucus by cilia.
\newblock {\em American Review of Respiratory Disease}, 137(3):726--741, March
  1988.

\bibitem{smith_regulation_1992}
E.~F. Smith and W.~S. Sale.
\newblock Regulation of dynein-driven microtubule sliding by the radial spokes
  in flagella.
\newblock {\em Science}, 257(5076):1557--1559, September 1992.

\bibitem{smith_microtubule_1991}
Elizabeth~F. Smith and Winfield~S. Sale.
\newblock Microtubule binding and translocation by inner dynein arm subtype i1.
\newblock {\em Cell Motility and the Cytoskeleton}, 18(4):258--268, January
  1991.

\bibitem{spungin_dynein_1987}
Bennet Spungin, Jock Avolio, Stuart Arden, and Peter Satir.
\newblock Dynein arm attachment probed with a non-hydrolyzable {ATP} analog:
  Structural evidence for patterns of activity.
\newblock {\em Journal of Molecular Biology}, 197(4):671--677, October 1987.

\bibitem{summers_adenosine_1971}
Keith~E. Summers and I.~R. Gibbons.
\newblock Adenosine triphosphate-induced sliding of tubules in trypsin-treated
  flagella of sea-urchin sperm.
\newblock {\em Proceedings of the National Academy of Sciences},
  68(12):3092--3096, December 1971.

\bibitem{svoboda_force_1994}
Karel Svoboda and Steven~M. Block.
\newblock Force and velocity measured for single kinesin molecules.
\newblock {\em Cell}, 77(5):773--784, June 1994.

\bibitem{taute_microtubule_2008}
Katja~M. Taute, Francesco Pampaloni, Erwin Frey, and Ernst-Ludwig Florin.
\newblock Microtubule dynamics depart from the wormlike chain model.
\newblock {\em Physical Review Letters}, 100(2):028102, January 2008.

\bibitem{tornberg_simulating_2004}
Anna-Karin Tornberg and Michael~J. Shelley.
\newblock Simulating the dynamics and interactions of flexible fibers in stokes
  flows.
\newblock {\em Journal of Computational Physics}, 196(1):8--40, May 2004.

\bibitem{tsuda_optimum_1977}
T.~Tsuda, H.~Noguchi, Y.~Takumi, and O.~Aochi.
\newblock {OPTIMUM} {HUMIDIFICATION} {OF} {AIR} {ADMINISTERED} {TO} a
  {TRACHEOSTOMY} {IN} {DOGS} scanning electron microscopy and surfactant
  studies.
\newblock {\em British Journal of Anaesthesia}, 49(10):965--977, October 1977.

\bibitem{vernon_basal_2004}
Geraint~G. Vernon and David~M. Woolley.
\newblock Basal sliding and the mechanics of oscillation in a mammalian sperm
  flagellum.
\newblock {\em Biophysical Journal}, 87(6):3934--3944, December 2004.

\bibitem{walker_dynamic_1988}
R.~A. Walker, E.~T. O'Brien, N.~K. Pryer, M.~F. Soboeiro, W.~A. Voter, H.~P.
  Erickson, and E.~D. Salmon.
\newblock Dynamic instability of individual microtubules analyzed by video
  light microscopy: rate constants and transition frequencies.
\newblock {\em The Journal of Cell Biology}, 107(4):1437--1448, October 1988.

\bibitem{witman_chlamydomonas_1978}
G.~B. Witman, J.~Plummer, and G.~Sander.
\newblock Chlamydomonas flagellar mutants lacking radial spokes and central
  tubules. structure, composition, and function of specific axonemal
  components.
\newblock {\em The Journal of Cell Biology}, 76(3):729--747, March 1978.

\bibitem{witman_chlamydomonas_2009}
George Witman.
\newblock {\em The Chlamydomonas Sourcebook: Cell Motility and Behavior}.
\newblock Academic Press, March 2009.

\bibitem{woolley_compliance_2008}
D.~M. Woolley, D.~A. Carter, and G.~N. Tilly.
\newblock Compliance in the neck structures of the guinea pig spermatozoon, as
  indicated by rapid freezing and electron microscopy.
\newblock {\em Journal of Anatomy}, 213(3):336--341, September 2008.

\bibitem{woolley_study_2001}
D.~M. Woolley and G.~G. Vernon.
\newblock A study of helical and planar waves on sea urchin sperm flagella,
  with a theory of how they are generated.
\newblock {\em Journal of Experimental Biology}, 204(7):1333--1345, April 2001.

\bibitem{woolley_evidence_2006}
David Woolley, Catarina Gadelha, and Keith Gull.
\newblock Evidence for a sliding-resistance at the tip of the trypanosome
  flagellum.
\newblock {\em Cell Motility and the Cytoskeleton}, 63(12):741--746, December
  2006.

\bibitem{woolley_flagellar_2010}
David~M. Woolley.
\newblock Flagellar oscillation: a commentary on proposed mechanisms.
\newblock {\em Biological Reviews}, 85(3):453--470, August 2010.

\bibitem{yagi_novel_1995}
Toshiki Yagi and Ritsu Kamiya.
\newblock Novel mode of hyper-oscillation in the paralyzed axoneme of a
  chlamydomonas mutant lacking the central-pair microtubules.
\newblock {\em Cell Motility and the Cytoskeleton}, 31(3):207--214, January
  1995.

\bibitem{yagi_vigorous_2000}
Toshiki Yagi and Ritsu Kamiya.
\newblock Vigorous beating of chlamydomonas axonemes lacking central
  pair/radial spoke structures in the presence of salts and organic compounds.
\newblock {\em Cell Motility and the Cytoskeleton}, 46(3):190--199, July 2000.

\bibitem{yagi_identification_2009}
Toshiki Yagi, Keigo Uematsu, Zhongmei Liu, and Ritsu Kamiya.
\newblock Identification of dyneins that localize exclusively to the proximal
  portion of chlamydomonas flagella.
\newblock {\em Journal of Cell Science}, 122(9):1306--1314, May 2009.

\bibitem{zanetti_effects_1979}
N.~C. Zanetti, D.~R. Mitchell, and F.~D. Warner.
\newblock Effects of divalent cations on dynein cross bridging and ciliary
  microtubule sliding.
\newblock {\em The Journal of Cell Biology}, 80(3):573--588, March 1979.

\end{thebibliography}



\end{document}